\documentclass[11pt]{article}
\usepackage{hyperref}
\usepackage{graphicx}
\usepackage{graphics}
\usepackage{dsfont}
\usepackage{epsfig}
\usepackage{amsmath,amssymb,amsthm,amscd}

\newcommand{\Tr}{\textrm{Tr}}
\def\ket{\rangle}
\def\bra{\langle}

\numberwithin{equation}{section}

\begin{document}
\begin{titlepage}
{}~ \hfill\vbox{ \hbox{} }\break

\rightline{IPMU10-0169}

\vskip 3 cm

\centerline{\Large \bf Higher Genus BMN Correlators: }  
\centerline{\Large \bf  Factorization and Recursion Relations }    \vskip 0.5 cm
\renewcommand{\thefootnote}{\fnsymbol{footnote}}
\vskip 30pt \centerline{ {\large \rm Min-xin Huang}
\footnote{minxin.huang@ipmu.jp} } \vskip .5cm \vskip 30pt

\begin{center}
{\em  Institute for the Physics and Mathematics of the Universe (IPMU),  \\ University of Tokyo, Kashiwa, Chiba 277-8582, Japan}\\ [3 mm]
\end{center}

\setcounter{footnote}{0}
\renewcommand{\thefootnote}{\arabic{footnote}}
\vskip 60pt
\begin{abstract}
We systematically study the factorization and recursion relations in higher genus correlation functions of BMN (Berenstein-Maldacena-Nastase) operators in free $\mathcal{N}=4$ super Yang Mills theory. These properties were found in a previous paper by the author, and were conjectured to result from the correspondence with type IIB string theory on the infinitely curved pp-wave background, where the strings become effectively infinitely long. Here we push the calculations to higher genus, provide more clarifications and verifications of the factorization and recursion relations.  Our calculations provide conjectural indirect tests of the AdS/CFT correspondence for multi-loop superstring amplitudes of stringy modes.

\end{abstract}

\end{titlepage}
\vfill \eject


\newpage

\baselineskip=16pt

\tableofcontents

\section{Introduction}
The AdS/CFT correspondence \cite{Maldacena, Gubser, Witten} has been a main direction of research in string theory for more than a decade. The original correspondence relates maximally supersymmetric string theories with gauge theories, for example the type IIB string theory on the $AdS_5\times S^5$ background with $\mathcal{N}=4$ $SU(N)$ super Yang-Mills theory in four dimensions. By now the correspondence has been applied many less symmetric cases, as well as other research areas such as QCD physics and condensed matter physics. The hope is that  gravity in the AdS space can provide useful effective descriptions for strong coupling dynamics which is difficult to deal with theoretically  but can be observed experimentally in QCD physics or condensed matter physics. 

In this paper we pursue the opposite direction, namely we try to study difficult string dynamics using gauge theory. This was made possible in a pp-wave limit of the AdS space, corresponding to the BMN (Berenstein-Maldacena-Nastase \cite{BMN}) sector in the $\mathcal{N}=4$ $SU(N)$ super Yang-Mills theory.  We will further restrict ourself to the case where the pp-wave background is infinitely curved comparing with the string scale, which corresponds to the case that the $\mathcal{N}=4$ super Yang-Mills theory is free. One may wonder how a free gauge theory could describe non-trivial dynamics. We note that it is quite common in string dualities that a weakly coupled theory on one side is equivalent to a strongly coupled theory on the other side, as well as that a free theory on one side is equivalent to an interacting theory on the other side. In fact it is well known that the free string spectrum in the pp-wave background is described by perturbative planar gauge interactions of the BMN operators. So we should not immediately dismiss the notion that the free gauge theory can describe non-trivial string interactions. Since we don't have supports of experiments as in the case of QCD physics or condensed matter physics, and string theory in highly curved Ramond-Ramond backgrounds is not well understood, we will use an indirect approach in our study of the duality. We conjecture that the string theory on the infinitely curved pp-wave background is extremely simplified, and the string interaction amplitudes can be computed by a diagrammatic approach, similar to those in string field theory, which we call the string diagrams. These string diagrams compose of three point couplings of the BMN operators, therefore the correspondence between string theory and gauge theory induces certain factorization and recursion relations \cite{Huang2} for the correlation functions of the BMN operators in gauge theory at higher genus. 

One may wonder whether these factorization and recursion relations are just properties of the gauge theory with no relation to string theory. But the string theory perspective helps to derive these relations, so it is useful. Furthermore, even if it may be possible to systematically prove these relations within gauge theory, as we mentioned we still expect free gauge theory to describe non-trivial string dynamics if the AdS/CFT correspondence is correct. The natural physical observable of string theory in the pp-wave limit would be described by the correlation functions of the BMN operators.   

The main purpose of this paper is to precisely formulate the factorization relation (\ref{factorization}) and test it systematically in many examples. The paper is organized as the followings. In Section \ref{reviewsection} we review the basic ideas of the pp-wave limit, and also try to clarify some puzzling issues in the literature. In Section \ref{planarsection} we study some simple examples of the factorization and recursion relations for planar correlation functions of multi-trace BMN operators. In Section \ref{factorizationsection} we state the precise rules for factorization and recursion relations, which we conjecture to result from the correspondence with string theory on pp-wave. In Section \ref{highergenussection} we test the factorization and recursion relation properties for BMN correlators at higher genus. In Section \ref{othertype} we consider general type of BMN operators with more string excitations.

\section{Reviews of the pp-wave geometry and BMN operators} \label{reviewsection}
We should review some basic terminologies of the pp-wave geometry and the corresponding BMN operators in $\mathcal{N}=4$ super Yang-Mills theory to refresh the memory of the readers, and we also clarity some confusing points in the literature from our perspective. A long time ago, Penrose pointed out when one zooms in a null geodesics of any geometry, one finds a pp-wave type of geometry \cite{Penrose}. This procedure is applied to the well-known $AdS_5\times S^5$ background of the type IIB string theory to obtain the corresponding pp-wave geometry \cite{BMN, Blau}. This is known as the pp-wave limit, or BMN limit. In this paper, we use the notation ``pp-wave geometry" to refer solely to the pp-wave geometry from the Penrose limit of  $AdS_5\times S^5$ background. The pp-wave geometry is rather special because it is the only maximally supersymmetric background of the type IIB string theory besides the flat 10-dimensional Minkowski space and the $AdS_5\times S^5$ space. The metric of the pp-wave geometry  is 
\begin{equation} \label{pp-wavegeometry}
ds^2=-4dx^{+}x^{-}-\mu^2(\vec{r}^{~2}+\vec{y}^{~2})(dx^{+})^2+d\vec{r}^{~2}+d\vec{y}^{~2}
\end{equation}
where the $x^+$, $x^{-}$ are the light cone coordinates and the $\vec{r}$ and $\vec{y}$ parameterize points in the two $R^4$'s  coming from $AdS_5$ and $S^5$. The mass parameter $\mu$ parametrizes the curvature, or the inverse of the length scale of the geometry. There is also a five form Ramond-Ramond background flux $F_{+1234}=F_{+5678}\sim \mu$.

The pp-wave background has the nice property that the free string spectrum is easily solved, which is a difficult problem in the $AdS_5\times S^5$. This can be achieved using the Green-Schwarz formalism in the light cone gauge. The vacuum string state is denoted by $|0,p^+\ket$  where $p^+$ is the light cone momentum. Here we consider the vacuum state with only the light cone momentum $p^+$ and zero momenta in all other 8 directions, because the BMN operators are constructed for these states. We can then construct a general excited string state by acting on it the string creation operators denoted $(a_{n}^I)^\dagger$ for bosonic excitations and $(S^b_n)^\dagger$ for fermionic excitations. Here $I,b=1,2,...8$ label the spacetime directions other than the light cone directions, and $n$ is the excitation level number. We use the notation that positive $n$'s denote the left-moving excitations of the type IIB closed string, negative $n$'s denote the right-moving excitations and $n=0$ denotes supergravity mode. The string states have to satisfy the level matching conditions with equal number of left-moving and right moving excitations. So with one creation operator we only have the supergravity modes $(a_0^I )^\dagger|0,p^+\ket$, and with two creation operators we can create string modes such as  $(a_{-n}^{I_1} )^\dagger(a_n^{I_2} )^\dagger|0,p^+\ket$. 

The mass of these string states have been studied using the Green-Schwarz formalism \cite{BMN, MT}. The vacuum state $|0,p^+\ket$ have a mass proportional to $p^+$, and a creation operator of level $n$ acting on the vacuum state contributes to the string mass 
\begin{eqnarray} \label{stringmass}
M_n =\mu\sqrt{1+\frac{n^2}{(\mu\alpha^\prime p^+)^2}}
\end{eqnarray}
where $\alpha^\prime=l_s^2$ is the string length square. There are two limits one can take. One is  $\mu\alpha^\prime p^+\gg 1$, which means the spacetime curvature parametrized by $\mu$ is very large comparing with the string scale. In this limit the stringy modes are almost degenerate for all excitations, so this is highly stringy regime. The other limit is $\mu\alpha^\prime p^+\ll 1$, which approach the flat space limit, and there is a clear mass gap between stringy modes. 

Now we turn to the BMN operators in $\mathcal{N}=4$ $SU(N)$ super Yang-Mills corresponding to the string states.  The $\mathcal{N}=4$ super Yang-Mills has six real scalar fields in the adjoint representation of $SU(N)$ gauge group and can be written in terms of 3 complex scalars
\begin{equation}
X=\frac{\phi^1+i\phi^2}{\sqrt{2}},~~Y=\frac{\phi^3+i\phi^4}{\sqrt{2}},
~~Z=\frac{\phi^5+i\phi^6}{\sqrt{2}}
\end{equation}
The light cone direction corresponds to one of the complex scalar fields which is picked as $Z$.  The vacuum string state corresponds to the operator $\Tr(Z^J)$, which we call the vacuum operator. Here $J$ is an integer equal to the R-charge of the vacuum operator in the $Z$ direction of the $SO(6)$ R-symmetry group. The AdS/CFT dictionary in the pp-wave/BMN limit relates the parameters of the two theories as 
\begin{eqnarray}
\mu\alpha^\prime p^+ = \frac{J}{g_{\textrm{YM}}\sqrt{N}}, ~~~4\pi g_s=g_{\textrm{YM}}^2
\end{eqnarray} 
where $g_s$  and $g_{\textrm{YM}}$ are the string coupling constant and the Yang-Mills coupling constant. In the BMN limit, $J\sim \sqrt{N}\sim+ \infty$, and we can define two finite dimensionless parameters $\lambda^\prime$ and $g$ as the followings 
\begin{equation} \label{dimensionlessparameters}
\lambda^{'}=\frac{g_{YM}^2 N}{J^2}, ~~~~~
g=\frac{J^2}{N}
\end{equation}

The stringy excitations correspond to inserting the operators $\phi^i$, $D_i$, ($i=1,2,3,4$) for the bosonic excitations and eight  components of the gaugino for the fermionic excitation. In this paper we mostly study bosonic modes with  $\phi^i$ insertions for simplicity. The level number of the string states are encoded by a complex phase. For example, the BMN operator  for one excitation and two excitations are 
\begin{eqnarray}
(a^{I}_n)^\dagger|0,p^{+}\ket &\longleftrightarrow &
\sum_{l=0}^{J-1} \Tr(Z^l \phi^{I}Z^{J-l}) e^{\frac{2\pi i nl}{J}} ,
\nonumber \\
(a^{I_1}_{n_1})^\dagger (a^{I_2}_{n_2})^\dagger |0,p^{+}\ket &\longleftrightarrow  &
\sum_{l_1=0}^{J-1} \sum_{l_2=0}^{J-1} \Tr(Z^{l_1} \phi^{I_1}Z^{l_2-l_1}\phi^{I_2} Z^{J-l_2}) e^{\frac{2\pi i n_1l_1}{J}} e^{\frac{2\pi i n_2l_2}{J}} 
\end{eqnarray} 
We see the BMN operators nicely take into account of the level-matching conditions. Due to the cyclicality of the trace, the operator with one excitation at level $n$ vanishes if $n\neq 0$, and the operators with two excitations level numbers $n_1, n_2$ vanishes if $n_1+n_2\neq 0$. In this paper we will mostly consider the BMN operators with two different $\phi^{I_1}, \phi^{I_2}$ ($I_1\neq I_2$) insertions, and discuss some general types of operators with more excitations in Section \ref{othertype}.  We use the properly normalized BMN operators for up to two excitations as the followings 
 \begin{eqnarray} \label{BMNoperators}
O^{J}  &=& \frac{1}{\sqrt{JN^J}}TrZ^J,   \nonumber \\  
O^{J}_{0} &=& \frac{1}{\sqrt{N^{J+1}}} Tr(\phi^{I} Z^{J}), \nonumber \\
O^J_{-m,m} &=& \frac1{\sqrt{JN^{J+2}}} \sum_{l=0}^{J-1}e^{\frac{2\pi iml}{J}}
Tr(\phi^{I_1} Z^l\phi^{I_2} Z^{J-l}).
\end{eqnarray}
These operators are normalized such that their free planar two point correlation functions are orthonormal to each others. The vacuum operator $O^{J}$ and the operator $O^{J}_0$ with one supergravity mode are half-BPS operators whose conformal dimensions receive no quantum correction and are simply the number of fields in the operators. The operator $O^J_{-m,m}$ with stringy modes are not BPS, but the quantum correction to its conformal dimension can be computed perturbatively for small $\lambda^\prime$.

It is well known that the field theory diagrams can be drawn in the t'Hooft double line notation, and it is generally assumed that in large N duality such as the AdS/CFT correspondence, and the genus of the field theory diagrams correspond to the genus of the string worldsheet. We mentioned there are two dimensionless parameters (\ref{dimensionlessparameters}) in the BMN limit. It turns out that in the BMN limit, the genus of the field theory diagrams is counted by the power of the parameter $g=\frac{J^2}{N}$. There are two limits one can take. Firstly, one can take a planar limit, set $g=0$ and $\lambda^\prime$ finite. This suppresses higher genus diagrams and we have a free string theory and a planar interacting gauge theory. The string spectrum (\ref{stringmass}) can be reproduced by the calculations of the conformal dimensions of the BMN operators, which was done at one loop in \cite{BMN} and exactly to all loops using a $\mathcal{N}=1$ superspace formalism in \cite{SZ}. Secondly, one can also take the limit $\lambda^\prime=0$ and keep the parameter $g=\frac{J^2}{N}$ finite. This makes the gauge theory free, but we have string interactions because of the higher genus diagrams. The second limit would be the focus of this paper.   
 
The free field limit $\lambda^\prime=0$ corresponds to $\mu\alpha^\prime p^+ =+\infty$, so on the string theory side, this is an infinitely curved space. The excited string states are tensionless, infinitely long and have degenerate mass. This raises a puzzling question of what the correct basis of physical states is. In the planar limit, one can see that the BMN operators (\ref{BMNoperators}) are orthogonal to each other when one computes the two point functions at one loop
\begin{eqnarray}
\bra \bar{O}_{-m,m}(x)O_{-n,n}(0)\ket = \frac{\delta_{mn}}{|x|^{2(J+2)}} (1- 2 \lambda^{\prime} n^2 \log(|x|\Lambda)) 
\end{eqnarray}
where $x$ is the 4-d spacetime coordinate. The one loop piece is proportional to $\lambda^\prime$ and gives rise to the anomalous conformal dimensions of the operators. Since the one loop pieces are different for different level numbers $n$ (except $\pm n$, which correspond to the left-moving and right-moving string modes of level $|n|$), an unitary transformation of the BMN basis is not allowed. So we see in the planar limit, the BMN operators form the correct physical basis that have well-defined conformal dimensions, or well-defined mass for the corresponding string states. However, once we turn on finite $g=\frac{J^2}{N}$, the two point functions are no longer orthogonal at free field or one loop level,  
\begin{eqnarray}
\bra \bar{O}_{-m,m}(x)O_{-n,n}(0)\ket = \frac{1}{|x|^{2(J+2)}} (C^{(0)}_{m,n}(g)+ C^{(1)}_{m,n}(g)\lambda^{\prime}  \log(|x|\Lambda)) 
\end{eqnarray}
Since the one loop contribution depends on the space time coordinate $x$, but a transformation of the BMN basis should be independent of the spacetime coordinate, we must simultaneously diagonalize two matrices $C^{(0)}_{m,n}$ and $C^{(1)}_{m,n}$. One would need to simultaneously diagonalize more matrices at higher loop levels. It is not clear how to do this systematically to all orders both in $\lambda^\prime$ and $g$, or whether it is possible to do so. 

Here we will not provide an answer to this puzzling question, but instead go to the free field limit $\lambda^\prime=0$ with finite $g$, where a nice situation emerges. Here the BMN operators are already orthogonal to each others at planar level, so we can use the BMN basis as physical basis and interpret the higher order corrections in $g=\frac{J^2}{N}$ as string loop corrections. One may ask why not diagonalize the free field two point functions to higher orders in $g$, and use the diagonalized basis as the physical basis of states. There are several reasons against doing this. Firstly, we don't know how to find a  diagonal basis for finite $\lambda^\prime$ and $g$, so there is no compelling reason to change the BMN basis for $\lambda^\prime=0$ either. Of course the planar free two point functions would still be orthonormal if we apply any unitary transformation to the BMN basis. But since at the planar limit we know that the BMN operators form the correct physical basis of states, it is possible that this remains the correct physical basis in the different limit of $\lambda^\prime=0$ and finite $g$, and we conjecture this is indeed the case. Secondly, general insights of large N duality tell us that non-planar diagrams in field theory should correspond to string loop interactions. It is not very helpful to simply diagonalize away the higher genus contributions, but it would be natural to study them as string loop contributions. Furthermore, as we see there is a natural interpretation of a single trace operators as a single string state, and we can multiply several single trace BMN operators into a multi-trace operator which corresponds to a multiple-string state. The correlation functions between a single trace BMN operator and a multi-trace BMN operator naturally represent the interaction processes of a single string splitting into several strings, or the reverse processes of several strings joining into one.  Diagonalizing the single trace BMN operators at non-planar level would lose these nice features. Including multi-trace operators for the diagonalization does not help.

We can compare the situation to those studied  in \cite{Balasubramanian, Corley}, where they consider BPS operators of very large R-charge of order $N$,  which is much larger than the $J\sim \sqrt{N}$ in the BMN limit. These operators are interpreted in the string theory side as D-branes, or giant gravitons in AdS space. The two point functions of the BPS operators receive no quantum correction, so here one simply can not rely on conformal dimension to find the correct physical basis of states. Actually, it was found that the Schur polynomials in terms of the scalar field $Z$ diagonalize the free two functions at finite $N$, not just in the planar limit \cite{Corley}. This is possible in this case because these are BPS operators. Since they are D-branes we do not need the higher genus corrections which correspond more naturally to string loop corrections, instead we can describe open strings attached to D-brane by attaching some small operators to the D-brane operator \cite{BHLN, BBFH, Koch}. For some constructions with sophisticated group theoretic aspects see \cite{Brown}. So it is good that a diagonal basis is available for the D-brane operators at finite $N$, however we should not try this to the BMN operators because as we mentioned, they are not BPS and we need the higher genus contributions which can be naturally interpreted in string perturbation theory.   

\section{Planar correlators of multi-trace BMN operators} \label{planarsection}

A nice property in the BMN sector is that there are essentially only two point functions, and no higher point functions. This is because the BMN operator have a large number of $Z$ fields, and in the gauge theory to compute the Feynman diagrams we contract $Z$ with $\bar{Z}$, so there must be equal numbers of $Z$ and $\bar{Z}$. The string states corresponding the operators composed of $Z$ have the same light cone momentum direction, which is opposite to that of the $\bar{Z}$ operators. String interactions are described by splitting or joining the trace of the $Z$ fields, and it is quite unnatural to contract the string excitation insertions in the BMN operators within the same light cone momentum direction, since the number of these insertions are very small comparing to the number of $Z$ fields. So in this paper we will only consider two point correlators, and represent the multiple string states by multi-trace BMN operators, which are just products of single trace BMN operators. We also conjecture that there is no contact interaction in the case we consider here, and the only interactions are cubic interactions, which represent a closed string splitting into two closed strings, or two closed strings joining into a closed string. We should note this is in general not true in closed string theory and there are indeed quartic and higher interactions \cite{GS}, but it is only possible here because we are in the free field limit and the corresponding strings are infinitely long. The space time dependence of the two functions in conformal field theory always takes the form $|x_1-x_2|^{-2\Delta}$. In this paper we are mostly interested in the coefficients of the two functions and for simplicity we will omit the spacetime dependence in the two point functions. 

\begin{figure}
  \begin{center}
  \includegraphics[width=6.5in]{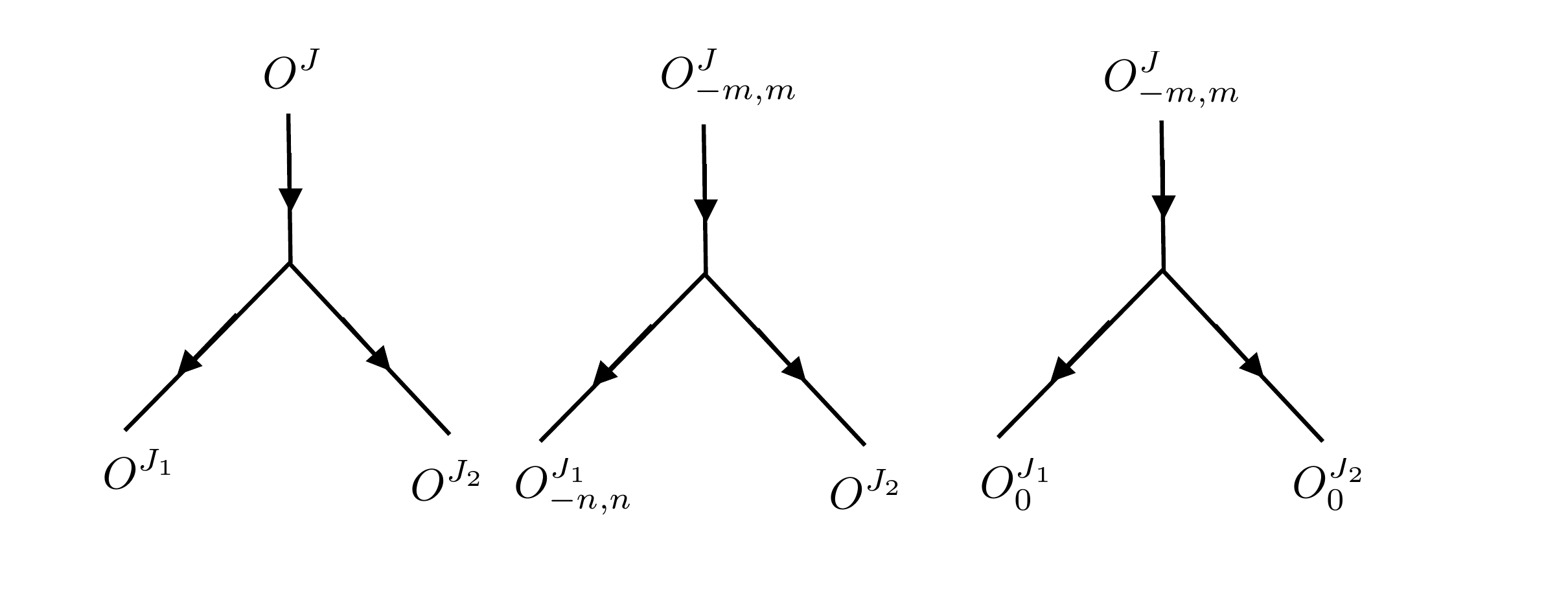} 
\end{center}
\caption{The 3-string vertices are represented by the correlators of a single trace operator with a double trace operator. We draw arrows at each edge of the vertex to represent the incoming or outgoing strings. These diagrams represent a long string splits into two short strings. We can also simply reverse the directions of the arrows to obtain the reverse processes of joining two strings into one.  }  \label{vertices}
\end{figure}

The basic ingredients of the string interaction are the 3-string interactions, which is described by the planar correlators of a single trace BMN operator with a double trace operator. The correlators were firstly computed e.g. in \cite{Constable1}, are listed below for BMN operators up to two insertions
\begin {eqnarray} \label{planar1}
\langle\bar{O}^JO^{J_1}O^{J_2}\rangle&=&\frac{g}{\sqrt{J}}\sqrt{x(1-x)} \nonumber \\
\langle\bar{O}^J_{0}O^{J_1}O^{J_2}_{0}\rangle&=&\frac{g}{\sqrt{J}}x^{\frac{1}{2}}(1-x) \nonumber \\
\langle\bar{O}^J_{00}O^{J_1}_{0}O^{J_2}_{0}\rangle&=&\frac{g}{\sqrt{J}}x(1-x) \nonumber \\
\langle\bar{O}^J_{00}O^{J_1}_{00}O^{J_2}\rangle&=&\frac{g}{\sqrt{J}}x^{\frac{3}{2}}(1-x)^{\frac{1}{2}}  \nonumber \\
\langle\bar{O}^J_{-m,m
}O^{J_1}_{0}O^{J_2}_{0}\rangle&=& - \frac{g}{\sqrt{J}}\frac{\sin^2(\pi mx)}{\pi^2m^2}  \nonumber \\
\langle\bar{O}^J_{-m,m}O^{J_1}_{-n,n}O^{J_2}\rangle&=&\frac{g}{\sqrt{J}}x^{\frac{3}{2}}(1-x)^{\frac{1}{2}}\frac{\sin^2(\pi mx)}{\pi^2 (mx-n)^2}
\end {eqnarray}
Here $g=\frac{J^2}{N}$, $x=\frac{J_1}{J}$,  and it is implicit that $J=J_1+J_2$. We will always assume $x$ is of a generic value so that $mx-n$ is not an integer. It is known that these correlators (\ref{planar1}) can be derived from the interaction vertex of the Green-Schwarz light cone string field theory on pp-wave backgrounds \cite{Huang1, SV}.  They will be the building blocks of string diagrams. In Fig. \ref{vertices} we depict some examples of the 3-string vertices.

\begin{figure}
  \begin{center}
\includegraphics[width=6.5in]{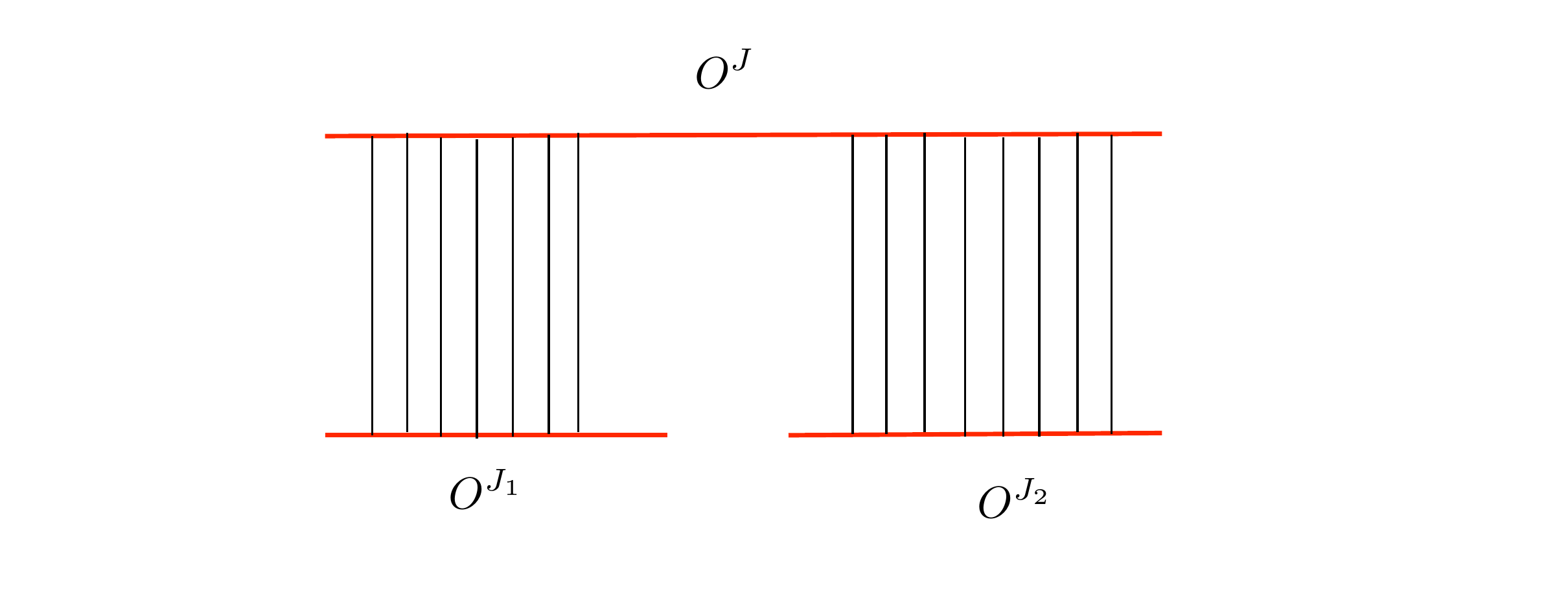} 
\end{center}
\caption{The field theory diagrams for calculating the planar vacuum correlator $\langle\bar{O}^JO^{J_1}O^{J_2}\rangle$. }  \label{3point}
\end{figure}

It turns out that sometimes it is more convenient to use the integral form of the 3-string vertex when we later sum over intermediate states in string diagrams. The 3-string vertices in (\ref{planar1}) are computed by inserting the scalar insertions into the vacuum operator with the BMN complex phases in Fig. \ref{3point}. We can write the 3-string vertices of the stringy mode in the integral form  
\begin{eqnarray} \label{integralform}
\langle\bar{O}^J_{-m,m
}O^{J_1}_{0}O^{J_2}_{0}\rangle&=& \frac{g}{\sqrt{J}} (\int_0^x dy_1 e^{2\pi i m y_1})  (\int_x^1 dy_2 e^{-2\pi i m y_2}) \\ \nonumber 
\langle\bar{O}^J_{-m,m
}O^{J_1}_{-n,n}O^{J_2}\rangle&=& \frac{g}{\sqrt{J}} ( \frac{1-x}{x} )^{\frac{1}{2}}  (\int_0^x dy_1 e^{2\pi i (m-\frac{n}{x})y_1})  (\int_0^x dy_2 e^{-2\pi i (m-\frac{n}{x})y_2}) 
\end{eqnarray}
where $x=\frac{J_1}{J}$. 

To understand the factorization and recursion relations, in this section we first consider two simple cases to exemplify the idea. The two examples are the planar correlators between a single trace and a triple trace operator, and the planar correlators between two double trace operators. 

\subsection{Correlators between a single trace and a triple trace operator}

There are two interesting cases to consider, namely the cases of $\bra \bar{O}_{-m,m}^J O_0^{J_1}O_0^{J_2}O^{J_3}\ket$ and  $\bra\bar{O}_{-m,m}^JO_{-n,n}^{J_1}O^{J_2}O^{J_3}\ket$, where $J=J_1+J_2+J_3$. We will discuss them respectively. We discuss the first case in more details and the second case is similar. The factorization relations we will discuss here also work for the non-stringy cases $\bra \bar{O}^J O^{J_1}O^{J_2}O^{J_3}\ket$ and $\bra \bar{O}_0^J O_0^{J_1}O^{J_2}O^{J_3}\ket$, but these cases are too trivial and we skip them to focus on the interesting cases that there is at least one non-zero stringy mode in the correlator.

\subsubsection{Case one: $\bra \bar{O}_{-m,m}^J O_0^{J_1}O_0^{J_2}O^{J_3}\ket$}

\begin{figure}
  \begin{center}
 \includegraphics[width=6.5in]{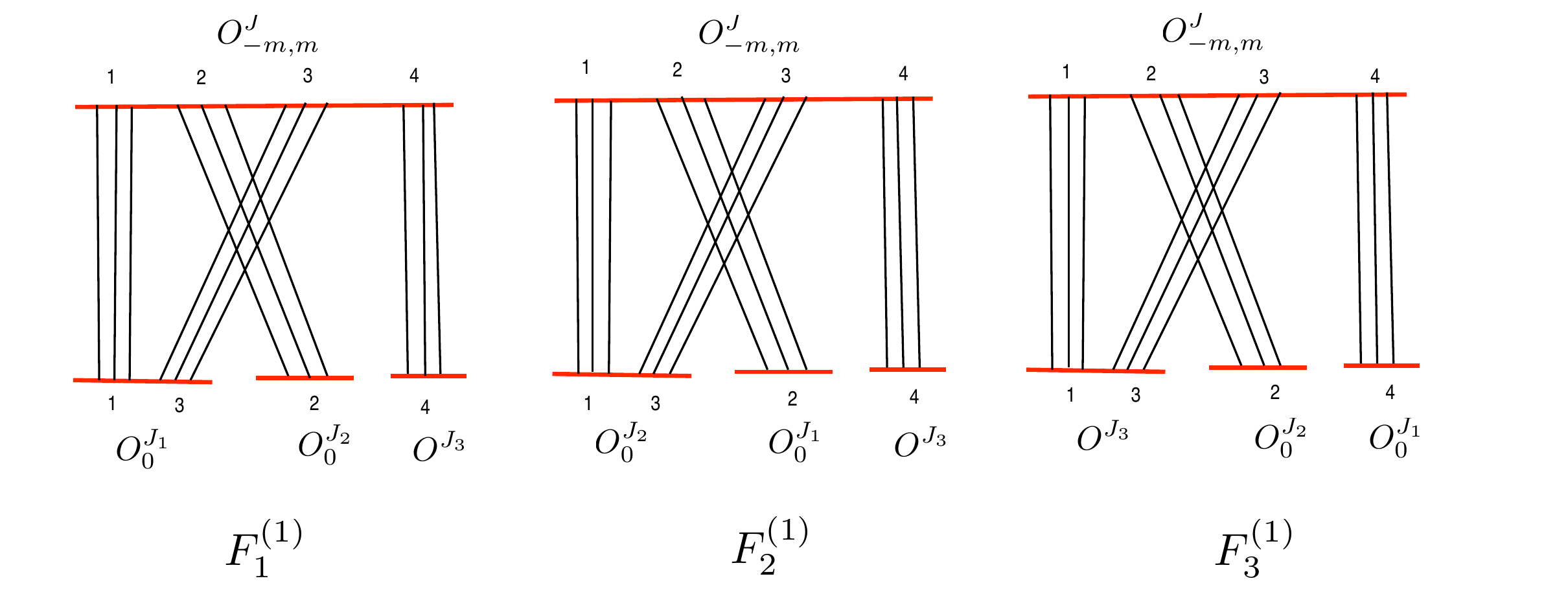} 
 \end{center}
\caption{The field theory diagrams for  $\bra \bar{O}_{-m,m}^J O_0^{J_1}O_0^{J_2}O^{J_3}\ket$. We denote the contributions of the 3 diagrams $F^{(1)}_1$,  $F^{(1)}_2$,  $F^{(1)}_3$.}  \label{F1}
\end{figure}

Here we assume the two scalar insertions in ${O}_{-m,m}^J$ are $\phi^1$, $\phi^2$, and the scalar insertions in $O^{J_1}_0$ and $O^{J_2}_0$ are $\phi^1$ and $\phi^2$ respectively. The field theory diagrams are depicted in Fig. \ref{F1}, and we denote their contributions $F^{(1)}_1$, $F^{(1)}_2$ and $F^{(1)}_3$ respectively. We will always denote the field theory diagram contributions by the letter $F$ with indices to indicate they are field theory contributions. Here the red lines represent the cyclic traces of $Z$ or $\bar{Z}$ fields, and the black lines connecting them represent the Wick contractions between fields. We draw the operators with traces of $\bar{Z}$ on the top and the operators with traces of $Z$ at the bottom. We divide the black lines from a operator into segments, and each segment represent a large number of Wick contractions between the scalar fields. The Wick contraction is actually a double line in t'Hooft double line notation since the fields are in the adjoint representation of $SU(N)$ and are $N\times N$ matrices. The genus of a diagram is the minimal of the genus of a Riemann surface where the diagram can be put on the Riemann surface without intersecting itself, but this is hard to visualize and it is more convenient to just count the power of $N$ with the double line notation in the fat graphs. 

We also label each segment by a number. For convenience we label the segments of the operators on the top by numerical order. The labels of the segments of the the bottom operators are the same as the label of the corresponding segment on the top connected by Wick contraction. If two segments are both adjacent to each other in the same order in a top operator and in a bottom operator, we can always combine them into one segment.  We remind the readers as an example that the segments 4 and 1 in the operator $O^J_{-m,m}$ in the diagrams in Fig. \ref{F1} are considered adjacent because of the cyclicality of the trace operator. So we will always combine segments which are both adjacent on the top and the bottom in the same order into one segments and use the minimal number of labels for a diagram. With the labels we can denote the field theory diagram by a process. For example, we can denote the 3 diagrams Fig. \ref{F1} as  
\begin{eqnarray} \label{F1shortprocess}
F^{(1)}_1&:~&   (1234)\rightarrow  (13)_1(2)_2(4)_3,   \nonumber \\
F^{(1)}_2&:~&   (1234)\rightarrow  (13)_2(2)_1(4)_3, \nonumber \\
F^{(1)}_3 &:~&   (1234)\rightarrow  (13)_3(2)_2(4)_1
\end{eqnarray}
Here each trace operator is denoted by a chain of numbers, and because of the cyclicality of the trace operator, the chain of  $(1,2,\cdots ,n)$ is equivalent to $(n,1,2,\cdots ,n-1)$. We also use a subscript to denote the operator when confusions may arise, for example the subscripts $1,2,3$ above denote the operators $O_0^{J_1}$, $O_0^{J_2}$, $O^{J_3}$. We call the processes above (\ref{F1shortprocess}) the `short processes"  which consist of a initial and final state. We will discuss in a moment how to extend a short process into a ``long process".

These diagrams in Fig. \ref{F1} look non-planar but they are actually planar, or leading order contributions in the BMN limit $J\sim \sqrt{N}\sim +\infty$. To see this for example for the first diagram whose contribution is $F^{(1)}_1$, we can pull the read line represented by the operator $O^{J_2}_0$ above $O^{J}_{-m,m}$. Here  for convenience we draw the incoming operators on the top and outgoing operators at the bottom . Of course the diagram would be also planar if we don't divide the operator  $O^{J_1}_0$ into two segments. The reason for the division is because the division into two segments make the diagram combinatorially  dominant over the one without the division. And this is because each operator has a large number of fields proportional to $J\sim \sqrt{N}$, we get to count an extra factor of $J $ if we divide the operator into two segments. We note this is the most we can do. If we further divide the operator $O^{J_1}_0$ into three segments, or divide another operator $O^{J_2}_0$ or $O^{J_3}$ into two segments. The diagram would become non-planar and has less power of $N$, which is not sufficiently compensated by the extra combinatoric power of $J\sim \sqrt{N}$. We will also explain another approach to determine the genus the field theory diagrams from the corresponding string diagrams, which we find more convenient. 
 
To compute the contributions $F^{(1)}_1$, $F^{(1)}_2$ and $F^{(1)}_3$ in Fig. \ref{F1}, we first count the combinatorics without the scalar insertions.  Let us look at $F^{(1)}_1$ for example. We need to choose the initial field in the traces of $Z$ or $\bar{Z}$ for the beginning of segments (1), (1), (2), (4) in the operators $O^J_{-m,m}$, $O^{J_1}_0$, $O^{J_2}_0$ and $O^{J_3}$, so this contribute a factor $JJ_1J_2J_3$. We also need to choose the beginning field for segment (3) in operator  $O^{J_1}_0$ which would contribute an extra factor of $J_1$. We note that there is an alternative diagram that we Wick contract $O^{J_2}_0$ with the segment (4) in $O^J_{-m,m}$ and $O^{J_3}$ with the segment (2) in  $O^J_{-m,m}$, however this is identical to the original diagram if we cyclically rotate the segment (1) to (3). So we have already accounted for this case when we choose the beginning field in $O^{J_1}_0$ to in one of $J_1$ $Z$'s, and don't need to consider it further. 
Next we put in the scalar insertions $\phi^1$ and $\phi^2$. The scalar field  $\phi^1$ can be inserted into any position in $O^{J_1}_0$, and the corresponding position of $\phi^1$ in $O^J_{-m,m}$ is fixed because we don't want the Wick contraction of the two $\phi^1$ fields to introduce negative powers of $N$. Similarly we can put $\phi^2$ into any position in $O^{J_2}_0$. Suppose the length of the segment (1) is $l$, we put the $\phi^1$ field in position $l_1$ in $O^{J_1}_0$, and $\phi^2$ field in position $l_2$ in $O^{J_2}_0$, then the complex phase factor from $O^J_{-m,m}$ would be $\exp(\frac{2\pi i m}{J}(l_1-l-l_2))$ if $\phi^1$ is in the segment (1), or  $\exp(\frac{2\pi i m}{J}(l_1+J_2-l-l_2))$ if $\phi^1$ is in the segment (3). We also note there are an extra factor of $1/J$, $1/J_1$ and $1/J_2$ for the normalization of the operators $O^J_{-m,m}$, $O^{J_1}_0$ and $O^{J_2}_0$ comparing with the normalization in (\ref{BMNoperators}) because we have allowed the $\phi^1$ fields to be at any position instead of fixing $\phi^1$ to be at the initial position of the trace using the cyclicality of trace in (\ref{BMNoperators}). Finally we also count the power of $N$. Since there is triple trace operator here, we should have an extra factor of $1/N^{2}$ comparing with two point functions of two single trace operators for planar diagrams. Putting things together, we find
\begin{eqnarray}
F^{(1)}_1 = \frac{1}{N^2}(\frac{J}{J_3})^{\frac{1}{2}}\sum_{l=0}^{J-1} [\sum_{l_1=0}^{l}\sum_{l_2=0}^{J_2-1} e^{\frac{2\pi i m}{J}(l_1-l-l_2)}+\sum_{l_1=l+1}^{J_1-1}\sum_{l_2=0}^{J_2-1} e^{\frac{2\pi i m}{J}(l_1+J_2-l-l_2)}]
\end{eqnarray}
We take the BMN limit $J\sim J_i\sim \sqrt{N}\sim \infty$, and the sum become an integral in the BMN limit
\begin{eqnarray}
\sum_{l=0}^{J_1-1}\rightarrow J_1\int_0^1dy, 
~~~\sum_{l_1=0}^{l}\rightarrow J_1\int_0^ydy_1, 
~~\sum_{l_2=0}^{J_2-1}\rightarrow J_2\int_0^1dy_2
\end{eqnarray} 
where we denote $l=J_1y$, $l_1=J_1y_1$, and $l_2=J_2y_2$.  Denoting $x_i=\frac{J_i}{J}$ so that $x_1+x_2+x_3=1$ and $g=\frac{J^2}{N}$,  we can compute 
\begin{eqnarray} \label{F11}
F^{(1)}_1 &=&  \frac{g^2}{J}x_1^2x_2x_3^{\frac{1}{2}} \int_0^1dy (\int_0^ydy_1+ e^{2\pi imx_2} \int_y^1dy_1)\int_0^1dy_2 e^{2\pi im[x_1(y_1-y)-x_2y_2]} \nonumber \\
&=& \frac{g^2}{J} \frac{x_3^{\frac{1}{2}}}{4\pi^3 m^3}[2m\pi x_1(\cos(2\pi mx_2)-1) \nonumber \\ && +\sin(2m\pi x_1)+\sin(2m\pi x_2)+\sin(2m\pi x_3)] 
\end{eqnarray}
Similarly we find 
\begin{eqnarray} \label{F12}
F^{(1)}_2 =  F^{(1)}_1( x_1\leftrightarrow x_2) 
\end{eqnarray}
The computation of $F^{(1)}_3$ is simpler because there is only one integral, and we find 
\begin{eqnarray} \label{F13}
F^{(1)}_3 = - \frac{g^2 }{J}  \frac{x_3^{\frac{1}{2}}}{4\pi^3 m^3}[\sin(2m\pi x_1)+\sin(2m\pi x_2)+\sin(2m\pi x_3)] 
\end{eqnarray}

Now we turn to the string diagram calculations. We mentioned that for each field theory diagram we associate a short process with it. The short process consists of an initial and final state. To extend the short process to a long process, we fill in the intermediate steps. In each step, we can cut one string into two strings, or join two strings into one string. For example, $(1,2,\cdots ,n)\rightarrow (1,2,\cdots ,i)(i+1,\cdots ,n)$ is a process of cutting a string into two. We call the process a long process after we fill in the intermediate steps. For a short process there may be many long processed associated with it, and we will need to find all of them. For example, the short processes in (\ref{F1shortprocess}) which represent the field theory diagrams in Fig. \ref{F1} can be extended to long processes as the following 
\begin{eqnarray} \label{F1longprocess}
F^{(1)}_1:~&&   (1234)\rightarrow (123)(4)_3 \rightarrow  (31)_1(2)_2(4)_3,   \nonumber \\
&& (1234)\rightarrow (341)(2)_2 \rightarrow  (13)_1(4)_3 (2)_2,   \nonumber \\
F^{(1)}_2:~&&   (1234)\rightarrow  (123)(4)_3   \rightarrow (31)_2(2)_1(4)_3, \nonumber \\
&&  (1234)\rightarrow  (341)(2)_1   \rightarrow (13)_2(4)_3 (2)_1, \nonumber \\
F^{(1)}_3 :~&&   (1234)\rightarrow  (123) (4)_1  \rightarrow  (31)_3(2)_2(4)_1, \nonumber \\
&&   (1234)\rightarrow  (341) (2)_2  \rightarrow  (13)_3(2)_2(4)_1
\end{eqnarray}
We note the ordering of the 3 strings in the final state are not important, also we have freely used the cyclicality in the cut and join processes (for example $(13)=(31)$). We see that for each short process there are two ways to fill in the intermediate steps and so there are two long processes associated to each short process. We also write the subscript for the string when it has reached the final state and no longer change in the subsequent steps. 

\begin{figure}
  \begin{center}
  \includegraphics[width=6.5in]{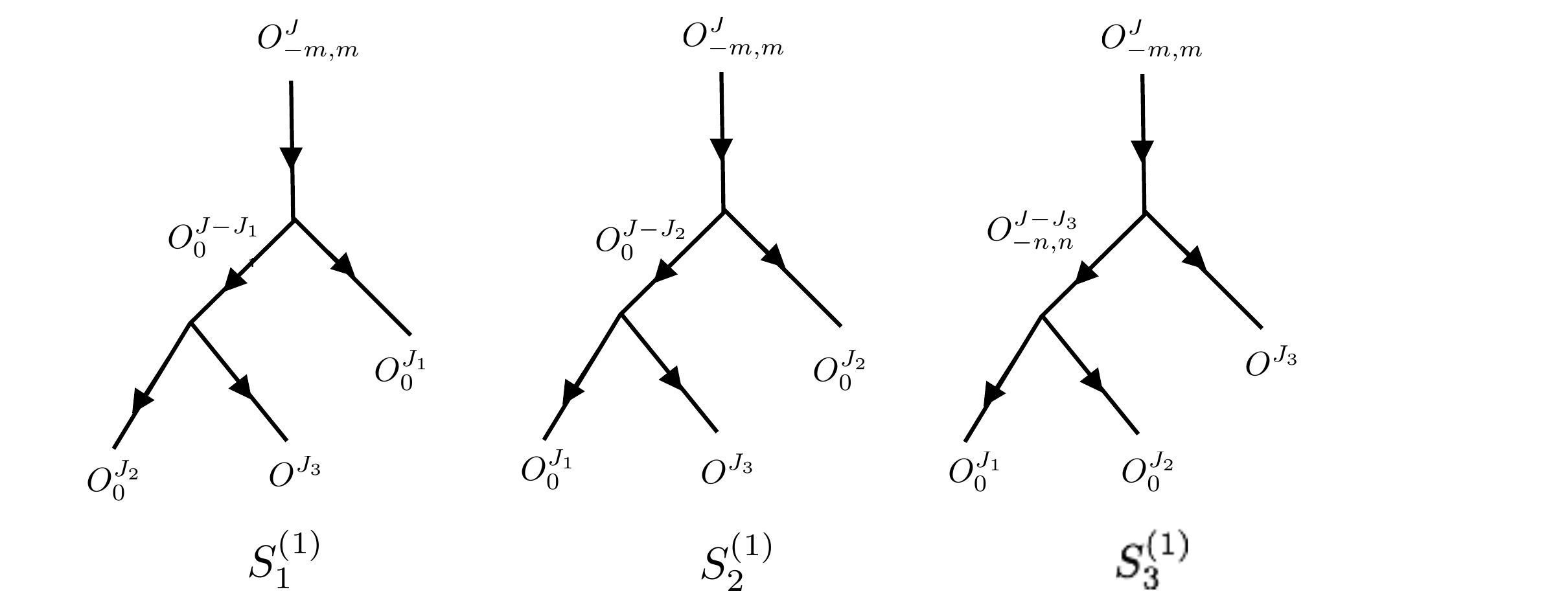} 
\end{center}
\caption{The string diagrams for  $\bra \bar{O}_{-m,m}^J O_0^{J_1}O_0^{J_2}O^{J_3}\ket$. We denote the contributions of the 3 diagrams $S^{(1)}_1$,  $S^{(1)}_2$,  $S^{(1)}_3$.}  \label{S1}
\end{figure}

Now for each long process we can draw a diagram for it which we call the string diagram. We represent the cut or join process by a 3-string vertex exemplified in Fig. \ref{vertices}. The string diagram is constructed by pasting together the 3-string vertices. We notice that different long processes can map to the same string diagram. For example, the second long process in $F^{(1)}_2$  and the first long process in  $F^{(1)}_3$ map to the same string diagram, the first diagram $S^{(1)}_1$ in Fig. \ref{S1} . Here it is the subscript that denotes the specific string and the labeling of segments is no longer distinguishable in the string diagrams. The string diagrams are depicted in Fig. \ref{S1}. We want to make two further points: 1. The long process only tells us how the strings split and join, but contains no information about the scalar insertions that represent string excitations. When we draw the string diagram for a long process, we will need to look for all possible ways to put in the sting excitations consistent with the 3-string vertices in (\ref{planar1}). 2.  There are other more complicated ways to fill in the intermediate steps, but we only consider string diagrams of the lowest order. For example, we see here that the string diagrams are tree level, which means that the corresponding field theory diagrams are planar, and we will not need to consider one-loop string diagrams here. This also give a convenient way to count the genus of the field theory diagrams by simply looking at the number of loops in the corresponding string diagrams, which turns out to be much easier at higher genus.  

We denote the contributions of the 3 diagrams in Fig. \ref{S1} as $S^{(1)}_1$,  $S^{(1)}_2$,  $S^{(1)}_3$. We see from (\ref{F1longprocess}) that $S^{(1)}_1$ represents the second long process of $F^{(1)}_2$ and the first long process of $F^{(1)}_3$,  $S^{(1)}_2$ represents the second long process of $F^{(1)}_1$ and the second long process of $F^{(1)}_3$,  and finally $S^{(1)}_3$ represents the first long process of $F^{(1)}_1$ and the first long process of $F^{(1)}_2$. 

The string diagrams are computed by simply multiplying the 3-string vertices in (\ref{planar1}), and sum up all possible intermediate states. Here the string theory is extremely simple and we do not need propagators between the vertices. For example, we can compute $S^{(1)}_1$ as the followings 
\begin{eqnarray} \label{S11}
S^{(1)}_1 &=& \bra \bar{O}^J_{-m,m} O_0^{J-J_1}O^{J_1}_0 \ket \bra  \bar{O}_0^{J-J_1} O^{J_2}_0 O^{J_3}\ket
\nonumber \\ &=& \frac{g^2}{J} x_2x_3^{\frac{1}{2}}\frac{\cos(2m\pi x_1)-1}{2\pi^2m^2}
\end{eqnarray}
Similarly for $S^{(1)}_2$,
\begin{eqnarray} \label{S12}
S^{(1)}_2 =  \frac{g^2}{J} x_1x_3^{\frac{1}{2}}\frac{\cos(2m\pi x_2)-1}{2\pi^2m^2}
\end{eqnarray}
The calculation of $S^{(1)}_3$ is a little more complicated as we need to sum over all possible $O^{J-J_3}_{-n,n}$ intermediate state 
\begin{eqnarray} \label{S13}
S^{(1)}_3 &=& \sum_{n=-\infty}^{\infty} \bra \bar{O}^J_{-m,m} O_{-n,n}^{J-J_3}O^{J_3} \ket \bra  \bar{O}_{-n,n}^{J-J_3} O^{J_1}_0 O^{J_2}_0\ket
\end{eqnarray}
The sum is convergent and the summation formulae (\ref{summation1}) and their derivatives in Appendix \ref{summationappendix} are useful for doing the summation.

We are now ready to state the \textit{factorization} rule. For a string diagram $S$ we count the number of its appearance in long processes associated with the short process of each field theory diagram $F$, and we call it the \textit{multiplicity} of the string diagram $S$ with respect to the field theory diagram $F$. For example, the multiplicity of $S^{(1)}_1$ is 0 with respect to $F^{(1)}_1$, and 1 with respect to $F^{(1)}_2$ and  $F^{(1)}_3$. Then the contribution of a string diagram is the sum of all field theory diagrams contributions weighted by the multiplicities. So for $S^{(1)}_1$ the factorization relation is 
\begin{eqnarray}
S^{(1)}_1&=& F^{(1)}_2+F^{(1)}_3
\end{eqnarray}
This is easily verified using (\ref{S11}, \ref{F12}, \ref{F13}). Similarly we also verify the factorization for $S^{(1)}_2$, $S^{(1)}_3$ 
\begin{eqnarray}
S^{(1)}_2 &=& F^{(1)}_1+F^{(1)}_3 \nonumber \\
S^{(1)}_3 &=& F^{(1)}_1+F^{(1)}_2 
\end{eqnarray}
We can then write the total contributions to the correlator as
\begin{eqnarray}
\bra \bar{O}_{-m,m}^J O_0^{J_1}O_0^{J_2}O^{J_3}\ket_{\textrm{\textrm{planar}}}  = F^{(1)}_1+F^{(1)}_2+F^{(1)}_3=\frac{1}{2}(S^{(1)}_1+S^{(1)}_2+S^{(1)}_3)
\end{eqnarray}

It may be illuminating to perform the sum in (\ref{S13}) with the integral form of the 3-string vertex (\ref{integralform}), and we can use the summation formula involving the delta function (\ref{deltafunction}) to perform the sum first before the integrals. We find (\ref{S13}) becomes
\begin{eqnarray}
S^{(1)}_3 & =& \frac{g^2}{J} x_1x_2x_3^{\frac{1}{2}}(1-x_3) \int_0^1 dy_1  \int_0^1 dy_2  \int_0^1 dy_3  \int_0^1 dy_4
e^{2\pi i m(1-x_3)(y_2-y_1) }
\nonumber \\ && \times [\sum_{k=-\infty}^{+\infty} \delta(y_1-y_2+\frac{x_1}{1-x_3}y_3+\frac{x_2}{1-x_3}y_4-k) ]
\end{eqnarray}
The delta function should be treated with cares. Since $x_1+x_2+x_3=1$, we see $0<\frac{x_1}{1-x_3}y_3+\frac{x_2}{1-x_3}y_4<1$, and we discuss two regions for the $y_2$ integration domain
\begin{itemize}
\item If $0<y_2<\frac{x_1}{1-x_3}y_3+\frac{x_2}{1-x_3}y_4$, then $-1<y_2-\frac{x_1}{1-x_3}y_3-\frac{x_2}{1-x_3}y_4<0$. The delta function fixes $y_1=y_2-\frac{x_1}{1-x_3}y_3-\frac{x_2}{1-x_3}y_4+1$ with $k=1$ in the sum. 
\item If $\frac{x_1}{1-x_3}y_3+\frac{x_2}{1-x_3}y_4<y_2<1$, then $0<y_2-\frac{x_1}{1-x_3}y_3-\frac{x_2}{1-x_3}y_4<1$. The delta function fixes $y_1=y_2-\frac{x_1}{1-x_3}y_3-\frac{x_2}{1-x_3}y_4$ with $k=0$ in the sum. 
\end{itemize}
We plug in the values of $y_1$ fixed by the delta function and also integrate the $y_2$ variable which no longer appears in the integrand. We find
\begin{eqnarray}
S^{(1)}_3 & =& \frac{g^2}{J} x_1x_2x_3^{\frac{1}{2}}  \int_0^1 dy_3  \int_0^1 dy_4
e^{2\pi i m(x_1y_3 +x_2y_4)}  \nonumber \\ & \times & [ e^{-2\pi i m (1-x_3)} (x_1y_3+x_2y_4)+ x_1(1-y_3)+x_2(1-y_4) ]
\end{eqnarray}
This integral can be identified with those of $F^{(1)}_1$, $F^{(1)}_2$ in (\ref{F11}, \ref{F12}) without evaluating them completely explicitly. To see this we first need to change the integration variables in (\ref{F11}) to  $y_4=1-y_2$, and $y_3=y_1-y$ if $y_1>y$ or $y_3=1+y_1-y$ if $y_1<y$. We also integrate the remaining variable which does not appear in the integrand. The integral for  $F^{(1)}_1$ becomes 
\begin{eqnarray}
F^{(1)}_1 =  \frac{g^2}{J}x_1^2x_2x_3^{\frac{1}{2}}  \int_0^1 dy_3  \int_0^1 dy_4
e^{2\pi i m(x_1y_3 +x_2y_4)}  [ e^{-2\pi i m (1-x_3)} y_3+ (1-y_3) ]
\end{eqnarray}
Similarly, 
\begin{eqnarray}
F^{(1)}_2 =  \frac{g^2}{J}x_1x_2^2x_3^{\frac{1}{2}}  \int_0^1 dy_3  \int_0^1 dy_4
e^{2\pi i m(x_1y_3 +x_2y_4)}  [ e^{-2\pi i m (1-x_3)} y_4+ (1-y_4) ]
\end{eqnarray}
Now we can see the factorization relation $S^{(1)}_3 = F^{(1)}_1 + F^{(1)}_2$ without the need to evaluating the integrals.  In Section \ref{integralsection} we will study in more details this method of summing over intermediate string modes using (\ref{deltafunction}) with the integral form (\ref{integralform}) of 3-string vertex in the case of torus correlator of two single trace operators. We will see that the dissection of the integration domain is quite tricky in higher genus. In most parts of the paper we will use the more straightforward methods of direct computations to check the factorization relation.

\subsubsection{Case two: $\bra \bar{O}_{-m,m}^J O_{-n,n}^{J_1}O^{J_2}O^{J_3}\ket$}

\begin{figure}
  \begin{center}
  \includegraphics[width=6.5in]{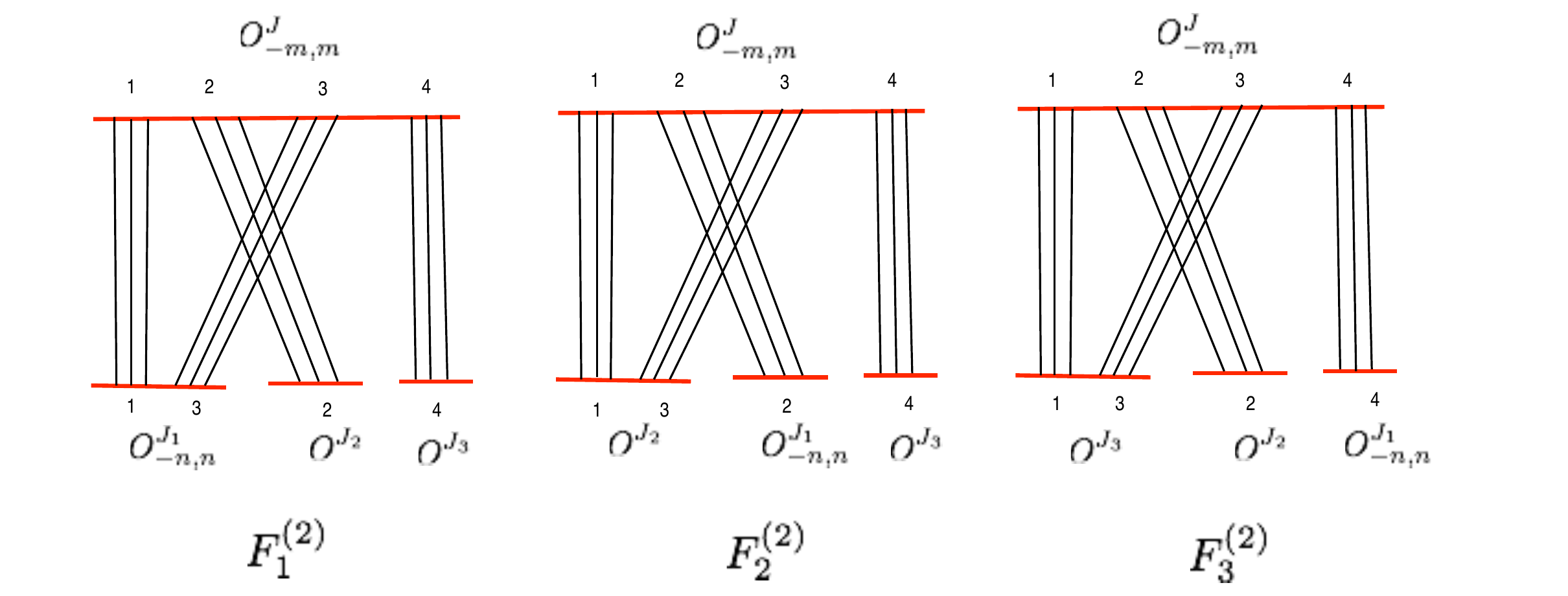} 
\end{center}
\caption{The field theory diagrams for  $\bra \bar{O}_{-m,m}^J O_{-n,n}^{J_1}O^{J_2}O^{J_3}\ket$. We denote the contributions of the 3 diagrams $F^{(2)}_1$,  $F^{(2)}_2$,  $F^{(2)}_3$.}  \label{F2}
\end{figure}

The computations for $\bra \bar{O}_{-m,m}^J O_{-n,n}^{J_1}O^{J_2}O^{J_3}\ket$ are similar to the previous case but a little more complicated since there are two stringy operators.  The field theory diagrams are listed in Fig. \ref{F2}. Again we denote $x_i=\frac{J_i}{J}$, and we find 
\begin{eqnarray}
F^{(2)}_1 &=& \frac{g^2}{J}\frac{x_1^{\frac{5}{2}}x_2^{\frac{1}{2}}x_3^{\frac{1}{2}}}{2\pi^2(n-mx_1)^2}[\frac{\sin(2\pi mx_1)+\sin(2\pi mx_2) +\sin(2\pi mx_3)}{\pi (n-mx_1)}\nonumber \\ && +2-\cos(2\pi m x_2)-\cos(2\pi m x_3)] \\
F^{(2)}_2 &=&  \frac{g^2}{J} x_1^{\frac{3}{2}}x_2^{\frac{3}{2}}x_3^{\frac{1}{2}} ~\frac{1-\cos(2\pi mx_1)}{2\pi^2(n-mx_1)^2} \\
F^{(2)}_3 &=&  \frac{g^2}{J} x_1^{\frac{3}{2}}x_2^{\frac{1}{2}}x_3^{\frac{3}{2}} ~\frac{1-\cos(2\pi mx_1)}{2\pi^2(n-mx_1)^2} 
\end{eqnarray}

\begin{figure}
  \begin{center}
  \includegraphics[width=6.5in]{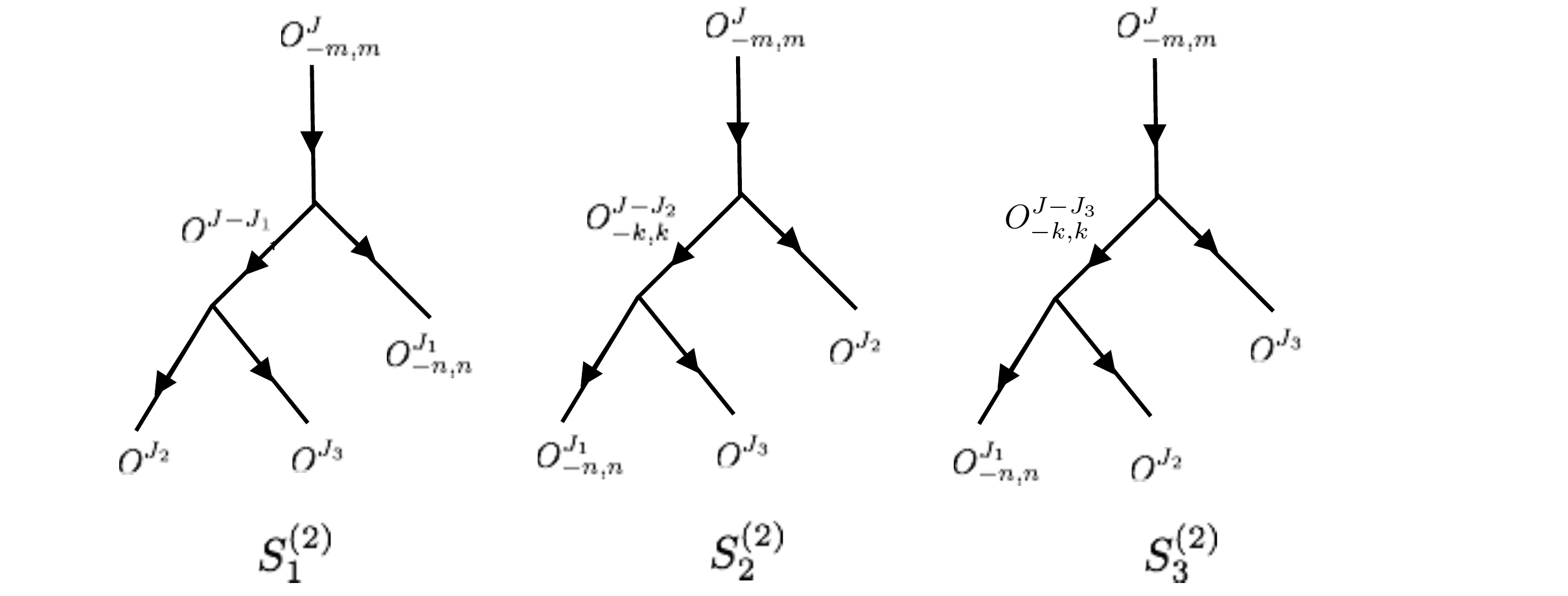} 
\end{center}
\caption{The string diagrams for  $\bra \bar{O}_{-m,m}^J O_{-n,n}^{J_1}O^{J_2}O^{J_3}\ket$. We denote the contributions of the 3 diagrams $S^{(2)}_1$,  $S^{(2)}_2$,  $S^{(2)}_3$.}  \label{S2}
\end{figure}

The string diagrams for $\bra \bar{O}_{-m,m}^J O_{-n,n}^{J_1}O^{J_2}O^{J_3}\ket$ is depicted in Fig. \ref{S2}. We can compute the contribution of each diagram
\begin{eqnarray} 
S^{(2)}_1 &=& \bra \bar{O}_{-m,m}^J O_{-n,n}^{J_1}O^{J-J_1}\ket \bra \bar{O}^{J-J_1} O^{J_2}O^{J_3}\ket \nonumber \\
&=&  \frac{g^2}{J}x_1^{\frac{3}{2}}(x_2x_3)^{\frac{1}{2}}(x_2+x_3) \frac{1-\cos(2\pi m x_1)}{2\pi^2 (n-mx_1)^2} \\
S^{(2)}_2 &=& \sum_{k=-\infty}^{\infty} \bra \bar{O}_{-m,m}^J O_{-k,k}^{J-J_2}O^{J_2}\ket \bra \bar{O}^{J-J_2}_{-k,k} O^{J_1}_{-n,n}O^{J_3}\ket \label{S22}\\
S^{(2)}_3 &=& \sum_{k=-\infty}^{\infty} \bra \bar{O}_{-m,m}^J O_{-k,k}^{J-J_3}O^{J_3}\ket \bra \bar{O}^{J-J_3}_{-k,k} O^{J_1}_{-n,n}O^{J_2}\ket \label{S23}
\end{eqnarray}

The derivation of the multiplicity of the string diagrams with respect to the field theory diagrams is the same as in the previous case. For (\ref{S22}, \ref{S23}) we need to perform the sum using the summation formulae in Appendix \ref{summationappendix}. Similar to the previous case, one can perform the sum either directly using the derivatives of  (\ref{summation1}), or using (\ref{deltafunction})  with the integral form (\ref{integralform}) of 3-string vertex.  We verify the factorization relations 
\begin{eqnarray}
S^{(2)}_1&=& F^{(2)}_2+F^{(2)}_3 \nonumber \\
S^{(2)}_2 &=& F^{(2)}_1+F^{(2)}_3 \nonumber \\
S^{(2)}_3 &=& F^{(2)}_1+F^{(2)}_2 
\end{eqnarray}
Similar the previous case, we can write the correlator as 
\begin{eqnarray}
\bra \bar{O}_{-m,m}^J O_{-n,n}^{J_1}O^{J_2}O^{J_3}\ket_{\textrm{\textrm{planar}}} = F^{(2)}_1+F^{(2)}_2+F^{(2)}_3=\frac{1}{2}(S^{(2)}_1+S^{(2)}_2+S^{(2)}_3)
\end{eqnarray}

\subsection{Correlators between two double trace operators} \label{22section}

This correlator is very similar to the $2\rightarrow 2$ scattering process familiar in the collider physics. The tree level string diagrams can be similarly classified as the $S$, $T$, $U$ channels.  Surprisingly, we discover a subtlety for the factorization rule. We will find that the factorization breaks down for the $S$ channel, while still holds for the $T$, $U$ channels. To illustrate the point, let us consider three cases.

\subsubsection{Case one: $\bra \bar{O}^{J_1} \bar{O}^{J_4}O^{J_2}O^{J_3}\ket$}

\begin{figure}
  \begin{center}
  \includegraphics[width=6.5in]{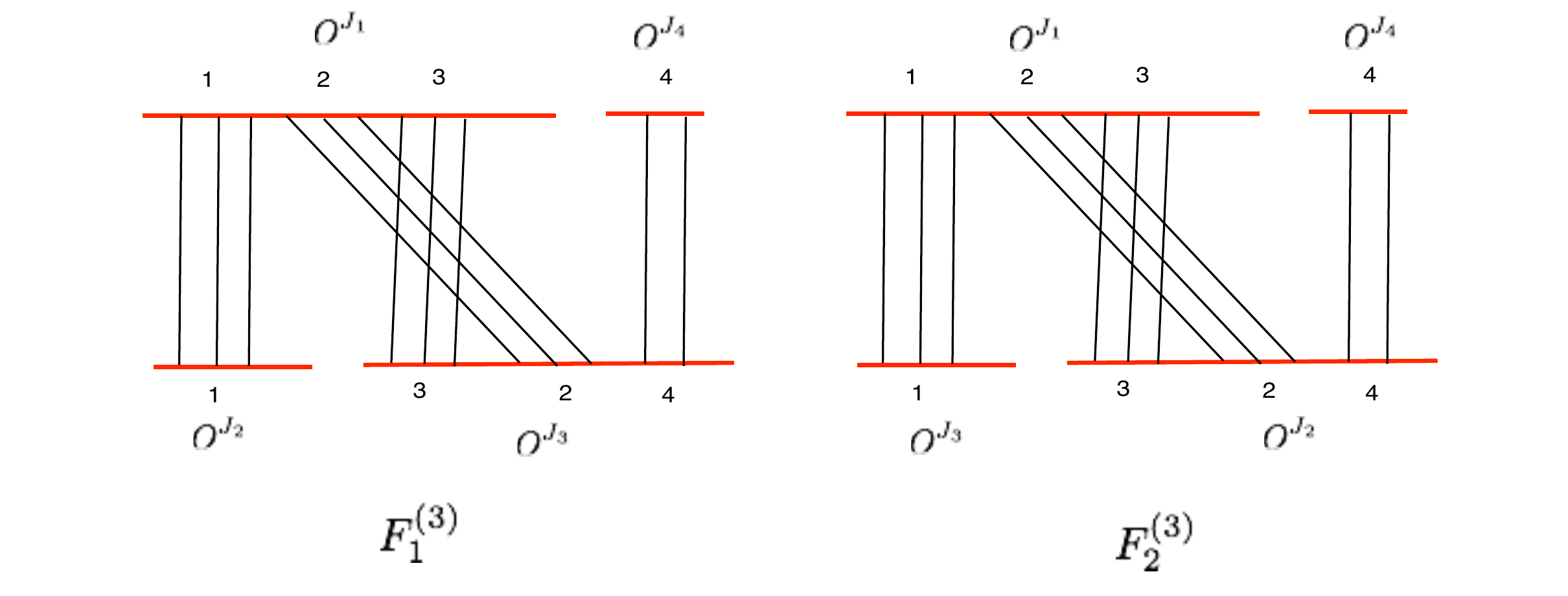} 
 \end{center}
\caption{The field theory diagrams for  $\bra \bar{O}^{J_1} \bar{O}^{J_4}O^{J_2}O^{J_3}\ket$. We denote the contributions of the 2 diagrams $F^{(3)}_1$,  $F^{(3)}_2$ respectively.}  \label{F3}
\end{figure}

This is the correlator of the vacuum operators, and it is implicit that $J_1+J_4=J_2+J_3$. Without loss of generality we assume $J_1>J_2>J_3>J_4$. At planar level there are two field theory diagrams, depicted in Fig. \ref{F3}. Again similar to previous cases, these diagrams look non-planar but are actually planar if we rearrange the operators. We have divided the operators into a maximal number of segments without violating planarity to obtain the combinatorially most dominant diagrams.  We will also see the corresponding string diagrams are tree level.

We denote $J=J_1+J_4=J_2+J_3$, and $x_i=\frac{J_i}{J}$. To count the combinatorics of the diagrams in Fig. \ref{F3}, we need to choose the beginning point for each of the operators, which contributes a factor of $J_1J_2J_3J_4$. Then for the longest operator $O^{J_1}$ we also need to fix a beginning point for segment (3), which contribute a factor $J_1-J_2$ for the first diagram, and a factor of $J_1-J_3$ for the second one. We use the normalization for operators in (\ref{BMNoperators}), and since each double trace operator contributes a negative power of $N$, we should have a total power of $1/N^2$ for each diagram. So the contributions of the diagrams are 
\begin{eqnarray}
F^{(3)}_1 &=& \frac{J_1J_2J_3J_4(J_1-J_2)}{N^2\sqrt{J_1J_2J_3J_4}}= \frac{g^2}{J}(x_1x_2x_3x_4)^{\frac{1}{2}}(x_1-x_2)  \\
F^{(3)}_2 &=& \frac{J_1J_2J_3J_4(J_1-J_3)}{N^2\sqrt{J_1J_2J_3J_4}}= \frac{g^2}{J}(x_1x_2x_3x_4)^{\frac{1}{2}}(x_1-x_3)
\end{eqnarray}

\begin{figure}
  \begin{center}
  \includegraphics[width=6.5in]{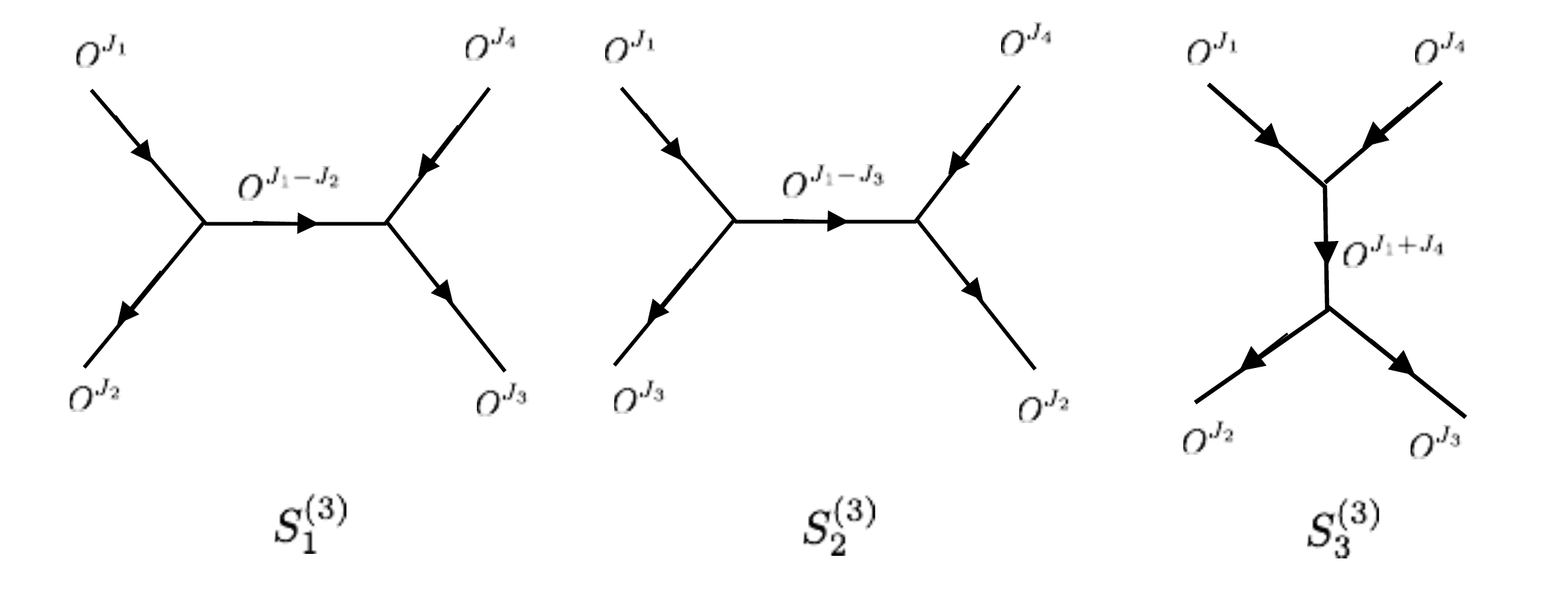} 
\end{center}
\caption{The string diagrams for  $\bra \bar{O}^{J_1} \bar{O}^{J_4}O^{J_2}O^{J_3}\ket$. We denote the contributions of the 3 diagrams $S^{(3)}_1$,  $S^{(3)}_2$, $S^{(3)}_3$ respectively. The 3 diagrams represent the $T$, $U$, $S$ channels in $2\rightarrow 2$ scattering. }  \label{S3}
\end{figure}

The string diagrams are depicted in Fig. \ref{S3}. It is simple to compute them using the 3-string vertex in (\ref{planar1}). We find 
\begin{eqnarray}
S^{(3)}_1 &=&  \frac{g^2}{J}(x_1x_2x_3x_4)^{\frac{1}{2}}(x_1-x_2) \\
S^{(3)}_2 &=&  \frac{g^2}{J}(x_1x_2x_3x_4)^{\frac{1}{2}}(x_1-x_3) \\
S^{(3)}_3 &=&  \frac{g^2}{J}(x_1x_2x_3x_4)^{\frac{1}{2}}
\end{eqnarray}

To count the multiplicity of the string diagrams. We expend the short process of the field theory diagrams into long processes as the followings
\begin{eqnarray}
F^{(3)}_1 : && (123)_1(4)_4 \rightarrow (1)_2(23)(4)_4 \rightarrow (1)_2(324)_3 \nonumber \\
&&  (123)_1(4)_4 \rightarrow (3124) \rightarrow (1)_2(243)_3 \\
F^{(3)}_1 : && (123)_1(4)_4 \rightarrow (1)_3(23)(4)_4 \rightarrow (1)_3(324)_2 \nonumber \\
&&  (123)_1(4)_4 \rightarrow (3124) \rightarrow (1)_3(243)_2
\end{eqnarray}
We find the string diagram $S^{(3)}_1$ has a multiplicity of 1 with respect to $F^{(3)}_1$, the string diagram $S^{(3)}_2$ has a multiplicity of 1 with respect to $F^{(3)}_2$, and the string diagram $S^{(3)}_3$ has the multiplicities of 1 with respect to both $F^{(3)}_1$ and $F^{(3)}_2$. We find that for  $S^{(3)}_1$  and  $S^{(3)}_2$, which represent the $T$ and $U$ channels of the $2\rightarrow 2$ scattering, the factorization relation holds, namely  
\begin{eqnarray}
 S^{(3)}_1 &=&  F^{(3)}_1 \nonumber \\
 S^{(3)}_2 &=&  F^{(3)}_2 
\end{eqnarray}
However, we find that the factorization breaks down for the S-channel process $S^{(3)}_3$. It is easy to see
\begin{eqnarray}
S^{(3)}_3 &\neq &  F^{(3)}_1+F^{(3)}_2
\end{eqnarray}
The total contribution to the correlator is 
\begin{eqnarray}
\bra \bar{O}^{J_1} \bar{O}^{J_4}O^{J_2}O^{J_3}\ket =  F^{(3)}_1+F^{(3)}_2
=  S^{(3)}_1+S^{(3)}_2
\end{eqnarray}

\subsubsection{Case two: $\bra \bar{O}^{J_1} \bar{O}^{J_4}_{-m,m} O^{J_2}_0O^{J_3}_0 \ket$ ($J_1>J_2>J_3>J_4$)}

\begin{figure}
  \begin{center}
\includegraphics[width=6.5in]{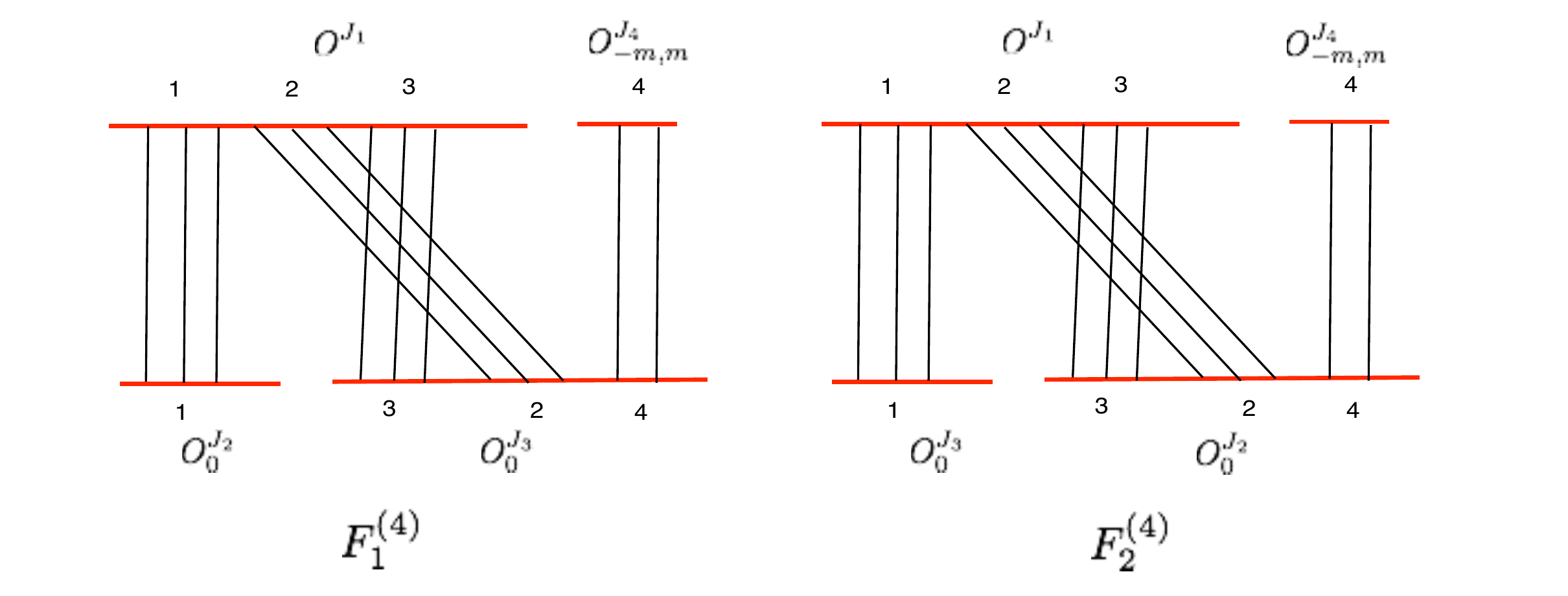} 
\end{center}
\caption{The field theory diagrams for  $\bra \bar{O}^{J_1} \bar{O}^{J_4}_{-m,m} O^{J_2}_0O^{J_3}_0 \ket$ ($J_1>J_2>J_3>J_4$ ). We denote the contributions of the 2 diagrams $F^{(4)}_1$,  $F^{(4)}_2$ respectively. These diagrams turn out to give vanishing contributions. }  \label{F4}
\end{figure}

We discuss an example where the S-channel factorization breaks down quite dramatically. We draw the field theory diagrams for $\bra \bar{O}^{J_1} \bar{O}^{J_4}_{-m,m} O^{J_2}_0O^{J_3}_0 \ket$ in Fig. \ref{F4}. These diagrams are structurally the same as those of vacuum operators in Fig. \ref{F3}, and we only need to insert scalar excitations into the trace operators. But since we assume the stringy operator with two scaler insertions $O^{J_4}_{-m,m}$ is the shortest, either $O^{J_2}_0$ or $O^{J_3}_0$ has no Wick contraction with $O^{J_4}_{-m,m}$. So it is impossible to put in the scalar insertions without violating planarity and these diagrams actually vanish
\begin{eqnarray} \label{F4eq}
F^{(4)}_1=0, ~~~~F^{(4)}_2=0
\end{eqnarray}
Consequently the correlator also vanishes  $\bra \bar{O}^{J_1} \bar{O}^{J_4}_{-m,m} O^{J_2}_0O^{J_3}_0 \ket =0$

\begin{figure}
  \begin{center}
  \includegraphics[width=6.5in]{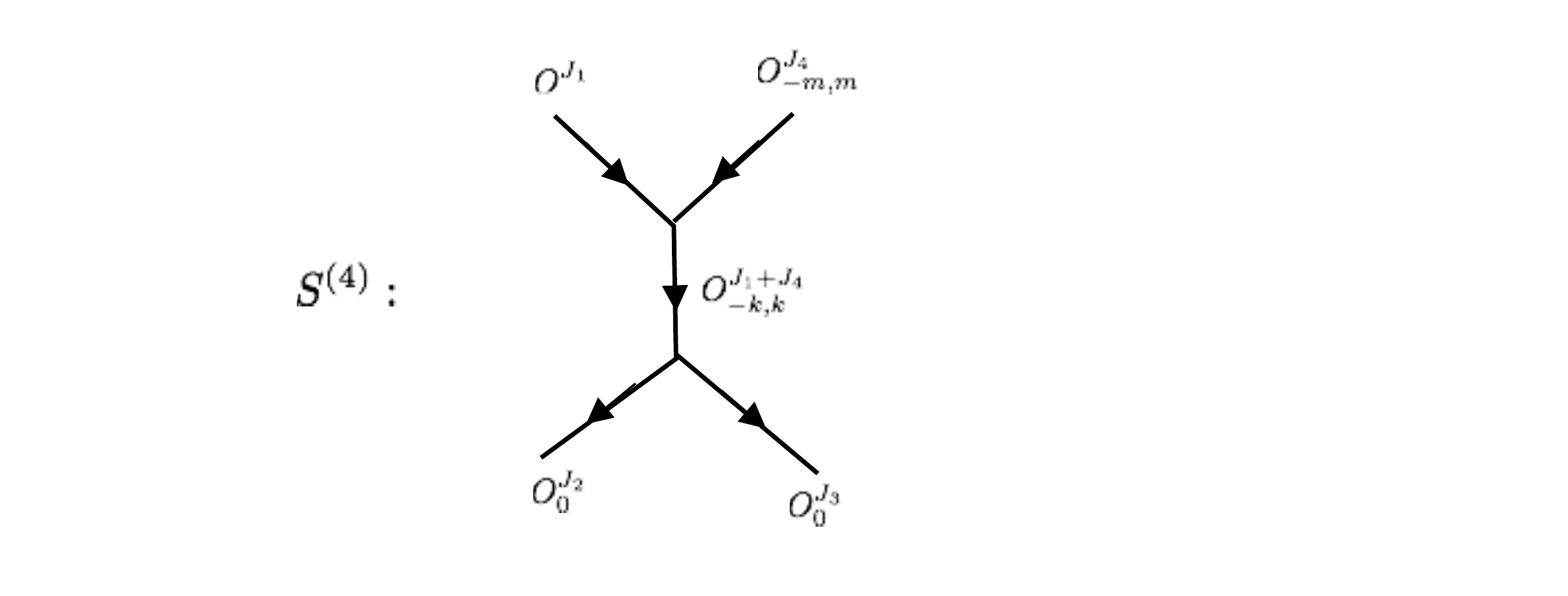} 
\end{center}
\caption{The string diagram for  $\bra \bar{O}^{J_1} \bar{O}^{J_4}_{-m,m} O^{J_2}_0O^{J_3}_0 \ket$ ($J_1>J_2>J_3>J_4$ ). This is the only non-vanishing S-channel diagram, which we denote  $S^{(4)}$. }  \label{S4}
\end{figure}

We look at the string diagrams. The longest operator is $O^{J_1}$, but it has no scalar insertion so it can not decay to $O^{J_2}_0$ or $O^{J_3}_0$. So the $T$, $U$ channels are impossible and we are left only with the S-channel contribution $S^{(4)}$ depicted in Fig. \ref{S4}. The vanishing of the $T$, $U$ channels is consistent with the factorization rules since the field theory diagram contributions vanish (\ref{F4eq}). The factorization rules would require the S-channel contribution also vanish. But this is not true, as we can calculate 
\begin{eqnarray}
S^{(4)} = \sum_{k=-\infty}^{+\infty}  \bra \bar{O}^{J_1} \bar{O}^{J_4}_{-m,m} O^{J_1+J_4}_{-k,k}\ket \bra \bar{O}^{J_1+J_4}_{-k,k} O^{J_2}_0O^{J_3}_0 \ket
\end{eqnarray}
But the 3-string vertices have definite signs $ \bra \bar{O}^{J_1} \bar{O}^{J_4}_{-m,m} O^{J_1+J_4}_{-k,k}\ket\geq 0$, $\bra \bar{O}^{J_1+J_4}_{-k,k} O^{J_2}_0O^{J_3}_0 \ket \leq 0$, and these vertices are not zero for $k\neq 0$. So it must be $S^{(4)} <0$, and we see quite explicitly the factorization does not hold for the S-channel string diagram.

\subsubsection{Case three: $\bra \bar{O}^{J_1}_{-m,m} \bar{O}^{J_4} O^{J_2}_{-n,n}O^{J_3} \ket$  ($J_1>J_2>J_3>J_4$)}

\begin{figure}
  \begin{center}
  \includegraphics[width=6.5in]{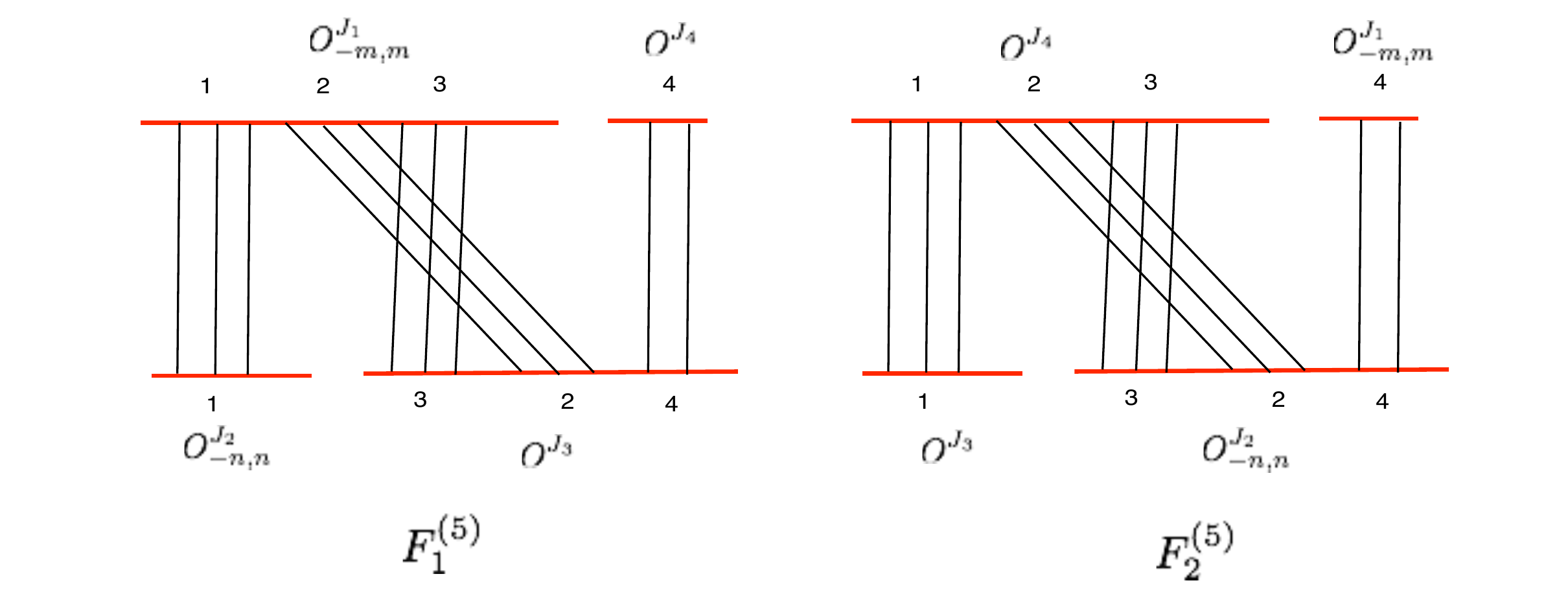} 
\end{center}
\caption{The field theory diagrams for  $\bra \bar{O}^{J_1}_{-m,m} \bar{O}^{J_4} O^{J_2}_{-n,n}O^{J_3} \ket$ ($J_1>J_2>J_3>J_4$ ). We denote the contributions of the 2 diagrams $F^{(5)}_1$,  $F^{(5)}_2$ respectively. }  \label{F5}
\end{figure}

Finally, let us consider an example where the $T,U$ channels factorization are less trivial than the previous cases. The field theory diagrams are depicted in Fig. \ref{F5}. The counting of the combinatorics is the same as that of  the vacuum operators depicted in Fig. \ref{F3}, and we just need to put in scalar insertions. Denoting again $J=J_1+J_4=J_2+J_3$, and $x_i=\frac{J_i}{J}$, we compute the diagrams as the followings 
\begin{eqnarray}
F^{(5)}_1 &=&  \frac{g^2}{J} (x_1)^{-\frac{1}{2}}x_2^{\frac{3}{2}} (x_3x_4)^{\frac{1}{2}} (x_1-x_2)\int_0^1 dy_1e^{-2\pi i (\frac{mx_2}{x_1}-n)y_1} \int_0^1 dy_2e^{2\pi i (\frac{mx_2}{x_1}-n)y_2} \nonumber \\
&=&  \frac{g^2}{J} (x_1x_2)^{\frac{3}{2}} (x_3x_4)^{\frac{1}{2}} (x_1-x_2)\frac{1-\cos(2\pi(\frac{mx_2}{x_1}-n))}{2\pi^2(mx_2-nx_1)^2}
\end{eqnarray}
For $F^{(5)}_2$ the calculations are more involved
\begin{eqnarray}
F^{(5)}_2 &=&  \frac{g^2}{J} (x_1x_2)^{-\frac{1}{2}} (x_3x_4)^{\frac{1}{2}} (x_1-x_3)^3 \int_0^1 dy  \times   \nonumber \\  &&
| (e^{- 2\pi i n\frac{x_1-x_3}{x_2}} \int_0^y dy_1+\int_y^1 dy_1) e ^{2\pi i (x_1-x_3)(\frac{m}{x_1}-\frac{n}{x_2})y_1} |^2  \nonumber \\
&=&  \frac{g^2}{J} \frac{(x_1x_2)^{\frac{3}{2}} (x_3x_4)^{\frac{1}{2}}}{2\pi^3(nx_1-mx_2)^3}\{ \pi(nx_1-mx_2)(x_1-x_3) 
[2-\cos(2m\pi\frac{x_3}{x_1})-\cos(2n\pi\frac{x_1-x_3}{x_2})] \nonumber \\ && 
+x_1x_2[\sin(2n\pi\frac{x_1-x_3}{x_2})+\sin(2m\pi \frac{x_3}{x_1})-\sin(2\pi\frac{nx_1(x_1-x_3)+mx_2x_3}{x_1x_2})] \}
\label{F52}
\end{eqnarray}

\begin{figure}
  \begin{center}
  \includegraphics[width=6.5in]{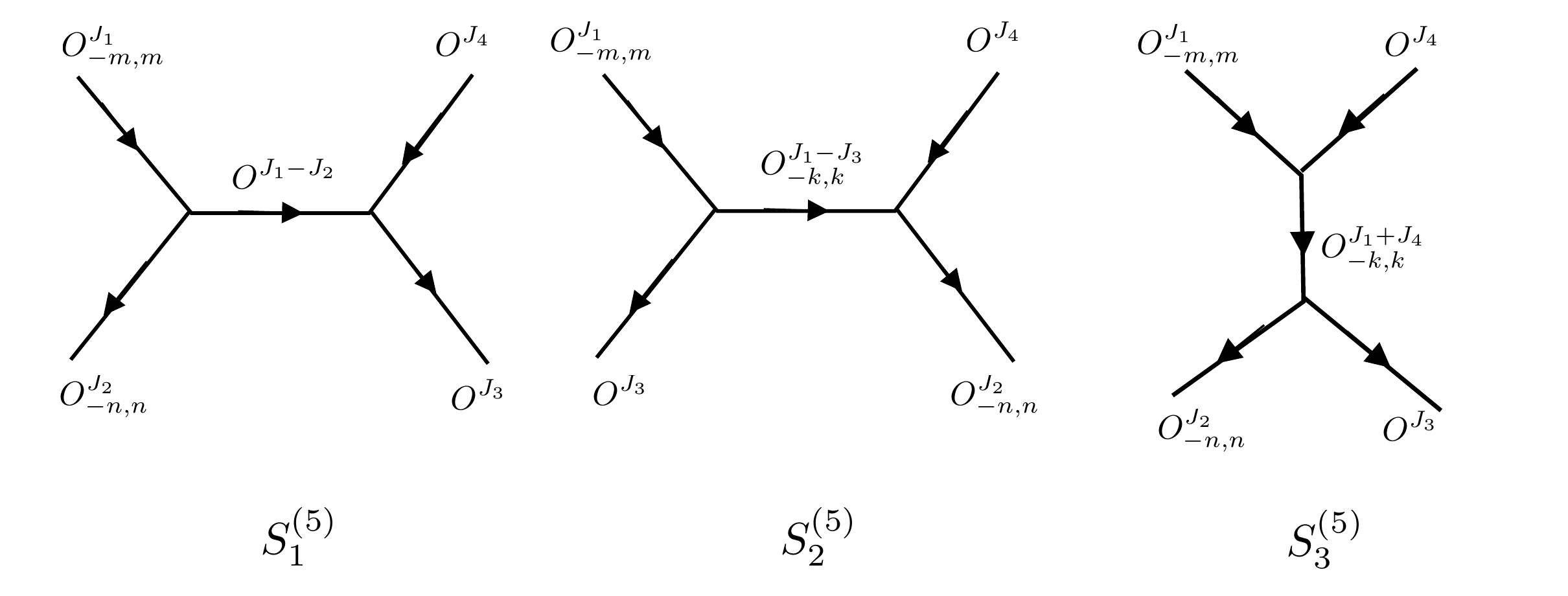} 
\end{center}
\caption{The string diagrams for  $\bra \bar{O}^{J_1}_{-m,m} \bar{O}^{J_4} O^{J_2}_{-n,n}O^{J_3} \ket$ ($J_1>J_2>J_3>J_4$ ). We denote the contributions of the 2 diagrams $S^{(5)}_1$,  $S^{(5)}_2$, $S^{(5)}_3$ respectively. }  \label{S5}
\end{figure}

Now we consider the string diagrams, depicted in Fig. \ref{S5}. Since we have learned the S-channel factorization does not hold, here we only compute the $T,U$ channels contributions denoted by $S^{(5)}_1$, $S^{(5)}_2$. The computation of $S^{(5)}_1$ is quite simple
\begin{eqnarray}
S^{(5)}_1= \bra \bar{O}^{J_1}_{-m,m} O^{J_2}_{-n,n}O^{J_1-J_2}\ket \bra \bar{O}^{J_1-J_2}\bar{O}^{J_4} O^{J_3} \ket
\end{eqnarray}
So using the 3-string vertex formulae (\ref{planar1}) we can easily see 
\begin{eqnarray}
S^{(5)}_1=F^{(5)}_1, 
\end{eqnarray}
consistent with the factorization rules.
We perform the sum in $S^{(5)}_2$ and check the agreement with (\ref{F52}) required by the factorization rules
\begin{eqnarray}
S^{(5)}_2 = \sum_{k=-\infty}^{+\infty} \bra \bar{O}^{J_1}_{-m,m} O^{J_3} O^{J_1-J_3}_{-k,k}\ket \bra \bar{O}^{J_1-J_3}_{-k,k}\bar{O}^{J_4} O^{J_2}_{-n,n} \ket = F^{(5)}_2
\end{eqnarray}
The total contributions to the correlator is 
\begin{eqnarray}
\bra \bar{O}^{J_1}_{-m,m} \bar{O}^{J_4} O^{J_2}_{-n,n}O^{J_3} \ket =F^{(5)}_1+F^{(5)}_2=S^{(5)}_1+S^{(5)}_2
\end{eqnarray}

The lesson of these exercises is that the factorization rules break down for the S-channel, but hold for the $T$, $U$ channels. This happens probably due to the fact that both initial and final states are multi-string states, and the combining of the strings in the intermediate steps is not captured by the field theory calculations. From now on, to avoid this subtlety we will without loss of generality focus on the cases that the initial state is a single string, or a single trace operator in the field theory side.

\section{Factorizations and recursion relations: the precise rules} \label{factorizationsection}

We have seen how the factorization worked in some examples, and we find that the factorization relations are non-trivial even for the tree level processes. We should now give some precise descriptions on the terminologies and the rules of the factorization property for general correlators at any genus level. To avoid the problem for the S-channel of $2\rightarrow 2$ process, we should consider only two-point correlators  where at least one operator is a single trace operator, or a single string state. We consider a general correlator $\bra \bar{O}_1O_2\ket $ where $O_1$ is a single trace BMN operator, and $O_2$ could be single trace or multi-trace. The operators $O_1$, $O_2$  are constructed by inserting scalar fields $\phi^i$ in the strings of $Z$'s with corresponding BMN phases. We denote the corresponding vacuum operators $O_1^{vacuum}=\Tr (Z^J)$, $O_{2}^{vacuum}=\Tr(Z^{J_1})\Tr(Z^{J_2})\cdots \Tr(Z^{J_n})$ and it is implicit that $J=J_1+J_2+\cdots J_n$. The derivations of the factorization rules for the correlator at genus $h$ follow three steps.

\subsection{Constructing the field theory diagrams}
First we should construct the field theory diagrams for the correlator of the vacuum operators $\bra \bar{O}_1^{vacuum} O_2^{vacuum}\ket$. We should divide each strings of $Z$'s in the traces into several parts which we call \textit{segments}. A segment consists of a large number of $Z$'s. There should be equal number of segments in $O_1^{vacuum}$ and in $O_2^{vacuum}$. The Wick contraction connects the $\bar{Z}$'s in $\bar{O}_{1}^{vacuum}$ with the $Z$'s in $O_2^{vacuum}$, and connects each segment in $\bar{O}_{1}^{vacuum}$ with a segment in $O_2^{vacuum}$.  If two segments are adjacent to each others in $\bar{O}_1^{vacuum}$ and their Wick contracted counterparts  in $O_2^{vacuum}$ are also adjacent in the same order, then we can combine them into one segment. We will always combine these unnecessary adjacent segments and we call the resulting diagram \textit{irreducible}. Each segment in an irreducible diagram generates a combinatorial factor of $J\sim \sqrt{N}\sim \infty$, so we will only need to  consider those diagrams with maximal numbers of segments at genus $h$. Since it is well known in large $N$ field theory that each additional genus generates a power of $1/N^2$, and non-planar diagrams in the BMN sector are perturbative in the powers of $g=\frac{J^2}{N}$, we should expect to  introduce 4 more segments for each additional genus. 

For each diagram we can write a \textit{short process} as the followings. We label the segments in $O_1^{vacuum}$ by numerical order as $1,2,\cdots l$. Then we also put the same label on the segment in $O_2^{vacuum}$ connected to $O_1^{vacuum}$ by Wick contraction. Then each trace operator becomes a finite chain of numbers $(a_1a_2\cdots )_i$, where $i=1,2,\cdots, n$ denote the trace operators in $O_2^{vacuum}$. So a short process can be written as 
\begin{eqnarray}
(12\cdots l)\rightarrow (a_{1,1}a_{1,2}\cdots)_1(a_{2,1}a_{2,2}\cdots)_2\cdots (a_{n,1}a_{n,2}\cdots)_n
\end{eqnarray}
for $O_2^{vacuum}$ a $n$-trace operator. Here the $a_{i,j}$'s is a permutation of $12\cdots l$, and each chain of numbers is considered to cyclic. As we mention we always combine unnecessary adjacent segments.  The short processes are in one to one correspondence with the field theory diagrams. 

Now we can put the scalar insertions into the trace operators. The scalars are inserted by pairs into both $O_1^{vacuum}$ and $O_2^{vacuum}$ and along the lines of the Wick contraction to preserve the genus of the diagrams. We sum over all these insertions with appropriate BMN phases to compute the contribution of a diagram to the correlator $\bra \bar{O}_1O_2\ket$. We denote the contribution $F_j$, where $j$ labels the field theory diagram, or a short process. 

\subsection{Constructing the string diagrams}

Similar to the field theory case, we first construct the string diagrams for the correlator of the vacuum operators $\bra \bar{O}_1^{vacuum} O_2^{vacuum}\ket$, which for convenience we call the vacuum diagram. The string diagrams are constructed by pasting the 3-string vertices, and for the vacuum operators we only need the first vertex in (\ref{planar1}). The first diagram in Fig. \ref{vertices} depicts the vacuum string splitting vertex, and the string joining vertex is obtained by just reversing the arrows.  We note the light cone momentum of the string states (which is proportional to the number of $Z$ fields in the BMN operators, and which we sometimes refer to as the length of the operator or the corresponding string and it goes like $\sqrt{N}\sim \infty$ in the BMN limit)  is conserved by the string vertex.  Each edge in a string diagram is represented by an operator propagating from one vertex to another, and we draw an arrow to denote the direction of propagation. We will draw an incoming arrow for $O_1^{vacuum}$ and outgoing arrows for each trace in $O_2^{vacuum}$ which are external edges of the string diagrams. We will distinguish between diagrams with different arrow directions on the edges.  For a correlator at genus $h$, we will consider string diagrams with $h$ loops. The number of string loops is $h=\frac{V-E}{2}+1$, where $V,E$ are the numbers of vertices and external edges. Since there are only cubic vertices in the string diagrams, we also have the formula for the number of vertices $3V=E+2I$ where $I$ is the number of internal edges. 

\begin{figure}
  \begin{center}
  \includegraphics[width=6.5in]{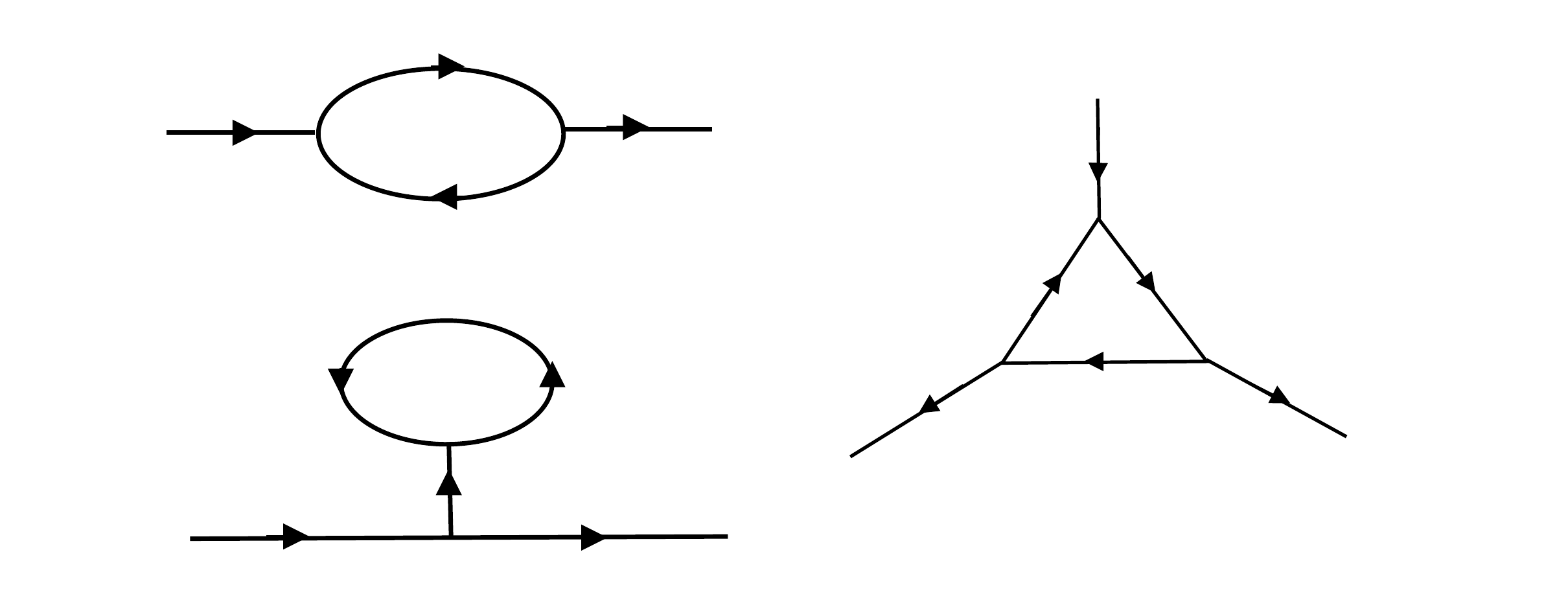} 
\end{center}
\caption{Examples of string diagrams not allowed because the arrows form a close loop. The diagram in lower left has a ``tadpole" part so it also violates momentum conservation. }  \label{closeloop}
\end{figure}

We note that we only consider connected string diagrams, however unlike the calculations of Feynman diagrams in conventional quantum field theory, we will need to calculate the un-amputated diagrams as well, i.e., the string diagrams do not need to be ``one-particle irreducible".  One special point to note is that in any parts of the string diagrams, we do not allow the arrow directions to form a closed loop. This kind of diagrams might not violate momentum conservation, but the operators propagating in the closed loop can have arbitrarily large number of  $Z$ fields  and make the contribution of diagram diverge. For example, the situations depicted in Fig.\ref{closeloop} are not allowed. We also note that a string diagram must have at least 2 external edges with both incoming and outgoing arrows, i.e. the ``vacuum bubble" and ``tadpole" diagrams are not possible. To see this point, we first note the conservation of light cone momentum of the string states rules out string diagrams with only incoming (or outgoing) external edges. For the vacuum bubble diagram, we can start from a vertex and move around the diagram following the arrow direction. This is always possible since there are only 2 types of string vertices, namely the joining vertex with two incoming and one outgoing arrows, and the splitting vertex with two outgoing and one incoming arrows. The path will eventually intersects itself and forms a closed loop if the string diagram is finite with no outgoing external edge, and as we mentioned a closed loop of arrows is not allowed.  Another consequence of the light cone momentum conservation and the rule of no closed loop of arrows is that no operator in the internal edges of a string diagram can be longer than the sum of the lengths of all outgoing operators (or equivalently all incoming operators). Otherwise this operator must have some numbers of $Z$ fields in the trace which are not present in the outgoing operators. We start from this longest operator and move around the string diagrams following the arrow direction that keeps those $Z$ fields which are not in the outgoing operators. Since we can always keep some of these $Z$ fields which can not go to an outgoing external edge, the path must eventually intersects itself and form a closed loop.  It is reassuring that we will see later this kind of situations will not appear in the correspondence with field theory diagrams.

The next step is to \textit{decorate} the vacuum diagrams with scalar excitations. When decorating the vacuum operators on the edge of the diagram with scalars, we make sure the vertices are still valid.  The string vertices up to two scalar insertions are described in (\ref{planar1}), and we will only need to use these vertices if the operator $O_1$ has no more than 2 scalar insertions. The same vacuum diagram could have many different decorations. The contribution of a string diagram is then computed by simply multiplying the vertices and summing over all possible ways of distributing the lengths of the intermediate trace operators. 

We note that a string state is characterized only by its length and string modes, in terms of the number of $Z$ fields and scalar insertions in the corresponding BMN operator. When we compute the contribution of a string diagram, we need to be careful in summing only different processes.  For example, when we consider the one-loop string propagation diagram in Fig. \ref{S6} in the next section, we see that in the undecorated vacuum string diagram, the string $O^J$ can split into $O^{xJ}O^{(1-x)J}$, while the excited state can split like $O^{J}_{-m,m}\rightarrow O^{xJ}_{0}O^{(1-x)J}_0$ or $O^{xJ}_{-k,k}O^{(1-x)J}$. For the vacuum diagram, we only need to sum over states with the integral $\int_0^{\frac{1}{2}} Jdx$, since the switch $x \rightarrow 1-x$ gives the same process. But for the decorated string diagrams, we need to integrate  $\int_0^{1} Jdx$, since the two smaller operators are distinguished by their scalar insertions. 

Two string diagrams are said to be the same shape if they are the decorations of the same vacuum diagram.  We group the string diagrams of the same shape together. We denote their total contribution $S_i$ where $i$ labels the undecorated diagram of vacuum operators, or a group of string diagrams with the same shape.

\subsection{Determine the multiplicity and factorization relations}

A short process consists of only an initial and a final state. We can \textit{extend} a short process into a \textit{long process} by filling in the intermediate steps. In each step we can either split a string into two strings or joining two strings into one. If the final state is a multi-string state, we use a subscript to denote the string once it has reached the final state and no longer changes. Since each step is a string splitting or joining process, we see that for a long process we can draw a string diagram of vacuum operators, or an undecorated string diagram.  We note that the string diagrams no longer contain information about the labeling of the segments of strings, so different long processes can map to the same string diagram. For example, in our calculations of the correlator $\bra O^{J}_{-m,m}O^{J_1}_0O^{J_2}_0O^{J_3}\ket$ in Sec. \ref{planarsection}, we find two long processes $(1234)\rightarrow  (341)(2)_1   \rightarrow (13)_2(4)_3 (2)_1$ and $(1234)\rightarrow  (123) (4)_1  \rightarrow  (31)_3(2)_2(4)_1$ correspond to the same string diagram, the first diagram in Fig. \ref{S1}.

A genus $h$ field theory diagram can be always extended into a long process that maps to an undecorated vacuum string diagram of $h$ loops. We will only consider string diagrams of minimal number of loops. This is an alternative ways of determining the genus of a field theory diagram by counting the minimal number of loops in the corresponding long process and (undecorated) string diagram, which seems less cumbersome than counting the power of $N$ in t'Hoof double line notation in field theory. 

It is easy to see the forbidden examples of string diagrams depicted in Fig. \ref{closeloop} can not appear when we extend a short process into a long process, because we can only combine and split strings already in the process. The first operator among a hypothetical closed loop of arrows to appear in a long process would have no where to come from.

\begin{figure}
  \begin{center}
  \includegraphics[width=6.5in]{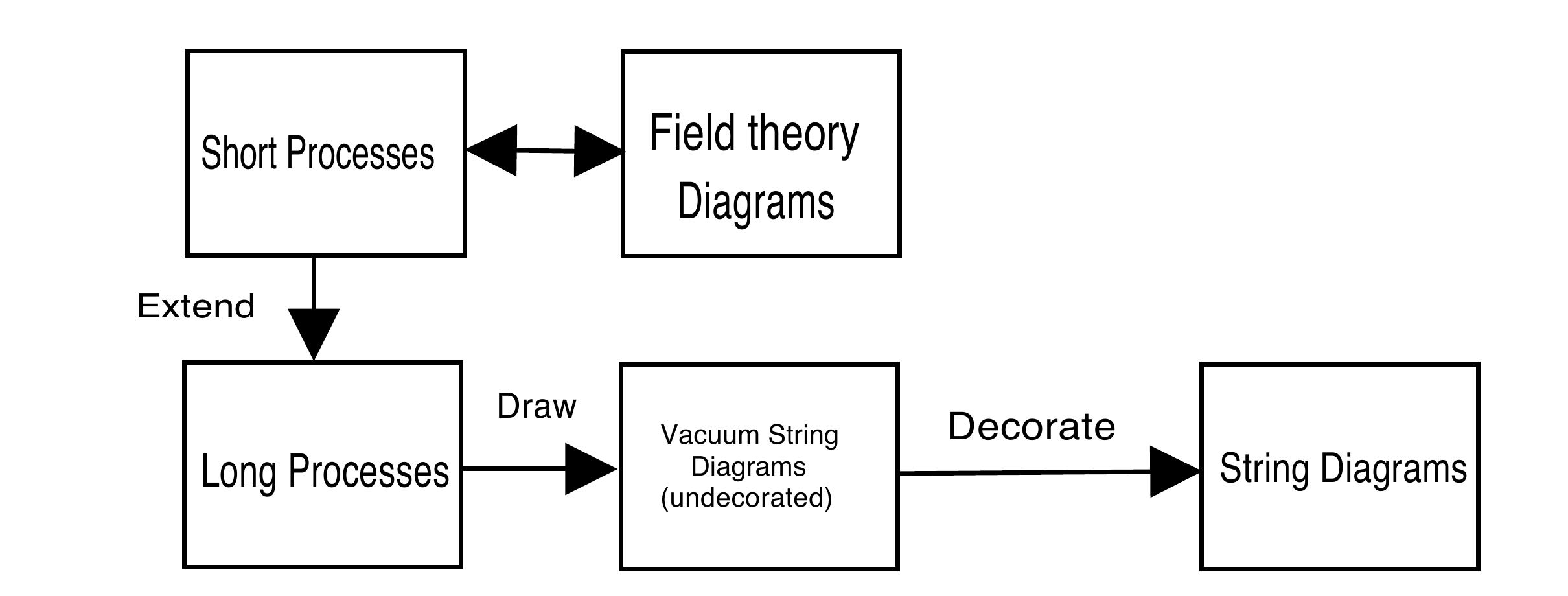} 
\end{center}
\caption{The logic between various components in the construction of factorization rules. Here the bidirectional arrow denotes the one-to-one correspondence between field theory diagrams and the short processes. The extension of short processes to long processes and the decoration are one-to-many operations, while the map from long processes to undecorated string diagrams is a many-to-one operation. }  \label{logic}
\end{figure}

Each short process may be extended into many long processes. We denote as $m_{ij}$ the number of appearance of an undecorated vacuum string diagram $i$ in the long processes associated with a short process $j$, and we call it the \textit{multiplicity} of string diagrams of shape $i$ with respect to field theory diagram $j$, which is a non-negative integer by definition. Then the statement of the \textit{factorization} is the relation 
\begin{eqnarray} \label{factorization}
S_i=\sum_j m_{ij}F_j, ~~~\textrm{for any}~ i
\end{eqnarray} 
We call it the factorization relation because in matrix form, the right hand side of the above equation is a product of two matrices. The factorization relation expresses the contributions of the string diagrams of the same shapes in terms of field theory diagrams. The reverse is not necessarily true. However, we find the total contributions are always proportional, namely,
\begin{eqnarray}
\sum_i S_i=m \sum_j F_j, 
\end{eqnarray} 
where $m \equiv \sum_i m_{ij}$ for any $j$. We can write the total contributions to the correlator at genus $h$   as 
\begin{eqnarray}
\bra \bar{O}_1 O_2 \ket_{\textrm{genus}~ h} =\ \sum_j F_j=\frac{1}{m} \sum_i S_i
\end{eqnarray}
Since the string diagrams are constructed by 3-string vertices, which are correlators of a single trace BMN operator with a double trace BMN operator, we see that the factorization induces recursion relations among the BMN correlators. We depict the logic between various components in the construction in Fig. \ref{logic}.

In the above constructions, we derive the multiplicity of string diagrams starting from field theory diagrams. One can also do this in reverse, and constructs the long processes associated with a string diagram to determine the multiplicity.  To do this, we start with a string $(1,2,\cdots, n)$ and perform the splitting and joining operations according to a string diagram, and keep those long processes whose end states are irreducible from combining segments. Actually this is much more convenient at higher genus as we will see that the number of field theory diagrams becomes much larger than that of the string diagrams at large genus. But we need to be careful of some redundant counting when the final state is a multi-string state. To illustrate, we consider the first string diagram $S^{(1)}_1$ in Fig. \ref{S1} in Section \ref{planarsection} as an example. Denoting the initial state as (1234), we can actually produce 8 long processes according to the string diagrams as the followings 
\begin{eqnarray} \label{8process}
1. && (1234)\rightarrow (234)(1)_1 \rightarrow (24)_2(3)_3(1)_1 \nonumber \\
2. && (1234)\rightarrow (234)(1)_1 \rightarrow (24)_3(3)_2(1)_1 \nonumber \\
3. && (1234)\rightarrow (134)(2)_1 \rightarrow (13)_2(4)_3(2)_1 \nonumber \\
4. && (1234)\rightarrow (134)(2)_1 \rightarrow (13)_3(4)_2(2)_1 \nonumber \\
5. && (1234)\rightarrow (124)(3)_1 \rightarrow (24)_2(1)_3(3)_1 \nonumber \\
6. && (1234)\rightarrow (124)(3)_1 \rightarrow (24)_3(1)_2(3)_1 \nonumber \\
7. && (1234)\rightarrow (123)(4)_1 \rightarrow (13)_2(2)_3(4)_1 \nonumber \\
8. && (1234)\rightarrow (123)(4)_1 \rightarrow (13)_3(2)_2(4)_1
\end{eqnarray}
We note that a process such as $(1234)\rightarrow (234)(1)_1 \rightarrow (23)_2(4)_3(1)_1$ is reducible because we can combine segments $(23)$ together, so it is not admissible. But out of the above 8 irreducible processes (\ref{8process}), we find only the final states of the 3rd and 8th processes can be identified with those of the field theory diagrams $F^{(1)}_2$ and $F^{(1)}_3$ in Fig. \ref{F1}. What about the other processes? The final states of the other processes could be also identified with those of $F^{(1)}_2$ or $F^{(1)}_3$ if we cyclically rotate the initial state. For example, the final state of the first process in (\ref{8process}) above could be identified with that of $F^{(1)}_2$ if we relabeling the initial state as $(2341)$ instead of $(1234)$. There are $n$ cyclic rotations for a $n$-segment initial state, but they are really the same state. We actually have already taken account for these contributions when we compute the field theory diagrams so we don't have to count them again. So in this example we only need to look at the 3rd and 8th processes whose final states can be exactly identified with the field theory diagrams $F^{(1)}_2$ and $F^{(1)}_3$, and disregard the other processes. Of course this issue will not appear when the final state is also a single string state because a cyclic  rotation of the initial state has the same effect as that of the final state in opposite direction.

\section{Higher genus BMN correlators} \label{highergenussection}

We now test the factorization relation (\ref{factorization}) for higher genus BMN correlators. We discuss several cases.

\subsection{Torus correlator between two single trace operators}

\begin{figure}
  \begin{center}
  \includegraphics[width=6.5in]{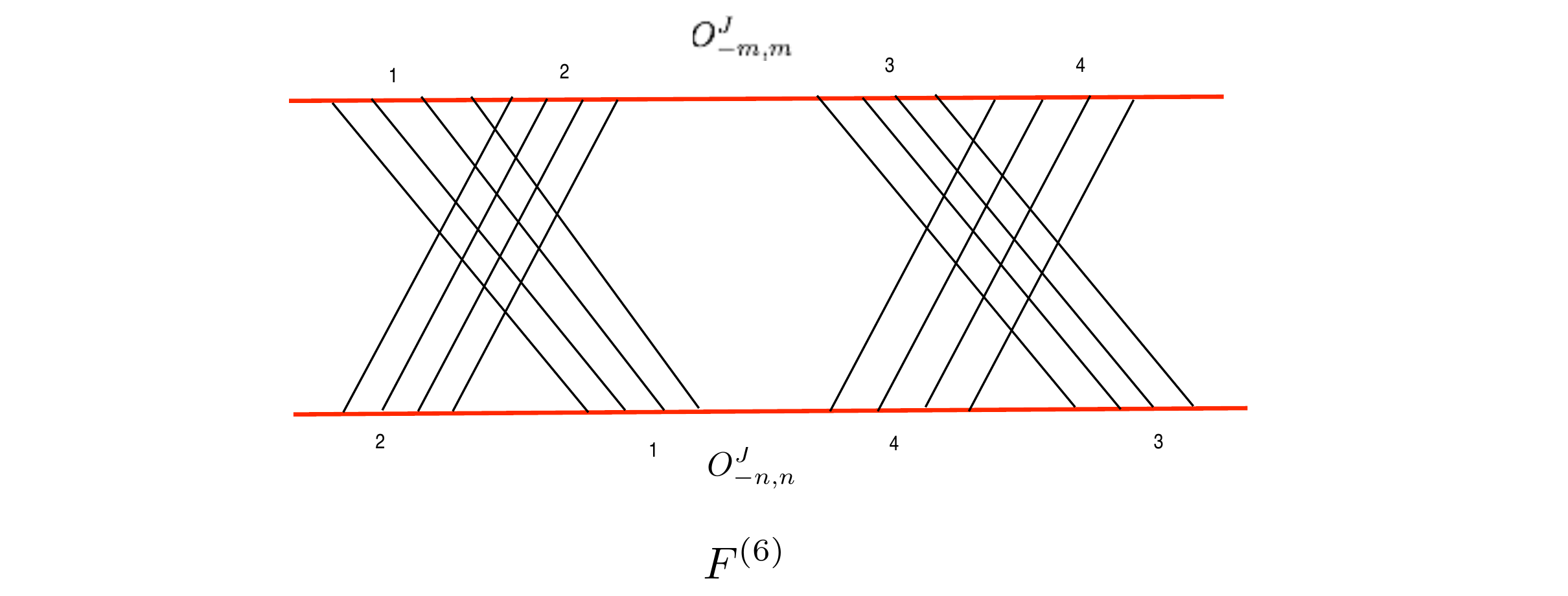} 
\end{center}
\caption{The torus correlator of $\bra \bar{O}_{-m,m}^J O_{-n,n}^J \ket_{\textrm{torus}}$. This is the only diagram and we denote its contribution $F^{(6)}$.}  \label{F6}
\end{figure}

This case describes the one loop string propagation process, and has been considered in \cite{Huang2}. We include it here for completeness. We consider the correlator between two BMN operators $\bra \bar{O}_{-m,m}^J O_{-n,n}^J \ket_{\textrm{torus}}$. For genus one the single trace  operator can be divided into at most 4 segments. The field theory diagram is depicted in Fig. \ref{F6}, and the corresponding short process 
\begin{eqnarray} \label{F6short}
(1234)\rightarrow (2143)
\end{eqnarray}
This is the only short process for the correlator. We note that one may also write e.g. a short process $(1234)\rightarrow (1432)$, but it is equivalent to (\ref{F6short}) due to the cyclicality of the trace. The calculations of the correlator are first done e.g. in \cite{Constable1}, and results are 

\begin{eqnarray}\label{F6torus}
 && F^{(6)}  \equiv \bra \bar{O}_{-m,m}^J O_{-n,n}^J \ket_{\textrm{torus}}  \\ 
&=& \left\{
\begin{array}{cl}
\frac{g^2}{24},    &   m=n=0;   \\
0,              &  m=0, n\neq0,   \\
&  \textrm{or}~n=0, m\neq0; \\
g^2(\frac{1}{60} - \frac{1}{24 \pi^2 m^2} + \frac{7}{16 \pi^4 m^4}),   &  m=n\neq0; \\
\frac{g^2}{16\pi^2m^2} ( \frac{1}{3}+\frac{35}{8\pi^2m^2}),  &  m=-n\neq0;    \nonumber \\
\frac{g^2}{4\pi ^{2}(m-n)^2} ( \frac{1}{3}+\frac{1}{\pi
^2n^2}+\frac{1}{\pi ^2m^2}-\frac{3}{2\pi ^2mn}-\frac{1}{2\pi
^2(m-n)^2}) & \textrm{all~other~cases} 
\end{array}
\right.
\end{eqnarray}

One can also calculate this result using matrix model method \cite{KPSS}. Some contributions from the connected diagrams in matrix models can be organized into generating functions known as resolvents in matrix models, which can be computed using loop equations in matrix model \cite{EK}. The loop equations provide recursion relations for the resolvents in matrix models, but they seem very different from the factorization relations studied here in this paper. Nevertheless it would be interesting to investigate whether there are some connections between these relations. 

\begin{figure}
  \begin{center}
  \includegraphics[width=6.5in]{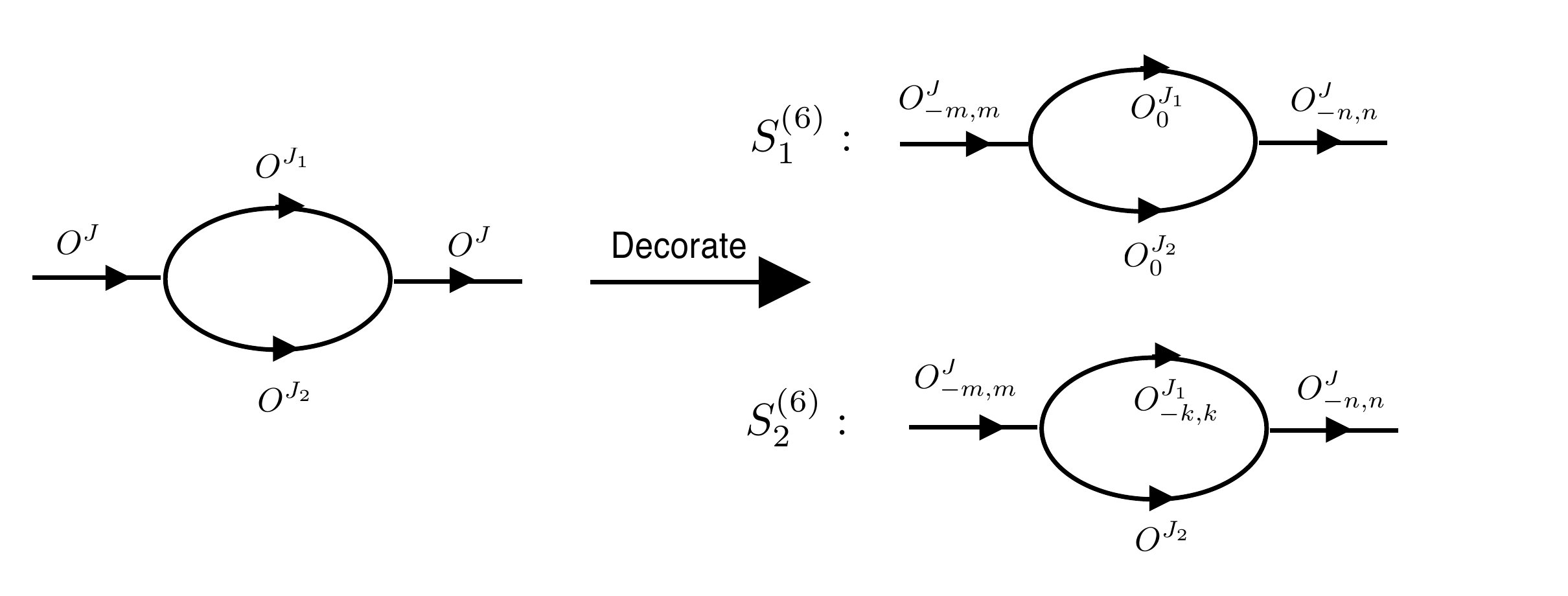} 
\end{center}
\caption{The string diagrams of $\bra \bar{O}_{-m,m}^J O_{-n,n}^J \ket_{\textrm{torus}}$. We start with diagram of vacuum operators and decorate it with scalar excitations. We denote the contributions of the two decorated diagrams $S^{(6)}_1$ and $S^{(6)}_2$ and their total contributions $S^{(6)}$.}  \label{S6}
\end{figure}

The string diagrams are drawn in Fig. \ref{S6}. There is only one undecorated diagram of vacuum operators, from which we generate two decorated string diagrams of the same shape. To derive its multiplicity, we extend the short process into long processes, and we find there are two ways of extension 
\begin{eqnarray} \label{F6long}
&&(1234)\rightarrow (12)(34) \rightarrow (2143) \nonumber \\
&&(1234) \rightarrow (41)(23) \rightarrow (1432)
\end{eqnarray}
where we have freely used the cyclicality of the strings.  So we find the multiplicity of the only (undecorated) string diagram is 2.  We also note that in the first decorated  diagram $S^{(6)}_1$ the two operators $O^{J_1}_0$ and $O^{J_2}_0$ have different scalar insertions $\phi^1$ and $\phi^2$, so the range $J_1$ can go from  $0$ to $J$. Using the vertex formulae (\ref{planar1}), the contributions of decorated diagrams are computed as  

\begin{eqnarray}\label{S6162}
S^{(6)}_1&=&\sum_{J_1=0}^{J}\langle \bar{O}^J_{-m,m} O^{J_1}_0 O^{J-J_1}_0
\rangle_{\textrm{planar}} \langle \bar{O}^{J_1}_0 \bar{O}^{J-J_1}_0
O^J_{-n,n} \rangle_{\textrm{planar}} \nonumber \\&&
=g^2\int_0^1dx\frac{\sin^2(m\pi x)}{m^2\pi^2}\frac{\sin^2(n\pi
x)}{n^2\pi^2} \nonumber \\
S^{(6)}_2&=&\sum_{J_1=0}^{J}\sum_{k=-\infty}^{+\infty} \langle
\bar{O}^J_{-m,m} O^{J_1}_{-k,k} O^{J_2} \rangle_{\textrm{planar}} \langle
\bar{O}^{J_1}_{-k,k} \bar{O}^{J_2} O^J_{-n,n}
\rangle_{\textrm{planar}}\\&&\nonumber =g^2\sum_{k=-\infty}^{+\infty}
\int_0^1dxx^{3}(1-x)\frac{\sin^2(m\pi
x)}{\pi^2(mx-k)^2}\frac{\sin^2(n\pi x)}{\pi^2(nx-k)^2}
\end{eqnarray}
Performing the sums and integrals we check the factorization relation 
\begin{eqnarray} \label{S6F6}
S^{(6)}=S^{(6)}_1+S^{(6)}_2=2F^{(6)}
\end{eqnarray}

Surprisingly, two additional identities similar to the factorization relation (\ref{S6F6}) were pointed out in \cite{Constable2}, and one of which involves field theory one-loop corrections to the correlator  $\bra \bar{O}_{-m,m}^J O_{-n,n}^J \ket_{\textrm{torus}}$. These two identities involve only the second decorated diagram in Fig. \ref{S6}, and amount to putting  a weight of $\frac{Jk}{J_1}$ or  $(\frac{Jk}{J_1})^2$ to the propagating edge of the operator $O^{J_1}_{-k,k}$. It is clearly interesting to see  whether it is possible to generalize these additional relations to higher genus, and to generalize the factorization rules we propose in Sec. \ref{factorizationsection} to include them.

\subsection{One loop string calculations with integral form of the vertex} \label{integralsection}

In the previous section we check the factorization relation for one loop string propagation by direct computations. But it seems a little mysterious how the factorization works, and it is not quite satisfying that we have to check each of the 5 cases in (\ref{F6torus}) separately. Here we provide a more illuminating and more unifying derivation by reducing the integrals into sums of some standard integrals (\ref{integral1}), which were used in \cite{Constable1} to calculate the higher genus correlators of 2 single trace operators. We provide some descriptions of the approach in Appendix \ref{2point}. 

For the genus one case, the correlator can be written as a sum of 5 standard integrals (\ref{integral1}) as the followings 
\begin{eqnarray} \label{F6integral} 
F^{(6)} &=& g^2 [I_{(1,5)}(2\pi i (m-n),0)+ I_{(1,5)}(-2\pi i (m-n),0)+I_{(2,2,2)}(2\pi im, 2\pi i n,0) \nonumber \\ && 
+I_{(2,2,2)}(-2\pi im, - 2\pi i n,0)
+I_{(2,1,1,2)}(2\pi i(m-n), 2\pi i m,-2\pi i n, 0) ], 
\end{eqnarray}
We note that the integral (\ref{integral1}) is invariant if we add an integer multiple of $2\pi i$ to the all the arguments. When some arguments in the standard integrals are identical, we need to combine them according to (\ref{combine}) before we can use (\ref{integral3}, \ref{integral4}) to compute them. Here the degeneracy happens when some of  the $m$, $n$, $m-n$ or $m+n$ vanish.

We then calculate the string diagram contributions (\ref{S6162}) using the integral form of the vertices (\ref{integralform}). To reduce the contributions to the standard integrals (\ref{integral1}), we need to carefully dissect the multi-dimensional integration domain and perform some tricky changes the integration variables, so that in each sector, the integration domain can be identified with that of a standard integral. For the first diagram we find 
\begin{eqnarray}
S^{(6)}_1 = g^2 \int_0^1dx (\int_0^x dy_1d\tilde{y}_1 e^{2\pi i (m y_1-n\tilde{y}_1) }) (\int_x^1 dy_2d\tilde{y}_2 e^{ - 2\pi i (m y_2-n\tilde{y}_2) })
\end{eqnarray}
We discuss several situations separately in the followings. 
\begin{enumerate}
\item $y_1<\tilde{y}_1$, $y_2<\tilde{y}_2 $. We change variables as $z_1=y_1, z_2=\tilde{y}_1-y_1, z_3=x-\tilde{y}_1$, $z_4=y_2-x$, $z_5=\tilde{y}_2-y_2$, $z_6=1-\tilde{y}_2$. Then the contribution becomes 
\begin{eqnarray}
S^{(6)}_{1,1} &=& g^2 \int_0^1 dz_1\cdots dz_6 \delta(\sum_{i=1}^6 z_i-1)  e^{2\pi i [(m-n)(z_1+z_6)-nz_2+mz_5] }
\nonumber \\
&=& g^2 I_{(2,1,1,2)}(2\pi i(m-n), 2\pi i m,-2\pi i n, 0)
\end{eqnarray}

\item $y_1<\tilde{y}_1$, $\tilde{y}_2<y_2 $. We change variables as $z_1=y_1, z_2=\tilde{y}_1-y_1, z_3=x-\tilde{y}_1$, $z_4=\tilde{y}_2-x$, $z_5=y_2-\tilde{y}_2$, $z_6=1-y_2$. Then the contribution becomes 
\begin{eqnarray}
S^{(6)}_{1,2} &=& g^2 \int_0^1 dz_1\cdots dz_6 \delta(\sum_{i=1}^6 z_i-1)  e^{2\pi i [(m-n)(z_1+z_6)-n(z_2+z_5)] }
\nonumber \\
&=& g^2 I_{(2,2,2)}(2\pi i(m-n),-2\pi i n, 0)
\nonumber \\
&=& g^2 I_{(2,2,2)}(2\pi im,2\pi i n, 0)
\end{eqnarray}

\item $\tilde{y}_1<{y}_1$, $y_2<\tilde{y}_2 $. We change variables as $z_1=\tilde{y}_1, z_2=y_1-\tilde{y}_1, z_3=x-{y}_1$, $z_4=y_2-x$, $z_5=\tilde{y}_2-y_2$, $z_6=1-\tilde{y}_2$. Then the contribution becomes 
\begin{eqnarray}
S^{(6)}_{1,3} &=& g^2 \int_0^1 dz_1\cdots dz_6 \delta(\sum_{i=1}^6 z_i-1)  e^{2\pi i [(m-n)(z_1+z_6)+m(z_2+z_5)] }
\nonumber \\
&=& g^2 I_{(2,2,2)}(2\pi i(m-n), 2\pi i m, 0)
\nonumber \\
&=& g^2 I_{(2,2,2)}(-2\pi i m, - 2\pi i n, 0)
\end{eqnarray}

\item $\tilde{y}_1<{y}_1$, $\tilde{y}_2<y_2 $. We change variables as $z_1=\tilde{y}_1, z_2=y_1-\tilde{y}_1, z_3=x-{y}_1$, $z_4=\tilde{y}_2-x$, $z_5=y_2-\tilde{y}_2$, $z_6=1-{y}_2$. Then the contribution becomes 
\begin{eqnarray}
S^{(6)}_{1,4} &=& g^2 \int_0^1 dz_1\cdots dz_6 \delta(\sum_{i=1}^6 z_i-1)  e^{2\pi i [(m-n)(z_1+z_6)+mz_2-nz_5] }
\nonumber \\
&=& g^2 I_{(2,1,1,2)}(2\pi i(m-n), 2\pi i m, -2\pi i n, 0)
\end{eqnarray}

\end{enumerate}
Summing up together the contributions, we find 
\begin{eqnarray} \label{S61integral}
S^{(6)}_1 &=&  S^{(6)}_{1,1}+S^{(6)}_{1,2}+S^{(6)}_{1,3}+S^{(6)}_{1,4} \nonumber \\
&=& g^2[I_{(2,2,2)}(2\pi i m, 2\pi i n, 0)+I_{(2,2,2)}(-2\pi im, -2\pi i n, 0) \nonumber \\ && 
+2 I_{(2,1,1,2)}(2\pi i(m-n), 2\pi i m, -2\pi i n, 0)]
\end{eqnarray}

For the second diagram $S^{(6)}_2$, we write the formula in (\ref{S6162}) using the integral form of the vertices and perform the summation over string mode using the summation formula (\ref{deltafunction}). The result is 
\begin{eqnarray}
S^{(6)}_2 &=& g^2 \int_0^1 (1-x)dx \int_0^x dy_1dy_2 dy_3 dy_4 e^{2\pi i [m(y_1-y_2)+n(y_3-y_4)]} \nonumber \\
&& \times \sum_{k=-\infty}^{\infty} \delta(y_1-y_2+y_3-y_4-kx)   
\end{eqnarray} 
We should integrate $y_1$ to cancel the delta function. Since $-x<y_2-y_3+y_4<2x$, we discuss several cases as the followings.

\begin{enumerate}
\item $-x<y_2-y_3+y_4<0$. The integral of $y_1$ over the delta function fixes $y_1=y_2-y_3+y_4+x$. We change variables $z_1=y_2$, $z_2=y_4$, $z_3=x-y_3$, such that $0<z_1,z_2,z_3, z_1+z_2+z_3<x$. Then we can also further change variable $z_4=x-z_1-z_2-z_3$ such that $0<z_4<x$. Furthermore, we can write $1-x= \int_0^{1-x}dz_5$, and $z_6=1-x-z_5$. The contribution becomes 
\begin{eqnarray}
S^{(6)}_{2,1} &=& g^2 \int_0^1 dz_1\cdots dz_6 \delta(\sum_{i=1}^6 z_i-1)  e^{2\pi i [m(z_2+z_3)+n(z_1+z_4)] }
\nonumber \\
&=& g^2 I_{(2,2,2)}(2\pi im, 2\pi i n, 0)
\end{eqnarray}
 
\item $x<y_2-y_3+y_4<2x$. The integral of $y_1$ over the delta function fixes $y_1=y_2-y_3+y_4-x$. We change variables $z_1=x-y_2$, $z_2=x-y_4$, $z_3=y_3$, such that $0<z_1,z_2,z_3, z_1+z_2+z_3<x$. Then we can also further change variable $z_4=x-z_1-z_2-z_3$ such that $0<z_4<x$. Furthermore, we can write $1-x= \int_0^{1-x}dz_5$, and $z_6=1-x-z_5$. The contribution becomes 
\begin{eqnarray}
S^{(6)}_{2,2} &=& g^2 \int_0^1 dz_1\cdots dz_6 \delta(\sum_{i=1}^6 z_i-1)  e^{2\pi i [-m(z_2+z_3)-n(z_1+z_4)] }
\nonumber \\
&=& g^2 I_{(2,2,2)}(-2\pi im, -2\pi i n, 0)
\end{eqnarray}
 
\item $0<y_2-y_3+y_4<x$. The integral of $y_1$ over the delta function fixes $y_1=y_2-y_3+y_4$. The range of $y_3$ is $y_2+y_4-x<y_3<y_2+y_4$. This does not quite fit into the integration domain of $y_3$ which is $[0,x]$ so makes this case more complicated. We further discuss several situations. 

\begin{enumerate}
\item $y_2+y_4<x$, $y_3<y_4$. Then we can change integration variables $z_1=y_3$, $z_2=y_4-y_3$, $z_3=y_2$ such that the integration domain of $z_1,z_2,z_3$ is $0<z_1,z_2,z_3, z_1+z_2+z_3<x$. The rests are similar to previous case $z_4=x-z_1-z_2-z_3$, $1-x= \int_0^{1-x}dz_5$, and $z_6=1-x-z_5$. The contributions in this case is 
\begin{eqnarray}
S^{(6)}_{2,3} &=& g^2 \int_0^1 dz_1\cdots dz_6 \delta(\sum_{i=1}^6 z_i-1)  e^{2\pi i (m-n)z_2}
\nonumber \\
&=& g^2 I_{(1,5)}(2\pi i(m-n), 0)
\end{eqnarray}

\item $y_2+y_4<x$, $y_3>y_4$. Then we can change integration variables $z_1=y_4$, $z_2=y_3-y_4$, $z_3=y_2-y_3+y_4$ such that the integration domain of $z_1,z_2,z_3$ is $0<z_1,z_2,z_3, z_1+z_2+z_3<x$. The rests are similar to previous case $z_4=x-z_1-z_2-z_3$, $1-x= \int_0^{1-x}dz_5$, and $z_6=1-x-z_5$. The contributions in this case is 
\begin{eqnarray}
S^{(6)}_{2,4} &=& g^2 \int_0^1 dz_1\cdots dz_6 \delta(\sum_{i=1}^6 z_i-1)  e^{-2\pi i (m-n)z_2 }
\nonumber \\
&=& g^2 I_{(1,5)}(-2\pi i(m-n), 0)
\end{eqnarray}

\item $y_2+y_4>x$, $y_3<y_4$. Then we can change integration variables $z_1=x-y_4$, $z_2=y_4-y_3$, $z_3=x-y_2+y_3-y_4$ such that the integration domain of $z_1,z_2,z_3$ is $0<z_1,z_2,z_3, z_1+z_2+z_3<x$. The rests are similar to previous case $z_4=x-z_1-z_2-z_3$, $1-x= \int_0^{1-x}dz_5$, and $z_6=1-x-z_5$. The contributions in this case is 
\begin{eqnarray}
S^{(6)}_{2,5} &=& g^2 \int_0^1 dz_1\cdots dz_6 \delta(\sum_{i=1}^6 z_i-1)  e^{2\pi i (m-n)z_2 }
\nonumber \\
&=& g^2 I_{(1,5)}(2\pi i(m-n), 0)
\end{eqnarray}

\item $y_2+y_4>x$, $y_3>y_4$. Then we can change integration variables $z_1=x-y_3$, $z_2=y_3-y_4$, $z_3=x-y_2$ such that the integration domain of $z_1,z_2,z_3$ is $0<z_1,z_2,z_3, z_1+z_2+z_3<x$. The rests are similar to previous case $z_4=x-z_1-z_2-z_3$, $1-x= \int_0^{1-x}dz_5$, and $z_6=1-x-z_5$. The contributions in this case is 
\begin{eqnarray}
S^{(6)}_{2,6} &=& g^2 \int_0^1 dz_1\cdots dz_6 \delta(\sum_{i=1}^6 z_i-1)  e^{-2\pi i (m-n)z_2 }
\nonumber \\
&=& g^2 I_{(1,5)}(-2\pi i(m-n), 0)
\end{eqnarray}

\end{enumerate}

\end{enumerate} 
Putting together the contributions 
\begin{eqnarray} \label{S62integral}
S^{(6)}_2 &=& g^2 [2I_{(1,5)}(2\pi i (m-n),0)+ 2 I_{(1,5)}(-2\pi i (m-n),0) \nonumber \\ && 
+I_{(2,2,2)}(2\pi im, 2\pi i n,0)  +I_{(2,2,2)}(-2\pi im, - 2\pi i n,0) ]
\end{eqnarray} 
Having written both the field theory diagram and string diagram contributions in terms of the standard integrals, we can easily check the factorization $S^{(6)}=2F^{(6)}$ using (\ref{F6integral}, \ref{S61integral}, \ref{S62integral}).

We can also derive the additional identities pointed out in \cite{Constable2} in this way. These identities modify the sum $S^{(6)}_2$ in (\ref{S6162}) by a factor of $\frac{k}{x}$ and $\frac{k^2}{x^2}$. We denote the modified sums as $S^{(6)}_3$ and  $S^{(6)}_4$, and they are
\begin{eqnarray} \label{S634}
S^{(6)}_3 &=&  \int_0^1 Jdx \sum_{k=-\infty}^{+\infty}\frac{k}{x} \langle
\bar{O}^J_{-m,m} O^{xJ}_{-k,k} O^{(1-x)J} \rangle \langle
\bar{O}^{xJ}_{-k,k} \bar{O}^{(1-x)J} O^J_{-n,n}
\rangle \\
S^{(6)}_4 &=&  \int_0^1 Jdx \sum_{k=-\infty}^{+\infty}\frac{k^2}{x^2} \langle
\bar{O}^J_{-m,m} O^{xJ}_{-k,k} O^{(1-x)J} \rangle \langle
\bar{O}^{xJ}_{-k,k} \bar{O}^{(1-x)J} O^J_{-n,n}
\rangle
\end{eqnarray}
Then the identities  are the followings 
\begin{eqnarray}
S^{(6)}_3 &=& (m+n) F^{(6)},  \label{S63factorization} \\ 
S^{(6)}_4 &=& (m^2+n^2) F^{(6)}+\frac{1}{4\pi^2} B_{m,n} \label{S64factorization}
\end{eqnarray}
where $B_{m,n}$ comes from the torus one-loop field theory correlator of BMN operators $O^{J}_{-m,m}$ and $O^{J}_{-n,n}$. In this paper we discuss mostly free field theory, and the higher genus correlators correspond to string loop amplitudes. But the last identity involves higher order contributions of  both genus and loop in field theory.  It was shown that the one-loop field theory contributions to the correlator $\bra \bar{O}^{J}_{-m,m} O^{J}_{-n,n} \ket$  at higher genus can be also written in terms of the standard integrals (\ref{integral1}). 

To derive these identities, we perform the sum over string modes with the derivatives of summation formula (\ref{deltafunction}). The results involve derivatives of the Dirac delta function
\begin{eqnarray} \label{S63}
S^{(6)}_3 &=&  g^2 \int_0^1 (1-x)dx \int_0^x dy_1dy_2 dy_3 dy_4 e^{2\pi i [m(y_1-y_2)+n(y_3-y_4)]} \nonumber \\
&& \times (-\frac{1}{2\pi i})  \sum_{k=-\infty}^{\infty} \delta^\prime (y_1-y_2+y_3-y_4-kx) ,   \\
S^{(6)}_4 &=& g^2 \int_0^1 (1-x)dx \int_0^x dy_1dy_2 dy_3 dy_4 e^{2\pi i [m(y_1-y_2)+n(y_3-y_4)]} \nonumber \\
&& \times \frac{1}{2\pi i} \sum_{k=-\infty}^{\infty}  \delta^{\prime\prime} (y_1-y_2+y_3-y_4-kx)  \label{S64}
\end{eqnarray} 

We can use integration by part for one of the variables $y_1$ to eliminate the derivatives in the delta function. It turn out the boundary terms of the integration by part also contribute. For the case of $S^{(6)}_3$ in (\ref{S63}) we find 
\begin{eqnarray}
S^{(6)}_3 &=&   g^2 \int_0^1 (1-x)dx \int_0^x dy_2 dy_3 dy_4\times \sum_{k=-\infty}^{\infty} \nonumber \\
&& [  \frac{1}{2\pi i}  (e^{2\pi i [-my_2+n(y_3-y_4)]}-e^{2\pi i [m(x-y_2)+n(y_3-y_4)]})  \delta (y_3-y_2-y_4-kx) \nonumber \\
&&  +  m \int_0^x dy_1 e^{2\pi i [m(y_1-y_2)+n(y_3-y_4)]}   \delta (y_1-y_2+y_3-y_4-kx) ]
\end{eqnarray}  
where the third line is exactly as we have done for $S^{(6)}_2$ before, and the second line comes from the boundary term of integration by part and can be computed similarly. The result of the computations are 
\begin{eqnarray} 
S^{(6)}_3 &=& g^2 \{ 2m I_{(1,5)}(2\pi i (m-n),0)+ 2m I_{(1,5)}(-2\pi i (m-n),0) \nonumber \\ && 
+m I_{(2,2,2)}(2\pi im, 2\pi i n,0)  +m I_{(2,2,2)}(-2\pi im, - 2\pi i n,0)  \nonumber \\ && 
+ \frac{1}{2\pi i} [ I_{(1,4)}(-2\pi i (m-n),0) - I_{(1,4)}(2\pi i (m-n),0) \nonumber \\ &&
+I_{(2,1,2)}(-2\pi im, -2\pi i n,0) - I_{(2,1,2)}(2\pi im, 2\pi i n,0)  ]\} 
\end{eqnarray} 
Using the recursion relation  (\ref{recursive})  for the standard integral one can easily check the identity (\ref{S63factorization}). This derivation is also valid regardless whether there are degeneracies in the parameters $m,n$ since the recursion relation (\ref{recursive}) is valid in the degenerate cases as well.  In this way one can also derive the last identity (\ref{S64factorization}) by performing the integration by part twice for (\ref{S64}) and noting that the torus one-loop field theory contribution $B_{m,n}$ can be also written in terms of the standard integrals.

Comparing to direct computations, this approach to the factorization relation is independent of whether there are some degeneracies when some of the $m$, $n$, $m-n$ or $m+n$ vanish, so we do not have to check each case separately. In this respect this approach of calculations using the integral form of vertices looks more promising for a systematic proof of the factorization at higher genus and for BMN operators with more string modes. However, we have seen that the dissections of  the integration domains are very tricky when we compute the string diagrams.  In most of the paper, we still use the more straightforward and explicit method of direct computations of both string and field theory diagrams to check the factorization relation.

\subsection{Torus correlators between a single trace operator and a double trace operator} \label{oneloopcubicsection}
The first two cases have been studied in \cite{Huang2}. Here we include them for completeness. 

\subsubsection{Case one: the vacuum diagrams}

\begin{figure}
  \begin{center}
  \includegraphics[width=6.5in]{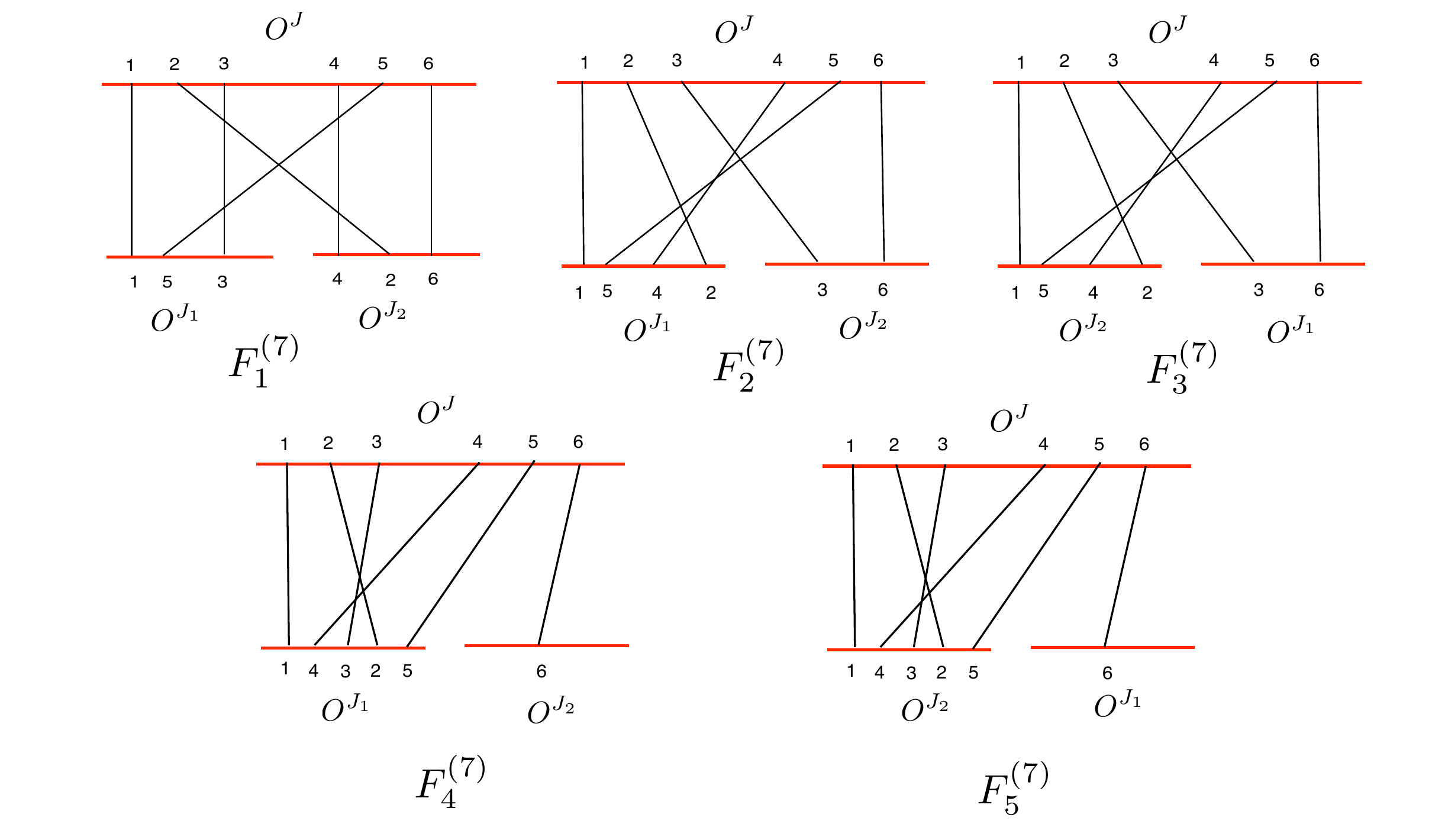} 
\end{center}
\caption{There are 5 diagrams contribute to the torus correlator 
 $\bra \bar{O}^J O^{J_1} O^{J_2} \ket$. We denote their contributions by $F^{(7)}_1$, $F^{(7)}_2$,
 $F^{(7)}_3$, $F^{(7)}_4$ and $F^{(7)}_5$ respectively. $F^{(7)}_2$, $F^{(7)}_4$ are related to $F^{(7)}_3$, $F^{(7)}_5$ by exchanging the operators $O^{J_1}$ and $O^{J_2}$. Here we use a single line to denote a segment of string consisting of a large number of $Z$ fields. We have checked these are the only torus diagrams.}
\label{F7}
\end{figure}

We first the case of vacuum operator $\bra \bar{O}^J O^{J_1} O^{J_2} \ket$ (where $J=J_1+J_2$) on the torus. There are 5 diagrams and they are depicted in Fig. \ref{F7}. This correlator is calculated in \cite{KPSS} using matrix model technique. Here we calculate the 5 diagrams separately to derive the multiplicity factors for the string diagrams.  The short processes of the 5 diagrams are the followings 
\begin{eqnarray} \label{F7short} 
F^{(7)}_1&:& (123456)\rightarrow (153)_1(426)_2 \nonumber \\
F^{(7)}_2&:& (123456)\rightarrow (1542)_1(36)_2 \nonumber \\
F^{(7)}_3&:& (123456)\rightarrow (1542)_2(36)_1 \nonumber \\
F^{(7)}_4&:& (123456)\rightarrow (14325)_1(6)_2 \nonumber \\
F^{(7)}_5&:& (123456)\rightarrow (14325)_2(6)_1
\end{eqnarray}

We count the combinatorics to compute the contribution of these diagrams.  We first pick the initial positions for segments in the three operators which contribute a factor of $JJ_1J_2$. Then for $F^{(7)}_1$ we divide both small operators $O^{J_1}$ and $O^{J_2}$ into 3 segments, so we have another factor of  $\frac{J_1^2J_2^2}{(2!)^2}$. However we have over-counted by a factor of $3$. To see the over-counting, we look at the short process for $F^{(7)}_1$  which is $(123456)\rightarrow (153)_1(426)_2 $. By cyclicality this is equivalent to $(345612)\rightarrow (315)_1(642)_2$, and if we relabel the initial operator by numerical order, we see this is the same as  before we do the cyclic rotation. The same is true for the cyclic rotations to  $(561236)\rightarrow  (531)_1(264)_2$. Putting together we find 
\begin{equation} \label{F71}
F^{(7)}_1=\frac{1}{12}\frac{g^{3}}{\sqrt{J}}\sqrt{x(1-x)}[x^2(1-x)^2]
\end{equation}
where we denote $x=J_1/J$. Similarly for the other contributions
\begin{eqnarray} \label{F72}
F^{(7)}_2 &=& \frac{1}{12}\frac{g^{3}}{\sqrt{J}}\sqrt{x(1-x)}x^3(1-x) \nonumber \\
F^{(7)}_3 &=& \frac{1}{12}\frac{g^{3}}{\sqrt{J}}\sqrt{x(1-x)}x(1-x)^3 \nonumber \\
F^{(7)}_4 &=& \frac{1}{24}\frac{g^{3}}{\sqrt{J}}\sqrt{x(1-x)}x^4 \nonumber \\
F^{(7)}_5 &=& \frac{1}{24}\frac{g^{3}}{\sqrt{J}}\sqrt{x(1-x)}(1-x)^4
\end{eqnarray}

\begin{figure}
  \begin{center}
  \includegraphics[width=6.5in]{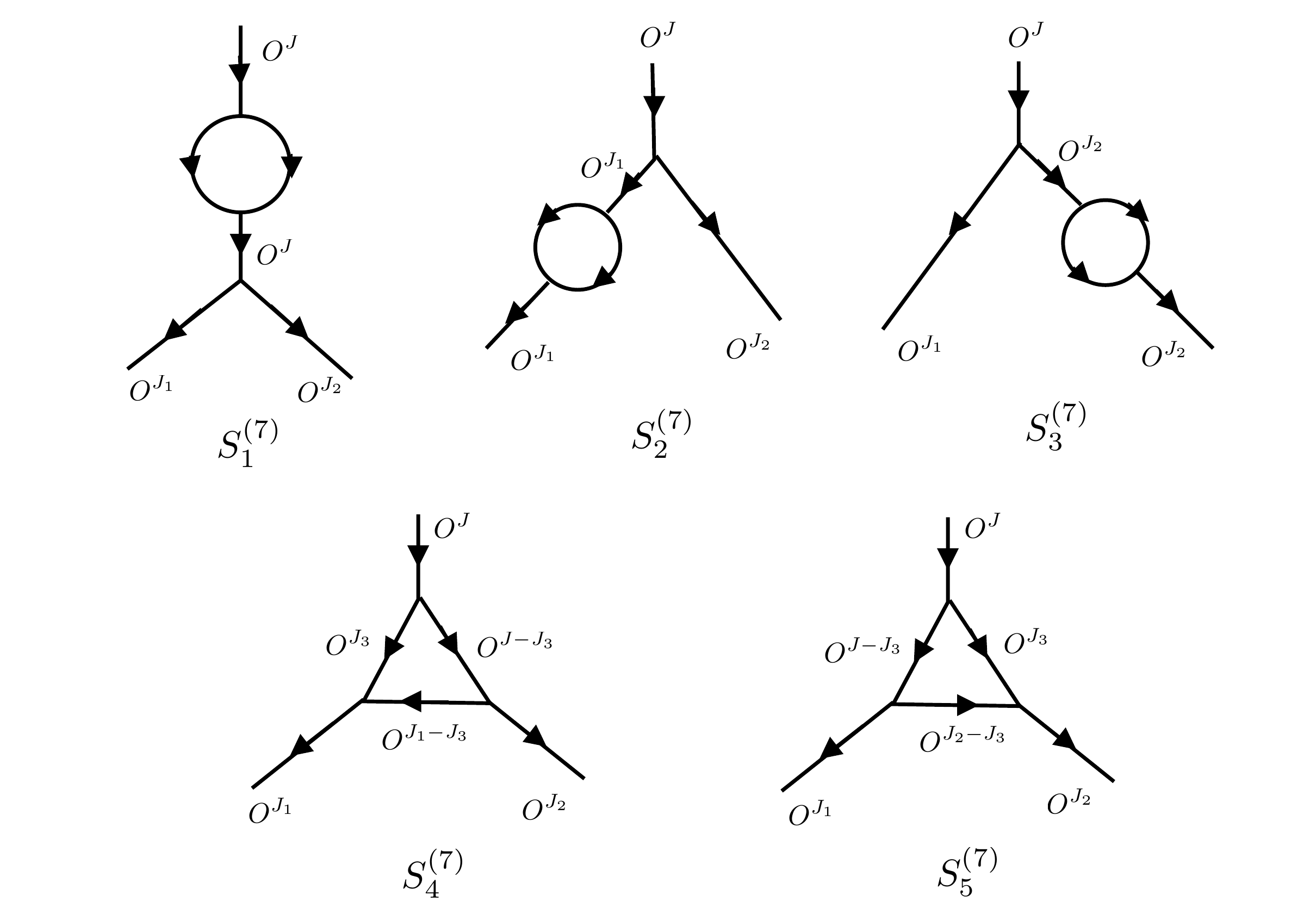} 
\end{center}
\caption{There are 5 one-loop string diagrams for correlator  $\bra \bar{O}^J O^{J_1} O^{J_2} \ket$. We denote their contributions by $S^{(7)}_1$, $S^{(7)}_2$, $S^{(7)}_3$, $S^{(7)}_4$ and $S^{(7)}_5$ respectively. $S^{(7)}_2$, $S^{(7)}_4$ are related to $S^{(7)}_3$, $S^{(7)}_5$ by exchanging the operators $O^{J_1}$ and $O^{J_2}$. }
\label{S7}
\end{figure}

We also draw the one-loop string diagrams in Fig. \ref{S7} and compute them as the followings
\begin{eqnarray} \label{S71}
S^{(7)}_1 &=& 2\bra \bar{O}^J O^J \ket_{\textrm{torus}}\bra \bar{O}^J O^{J_1}O^{J_2}\ket_{\textrm{\textrm{planar}}} \nonumber \\
&=& \frac{g^3}{\sqrt{J}}\frac{1}{12}\sqrt{x(1-x)} \nonumber \\
S^{(7)}_2 &=& 2\bra \bar{O}^J O^{J_1}O^{J_2}\ket_{\textrm{\textrm{planar}}} \bra \bar{O}^{J_1} O^{J_1} \ket_{\textrm{torus}} \nonumber \\
&=& \frac{g^3}{\sqrt{J}}\frac{1}{12}\sqrt{x(1-x)}x^4 \nonumber \\
S^{(7)}_3 &=& S^{(7)}_2(x\rightarrow 1-x) 
\end{eqnarray}
where the factor of $2$ is the multiplicity of one-loop string propagation with respect to the field theory torus correlator of two single trace operators discussed in (\ref{F6long}). And the  diagrams $S^{(7)}_4$ and $S^{(7)}_5$ are constructed only out of tree level 3-string vertices  
\begin{eqnarray} \label{S74}
S^{(7)}_4 &=& \int_0^x Jdy \bra \bar{O}^J O^{yJ}O^{(1-y)J}\ket_{\textrm{\textrm{planar}}}   \bra \bar{O}^{yJ} \bar{O}^{(x-y)J}O^{xJ}\ket_{\textrm{\textrm{planar}}}  \nonumber \\ && \times \bra \bar{O}^{(1-y)J} {O}^{(x-y)J}O^{(1-x)J}\ket_{\textrm{\textrm{planar}}}  \nonumber \\
&=& \frac{g^3}{\sqrt{J}}\frac{1}{12}\sqrt{x(1-x)} x^3(2-x)  \nonumber \\
S^{(7)}_5 &=& S^{(7)}_4(x\rightarrow 1-x) 
\end{eqnarray}

Now we count the multiplicity for the string diagrams. We extend the short processes (\ref{F7short}) into long processes, then determine the corresponding string diagrams of the long processes. This is done in Table \ref{F7longprocess}, and we write the multiplicity matrix in Table \ref{S7multiplicity}. We check that (\ref{F71}, \ref{F72}, \ref{S71}, \ref{S74}) satisfy the factorization relations according to the multiplicity matrix 
\begin{eqnarray}
S^{(7)}_1&=& 6 F^{(7)}_1+ 4 (F^{(7)}_2+F^{(7)}_3)+2 (F^{(7)}_4+F^{(7)}_5) \nonumber \\
S^{(7)}_2&=& 2 F^{(7)}_4 \nonumber \\
S^{(7)}_3&=& 2 F^{(7)}_5 \nonumber \\
S^{(7)}_4&=& 2F^{(7)}_2+2 F^{(7)}_4 \nonumber \\
S^{(7)}_5&=& 2F^{(7)}_3+2 F^{(7)}_5 \nonumber
\end{eqnarray}

\begin{table} 
\begin{center}
\begin{tabular}{| c | c | c | c | c | c | }
 \hline 
$m_{ij}$ & $S^{(7)}_1 $ & $S^{(7)}_2 $ & $S^{(7)}_3 $ & $S^{(7)}_4 $ &$S^{(7)}_5 $  Ê\\ \hline
$F^{(7)}_1$ & 6 & 0 & 0& 0 & 0 \\ \hline
$F^{(7)}_2$ & 4 & 0 & 0& 2 & 0 \\ \hline
$F^{(7)}_3$ & 4 & 0 & 0& 0 & 2 \\ \hline
$F^{(7)}_4$ & 2 & 2 & 0& 2 & 0 \\ \hline
$F^{(7)}_5$ & 2 & 0 & 2& 0 & 2 \\ \hline
 \end{tabular} 
  \caption{The multiplicity matrix of string diagrams in Fig. \ref{S7} with respect to the short processes (\ref{F7short}). }
 \label{S7multiplicity}
 \end{center}
 \end{table}

\begin{table} 
\begin{center}
\begin{tabular}{| c | l | c | }
 \hline 
 $F^{(7)}_1$ & 1.~~ $(123456)\rightarrow(123)(456)\rightarrow(312645)\rightarrow(531)_1(264)_2$~~   & $\in S^{(7)}_1$ \\ 
& 2.~~ $(123456)\rightarrow(123)(456)\rightarrow(231564)\rightarrow(315)_1(264)_2$ ~~&  $\in S^{(7)}_1$ \\ 
& 3.~~$ (123456)\rightarrow(234)(561)\rightarrow(342615)\rightarrow(315)_1(426)_2 $  ~~&  $\in S^{(7)}_1$ \\  
& 4.~~$(123456)\rightarrow(234)(561)\rightarrow(423156)\rightarrow(315)_1(426)_2 $ ~~&  $\in S^{(7)}_1$ \\ 
& 5.~~$(123456)\rightarrow(345)(126)\rightarrow(534261)\rightarrow(531)_1(426)_2  $~~&  $\in S^{(7)}_1$ \\ 
& 6.~~$(123456)\rightarrow(345)(126)\rightarrow(453126)\rightarrow(531)_1(426)_2   $~~&  $\in S^{(7)}_1$ \\  
  \hline 
 $F^{(7)}_2$ & 1.~~$(123456)\rightarrow(234)(156)\rightarrow(423615)\rightarrow(36)_2(4215)_1 $  ~~& $\in S^{(7)}_1$ \\ 
& 2.~~$(123456)\rightarrow(234)(156)\rightarrow(342156)\rightarrow(36)_2(4215)_1  $     ~~&  $\in S^{(7)}_1$ \\ 
& 3.~~$(123456)\rightarrow(12)(3456)\rightarrow(215634)\rightarrow(63)_2(2154)_1  $  ~~&  $\in S^{(7)}_1$ \\  
& 4.~~$(123456)\rightarrow(12)(3456)\rightarrow(12)(45)(36)_2\rightarrow(36)_2(2154)_1 $ ~~&  $\in S^{(7)}_4$ \\ 
& 5.~~$(123456)\rightarrow(45)(1236)\rightarrow(542361)\rightarrow(36)_2(5421)_1 $~~&  $\in S^{(7)}_1$ \\ 
& 6.~~$(123456)\rightarrow(45)(1236)\rightarrow(45)(12)(36)_2\rightarrow(36)_2(5421)_1   $~~&  $\in S^{(7)}_4$ \\ 
 \hline 
 $F^{(7)}_3$ & 1.~~$(123456)\rightarrow(234)(156)\rightarrow(423615)\rightarrow(36)_1(4215)_2 $  ~~& $\in S^{(7)}_1$ \\ 
& 2.~~$(123456)\rightarrow(234)(156)\rightarrow(342156)\rightarrow(36)_1(4215)_2  $     ~~&  $\in S^{(7)}_1$ \\ 
& 3.~~$(123456)\rightarrow(12)(3456)\rightarrow(215634)\rightarrow(63)_1(2154)_2  $  ~~&  $\in S^{(7)}_1$ \\  
& 4.~~$(123456)\rightarrow(12)(3456)\rightarrow(12)(45)(36)_1\rightarrow(36)_1(2154)_2 $ ~~&  $\in S^{(7)}_5$ \\ 
& 5.~~$(123456)\rightarrow(45)(1236)\rightarrow(542361)\rightarrow(36)_1(5421)_2 $~~&  $\in S^{(7)}_1$ \\ 
& 6.~~$(123456)\rightarrow(45)(1236)\rightarrow(45)(12)(36)_1\rightarrow(36)_1(5421)_2   $~~&  $\in S^{(7)}_5$ \\ 
\hline
 $F^{(7)}_4$ & 1.~~$(123456)\rightarrow(23)(4561)\rightarrow(325614)\rightarrow(14325)_1(6)_2 $  ~~& $\in S^{(7)}_1$ \\ 
& 2.~~$(123456)\rightarrow(34)(5612)\rightarrow(432561)\rightarrow(14325)_1(6)_2  $     ~~&  $\in S^{(7)}_1$ \\ 
& 3.~~$(123456)\rightarrow(12345)(6)_2\rightarrow(23)(451)(6)_2\rightarrow(51432)_1(6)_2  $  ~~&  $\in S^{(7)}_2$ \\  
& 4.~~$(123456)\rightarrow(12345)(6)_2\rightarrow(34)(512)(6)_2\rightarrow(43251)_1(6)_2  $ ~~&  $\in S^{(7)}_2$ \\ 
& 5.~~$(123456)\rightarrow(34)(5612)\rightarrow(34)(125)(6)_2\rightarrow(43251)_1(6)_2 $~~&  $\in S^{(7)}_4$ \\ 
& 6.~~$(123456)\rightarrow(23)(4561)\rightarrow(23)(145)(6)_2\rightarrow(51432)_1(6)_2   $~~&  $\in S^{(7)}_4$ \\ 
\hline
 $F^{(7)}_5$ & 1.~~$(123456)\rightarrow(23)(4561)\rightarrow(325614)\rightarrow(14325)_2(6)_1 $  ~~& $\in S^{(7)}_1$ \\ 
& 2.~~$(123456)\rightarrow(34)(5612)\rightarrow(432561)\rightarrow(14325)_2(6)_1  $     ~~&  $\in S^{(7)}_1$ \\ 
& 3.~~$(123456)\rightarrow(12345)(6)_1\rightarrow(23)(451)(6)_1\rightarrow(51432)_2(6)_1  $  ~~&  $\in S^{(7)}_3$ \\  
& 4.~~$(123456)\rightarrow(12345)(6)_1\rightarrow(34)(512)(6)_1\rightarrow(43251)_2(6)_1  $ ~~&  $\in S^{(7)}_3$ \\ 
& 5.~~$(123456)\rightarrow(34)(5612)\rightarrow(34)(125)(6)_1\rightarrow(43251)_2(6)_1 $~~&  $\in S^{(7)}_5$ \\ 
& 6.~~$(123456)\rightarrow(23)(4561)\rightarrow(23)(145)(6)_1\rightarrow(51432)_2(6)_1   $~~&  $\in S^{(7)}_5$ \\ 
\hline
 \end{tabular} 
  \caption{The extension of the short processes in (\ref{F7short}) into long processes. We also determine the corresponding string diagrams. }
 \label{F7longprocess}
 \end{center}
 \end{table}

\subsubsection{Case two: $ \bra \bar{O}^J_{-m,m} O^{J_1}_0 O^{J_2}_0 \ket_{\textrm{torus}}$}

For simplicity we discuss the generic case of $m\neq 0$ and also assume $x=\frac{J_1}{J}$ is a generic value. The special case of $m=0$ is much simpler and can be considered separately. The field theory diagrams are the same as in the case of vacuum operators, except we will insert scalar fields with phases into the trace operators.  Similarly we denote the contributions of the 5 diagrams $F^{(8)}_i$, where $i=1,2,3,4,5$. Denoting the number of $Z$'s in the segment (i) to be $x_iJ$ where $i=1,2,\cdots 6$, we find the contributions
\begin{eqnarray} \label{F81}
F^{(8)}_1&=&\frac{1}{3}\frac{g^3}{\sqrt{J}}\int_0^{1}dx_1dx_2dx_3dx_4dx_5dx_6\delta(x_1+x_3+x_5-x)\delta(x_2+x_4+x_6-(1-x))
 \nonumber \\&& \nonumber
(\int_0^{x_1}+\int_{x_1+x_2}^{x_1+x_2+x_3}+\int_{x_1+x_2+x_3+x_4}^{x_1+x_2+x_3+x_4+x_5})
e^{2\pi imy_1}dy_1 \\&&
(\int_{x_1}^{x_1+x_2}+\int_{x_1+x_2+x_3}^{x_1+x_2+x_3+x_4}+\int_{x_1+x_2+x_3+x_4+x_5}^{1})e^{-2\pi
imy_2}dy_2 \nonumber \\
&=&\frac{g^3}{\sqrt{J}}\frac{1}{16m^6\pi^6}
[-3+(1+2x-2x^2)m^2\pi^2-2x^2(1-x)^2m^4\pi^4)\nonumber\\&&+(3-(1-2x)^2m^2\pi^2)\cos(2m\pi
x)-3(1-2x)m\pi\sin(2m\pi x)],
\end{eqnarray}

\begin{eqnarray} \label{F82}
F^{(8)}_2&=&\frac{1}{2}\frac{g^3}{\sqrt{J}}\int_0^{1}dx_1dx_2dx_3dx_4dx_5dx_6\delta(x_1+x_2+x_4+x_5-x)\delta(x_3+x_6-(1-x))
\nonumber  \\&& \nonumber
(\int_0^{x_1+x_2}+\int_{x_1+x_2+x_3}^{x_1+x_2+x_3+x_4+x_5})
e^{2\pi imy_1}dy_1
(\int_{x_1+x_2}^{x_1+x_2+x_3}+\int_{x_1+x_2+x_3+x_4+x_5}^{1})e^{-2\pi
imy_2}dy_2 \nonumber \\
&=&\frac{g^3}{\sqrt{J}}\frac{1}{24m^6\pi^6}
[3-3m^2\pi^2x(1-x)-2m^4\pi^4(1-x)x^3\nonumber \\&&
-(3+3x(1-x)m^2\pi^2)\cos(2m\pi x)+(3(1-2x)m\pi-m^3\pi^3
x^3)\sin(2m\pi x)] ,\nonumber \\
F^{(8)}_3 &=& F^{(8)}_2(x\rightarrow 1-x) ,
\end{eqnarray}

\begin{eqnarray}
F^{(8)}_4&=&\frac{g^3}{\sqrt{J}}\int_0^{1}dx_1dx_2dx_3dx_4dx_5dx_6\delta(x_1+x_2+x_3+x_4+x_5-x)\delta(x_6-(1-x))
\nonumber \\&& \int_0^{x}dy_1 e^{2\pi imy_1} \int_x^{1}dy_2
e^{-2\pi imy_2}  \nonumber \\
&=& \frac{g^3}{\sqrt{J}}\frac{\cos(2m\pi
x)-1}{48m^2\pi^2}(x^4+(1-x)^4), \nonumber \\
F^{(8)}_5 &=& F^{(8)}_4(x\rightarrow 1-x) ,
\end{eqnarray}

\begin{figure}
  \begin{center}
  \includegraphics[width=6.5in]{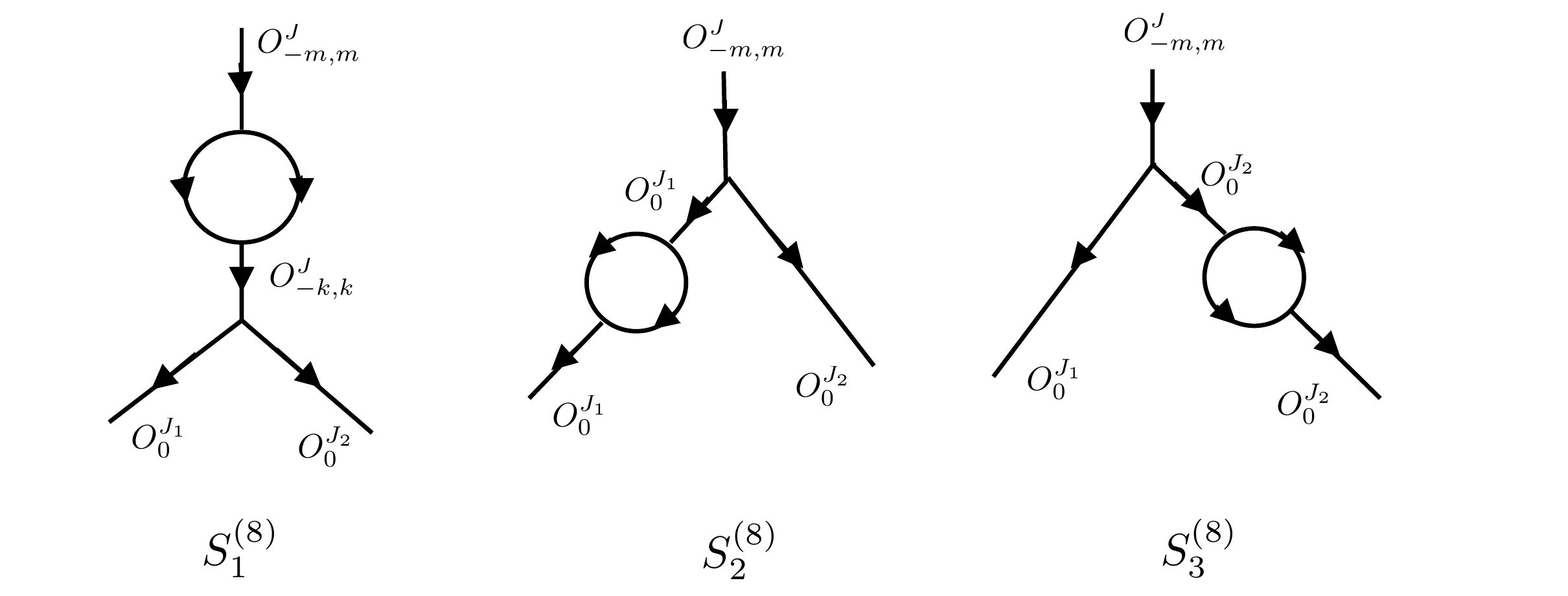} 
\end{center}
\caption{The decorated string diagrams of $S^{(7)}_1$, $S^{(7)}_2$ and $S^{(7)}_3$ in Fig. \ref{S7}.  We denote these contributions $S^{(8)}_1$, $S^{(8)}_2$, $S^{(8)}_3$ respectively. }
\label{S8}
\end{figure}

\begin{figure}
  \begin{center}
  \includegraphics[width=6.5in]{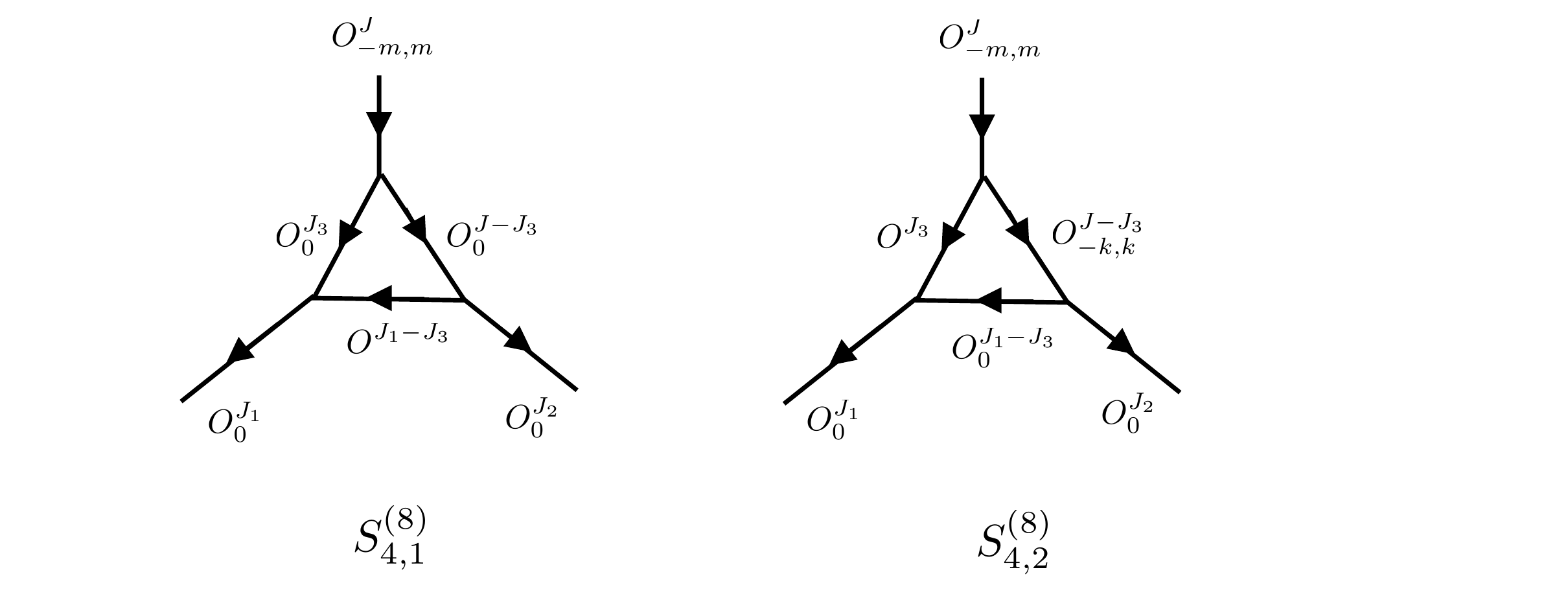} 
\end{center}
\caption{The decorated string diagrams of $S^{(7)}_4$ in Fig. \ref{S7}.  There are 2 diagrams and we denote the total contributions by $S^{(8)}_4$. }
\label{S84}
\end{figure}

The string diagrams are constructed by decorating the vacuum diagrams in Fig. \ref{S7} with scalar excitations, and we denote the corresponding contributions here $S^{(8)}_i$, $i=1,2,3,4,5$ accordingly. For the un-amputated diagrams $S^{(7)}_1$, $S^{(7)}_2$ and $S^{(7)}_3$, we find there is only one way to decorate the diagrams (without concerning the details of the one-loop string propagations in the diagrams).  We draw these decorated diagrams in Fig. \ref{S8}. For the diagrams   $S^{(7)}_4$ and $S^{(7)}_5$, there are 2 ways to decorate for each of them, and we draw the decorated diagrams of $S^{(7)}_4$ in Fig. \ref{S84}. The ones for $S^{(7)}_5$ are obtained from those of $S^{(7)}_4$ by simply exchanging the two operators $O^{J_1}_0$ and $O^{J_2}_0$. 

We compute these diagrams similarly 
\begin{eqnarray} \label{S81}
S^{(8)}_1 &=& \sum_{k=-\infty}^{\infty} 2\bra \bar{O}^J_{-m,m} O^J_{-k,k} \ket_{\textrm{torus}}\bra \bar{O}^J_{-k,k} O^{J_1}_0O^{J_2}_0\ket_{\textrm{\textrm{planar}}} \nonumber \\
S^{(8)}_2 &=& 2\bra \bar{O}^J_{-m,m} O^{J_1}_0O^{J_2}_0\ket_{\textrm{\textrm{planar}}} \bra \bar{O}^{J_1}_0 O^{J_1}_0 \ket_{\textrm{torus}} \nonumber \\
S^{(8)}_3 &=& S^{(8)}_2(x\rightarrow 1-x) \nonumber \\
S^{(8)}_4 &=& \int_0^x Jdy [\bra \bar{O}^J_{-m,m} O^{yJ}_0O^{(1-y)J}_0 \ket  \bra \bar{O}^{(1-y)J}_0 {O}^{(x-y)J}O^{(1-x)J}_0 \ket  \bra \bar{O}^{yJ}_0 \bar{O}^{(x-y)J}O^{xJ}_0\ket   \nonumber \\ &&
+\sum_{k=-\infty}^{\infty} \bra \bar{O}^J_{-m,m} O^{yJ}O^{(1-y)J}_{-k,k} \ket  \bra \bar{O}^{(1-y)J}_{-k,k} {O}^{(x-y)J}_0O^{(1-x)J}_0 \ket  \bra \bar{O}^{yJ} \bar{O}^{(x-y)J}_0O^{xJ}_0\ket  ] \nonumber \\
S^{(8)}_5 &=& S^{(8)}_4(x\rightarrow 1-x) 
\end{eqnarray}
Using the vertex formulae (\ref{planar1}) one can perform the sums and integrals to check the factorization relations
\begin{eqnarray}
S^{(8)}_1&=& 6 F^{(8)}_1+ 4 (F^{(8)}_2+F^{(8)}_3)+2 (F^{(8)}_4+F^{(8)}_5) \nonumber \\
S^{(8)}_2&=& 2 F^{(8)}_4 \nonumber \\
S^{(8)}_3&=& 2 F^{(8)}_5 \nonumber \\
S^{(8)}_4&=& 2F^{(8)}_2+2 F^{(8)}_4 \nonumber \\
S^{(8)}_5&=& 2F^{(8)}_3+2 F^{(8)}_5 \nonumber
\end{eqnarray}

\subsubsection{Case three: $ \bra \bar{O}^J_{-m,m} O^{J_1}_{-n,n} O^{J_2} \ket_{\textrm{torus}}$} \label{subsubcase3}

As in the previous case, we discuss the generic case of $m\neq 0$ and $n\neq 0$, and also assume $x=\frac{J_1}{J}$ is a generic value such that $mx-n$ and $mx+n$ are not zero. The special cases of $m=0$ or $n=0$ are much simpler and can be considered separately. The field theory diagram contributions are basically computed by looking at the diagrams for vacuum operators in Fig. \ref{F7}, and inserting the scalar excitations and summing them with phases. We denote the corresponding contributions $F^{(9)}_i$ where $i=1,2,3,4,5$. Denoting the number of $Z$'s in the 6 segments in the single trace operator $O^{J}_{-m,m}$ by $x_iJ$ where $i=1,2,3,4,5,6$, the calculations go as the followings 
\begin{eqnarray}
F^{(9)}_1 &=& \frac{g^3}{3\sqrt{J}} (\frac{1-x}{x})^{\frac{1}{2}}  \int_0^1 dx_1dx_2\cdots dx_6 \delta(x_1+x_3+x_5-x) \delta(x_2+x_4+x_6-(1-x)) 
\nonumber \\ && |(\int_0^{x_1} dy +e^{2\pi i n\frac{x_2-x_5}{x}}\int_{x_1+x_2}^{x_1+x_2+x_3}dy +
e^{2\pi i n\frac{x_2+x_3+x_4}{x}}\int_{ x_1+x_2+x_3+x_4}^{ 1-x_6}dy)e^{2\pi i (m-\frac{n}{x}) y}  |^2 \nonumber \\
 \end{eqnarray}
\begin{eqnarray}
F^{(9)}_2 &=& \frac{g^3}{2\sqrt{J}} (\frac{1-x}{x})^{\frac{1}{2}} \int_0^1 dx_1dx_2\cdots dx_6 \delta(x_1+x_2+x_4+x_5-x) \delta(x_3+x_6-(1-x)) 
\nonumber \\ && |(\int_0^{x_1} dy +e^{-2\pi i n\frac{x_4+x_5}{x}}\int_{x_1}^{x_1+x_2}dy
+e^{2\pi i n\frac{x_2+x_3-x_5}{x}}\int_{x_1+x_2+x_3}^{x_1+x_2+x_3+x_4}dy
 \nonumber \\ && +
e^{2\pi i n\frac{x_2+x_3+x_4}{x}}\int_{ x_1+x_2+x_3+x_4}^{ x+x_3}dy)e^{2\pi i (m-\frac{n}{x}) y}  |^2
\end{eqnarray}
\begin{eqnarray}
F^{(9)}_3 &=& \frac{g^3}{2\sqrt{J}} (\frac{1-x}{x})^{\frac{1}{2}} \int_0^1 dx_1dx_2\cdots dx_6 \delta(x_1+x_2+x_4+x_5-(1-x)) \delta(x_3+x_6-x) 
\nonumber \\ && |(\int_{x_1+x_2}^{x_1+x_2+x_3}dy
+e^{2\pi i n\frac{x_4+x_5}{x}}\int_{1-x_6}^{1} dy ) e^{2\pi i (m-\frac{n}{x}) y}  |^2
\end{eqnarray}
\begin{eqnarray}
F^{(9)}_4 &=& \frac{g^3}{\sqrt{J}} (\frac{1-x}{x})^{\frac{1}{2}} \int_0^1 dx_1dx_2\cdots 
dx_6 \delta(x_1+x_2+x_3+x_4+x_5-x) \delta(x_6-(1-x)) 
\nonumber \\ && |(\int_0^{x_1} dy +e^{-2\pi i n\frac{x_3+x_4}{x}}\int_{x_1}^{x_1+x_2}dy
+e^{2\pi i n\frac{x_2-x_4}{x}}\int_{x_1+x_2}^{x_1+x_2+x_3}dy
 \nonumber \\ && +
e^{2\pi i n\frac{x_2+x_3}{x}}\int_{ x_1+x_2+x_3}^{ x-x_5}dy
+\int_{x-x_5}^{ x}dy
)e^{2\pi i (m-\frac{n}{x}) y}  |^2
\end{eqnarray}
\begin{eqnarray}
F^{(9)}_5 &=& \frac{g^3}{\sqrt{J}} (\frac{1-x}{x})^{\frac{1}{2}} \int_0^1 dx_1dx_2\cdots 
dx_6 \delta(x_1+x_2+x_3+x_4+x_5-(1-x)) \nonumber \\ &&  \times \delta(x_6-x) 
 |(\int_0^{x} dy
)e^{2\pi i (m-\frac{n}{x}) y}  |^2
\end{eqnarray} 
We calculate these integrals respectively. We find the case of $F^{(9)}_5$ is the simplest and the case of $F^{(9)}_4$  is the most difficult. The results are the followings 
{\footnotesize 
\begin{eqnarray}
F^{(9)}_1   &=& 
\frac{g^3}{\sqrt{J}} \frac{x^{\frac{3}{2}}(1-x)^{\frac{1}{2}}}{ 16 \pi ^6 m^4 n (n-m x)^4}  \{ -2 \pi ^2 m^5 (1-x)^2 x^3 \sin ^2(\pi  m x)
 \nonumber \\ && -2 m^2 x^2 (n-m x) [(m^2 \pi ^2  (1-x)^2+2) \sin ^2(\pi  m x)+  \pi  m (1-x) \sin (2\pi  m x)]
 \nonumber \\ && 
+ m x (n-m x)^2 [2 m^4 \pi ^4 x^2(1-x)^2-2 m^2 \pi ^2 x(1-x)-1 \nonumber \\ && 
+(2 m^2 \pi ^2  (x-1) x+1) \cos (2 \pi  m x)-m \pi  (x+1) \sin (2 \pi  m x)]  \nonumber \\ &&
+(n-m x)^3[1+2 m^4 \pi ^4 x^2(1-x)^2+(2 m^2 \pi ^2 (x-1) x-1) \cos (2 \pi  m x) \nonumber \\ &&
-m \pi  (2 x-1) \sin (2 \pi  m x)] \}, \\
F^{(9)}_2   &=& \frac{x^{\frac{1}{2}}(1-x)^{\frac{1}{2}} }{48 \pi ^6} \{-\frac{6 \sin ^2(\pi  m x)}{m^5 (n+m x)}-\frac{6 \pi  x^2 [2 \pi  m (x-1)+\sin (2 \pi  m x)]}{m^3
   n^2} \nonumber \\ && +\frac{2 \pi  (1-x) x [4 \pi ^3 m^3 x^3+3 \pi  m x-3 \sin (2 \pi  m x)+3 \pi  m x \cos (2 \pi  m x)]}{m^3
   (n-m x)^2}\nonumber \\ && +\frac{12 \pi ^2 (1-x) x}{m^3 n}+\frac{6 \pi  (1-x) x^2 \sin (2 \pi  m x)}{m^2 (n-m x)^3}+\frac{3 [4 \pi
   ^2 m^2 (x-1) x+1-\cos (2 \pi  m x)]}{m^5 (n-m x)} \nonumber \\ && +\frac{12 \pi ^2 (x-1) x^4}{(n-m x)^4}+\frac{6 \pi  (x-1) x^4
   \sin (2 \pi  m x)}{(n-m x)^5} \}, \\
F^{(9)}_3   &=&
   \frac{x^{\frac{5}{2}}(1-x)^{\frac{1}{2}} }{24 \pi ^6 m^3 (m x-n)^3} \{-2 \pi ^4 m^4 x^4+6 \pi ^4 m^4 x^3-6 \pi ^4 m^4 x^2+2 \pi ^4 m^4 x+2 \pi ^4 m^3 n x^3\nonumber \\ && -6 \pi ^4 m^3 n x^2+6
   \pi ^4 m^3 n x -2 \pi ^4 m^3 n -3 \pi ^2 m^2 x^2+3 \pi ^2 m^2 x+3 \pi ^2 m n x  \nonumber \\ && -3 \pi ^2 m
   n-3    +\pi  [\pi ^2 m^3 (x-1)^3+m (6 x-3)-3 n] \sin (2 \pi  m x)   \nonumber \\ && +[-3 \pi ^2 m^2
   (x-1) x+3 \pi ^2 m n (x-1)+3] \cos (2 \pi  m x)   \},   
\end{eqnarray} 
\begin{eqnarray}   
F^{(9)}_4   &=& \frac{x^{\frac{1}{2}}(1-x)^{\frac{1}{2}} }{96 \pi ^6}  \{  \frac{24 \sin ^2(\pi  m x)}{m^5 (m x+n)}+\frac{6 x \sin ^2(\pi  m x)}{m^4 (m x+n)^2}+\frac{6 \pi  x [2 \pi  m
   x+3 \sin (2 \pi  m x)]}{m^4 n}  \nonumber \\ &&  +\frac{24 \pi  x^2 [\pi  m x+\sin (2 \pi  m x)]}{m^3 n^2}+\frac{6 x^3 [-5 \pi ^2 m^2
   x^2+3 (\pi ^2 m^2 x^2+1) \cos (2 \pi  m x)-3]  }{m^2 (n-m x)^4}\nonumber \\ &&  +\frac{6 [2 \pi ^2 m^2 x^2+3 \pi  m x
   \sin (2 \pi  m x)-2 \cos (2 \pi  m x)+2] }{m^5 (m x-n)}\nonumber \\ &&  +\frac{x [10 \pi ^4 m^4 x^4+(15-2 \pi ^4 m^4
   x^4 ) \cos (2 \pi  m x)+12 \pi ^2 m^2 x^2+18 \pi  m x \sin (2 \pi  m x)-15] }{m^4 (n-m x)^2} \nonumber \\ &&  +\frac{4 x^2
   [\pi  m x (2 \pi ^2 m^2 x^2+3) \sin (2 \pi  m x)+6 \cos (2 \pi  m x)-6]  }{m^3 (m x-n)^3}+\frac{60
   x^5 \sin ^2(\pi  m x)}{(n-m x)^6}  \nonumber \\ &&  +\frac{24 \pi  x^5 \sin (2 \pi  m x)}{(n-m x)^5} \},
  \\
F^{(9)}_5 &=& \frac{g^3}{ \sqrt{J}} x^{\frac{3}{2}}  (1-x)^{\frac{9}{2}}   \frac{\sin( m\pi x)^2} {24 \pi^2 (n-mx)^2}
\end{eqnarray} }

\begin{figure}
  \begin{center}
  \includegraphics[width=6.5in]{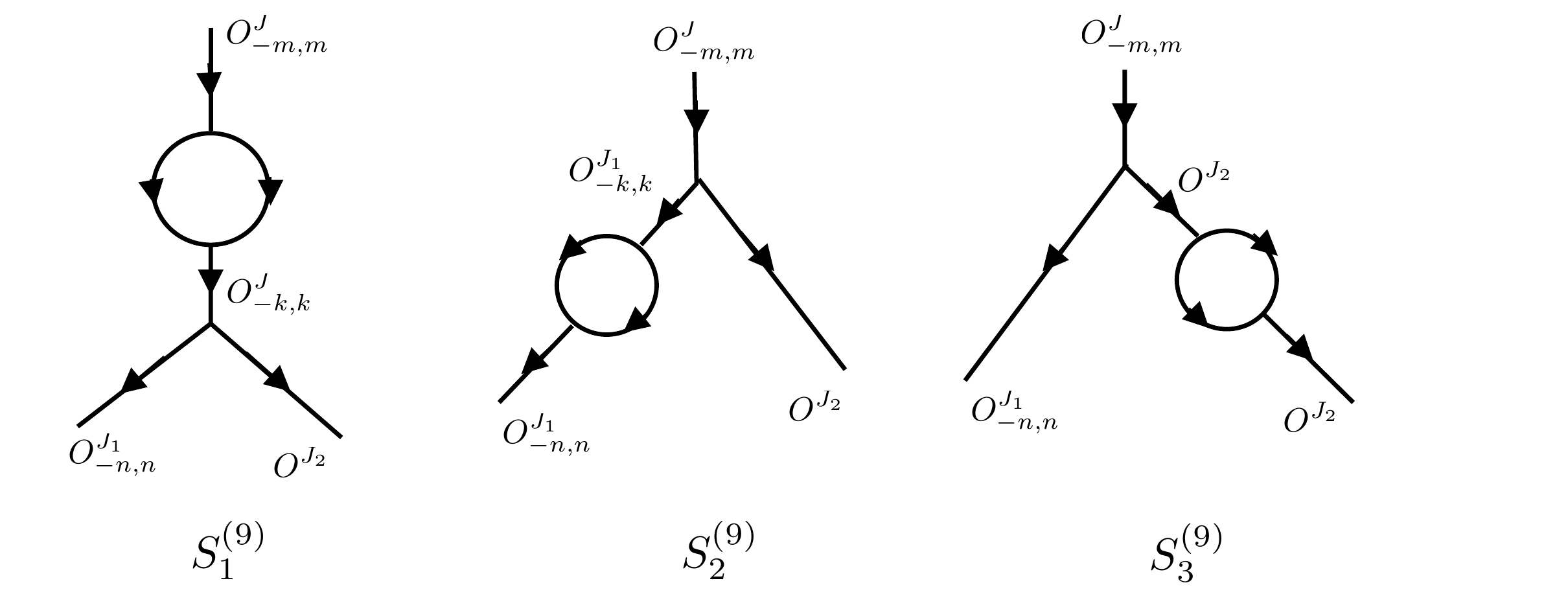} 
\end{center}
\caption{The decorated string diagrams of $S^{(7)}_1$, $S^{(7)}_2$ and $S^{(7)}_3$ in Fig. \ref{S7}.  We denote these contributions $S^{(9)}_1$, $S^{(9)}_2$, $S^{(9)}_3$ respectively. }
\label{S9}
\end{figure}

\begin{figure}
  \begin{center}
  \includegraphics[width=6.5in]{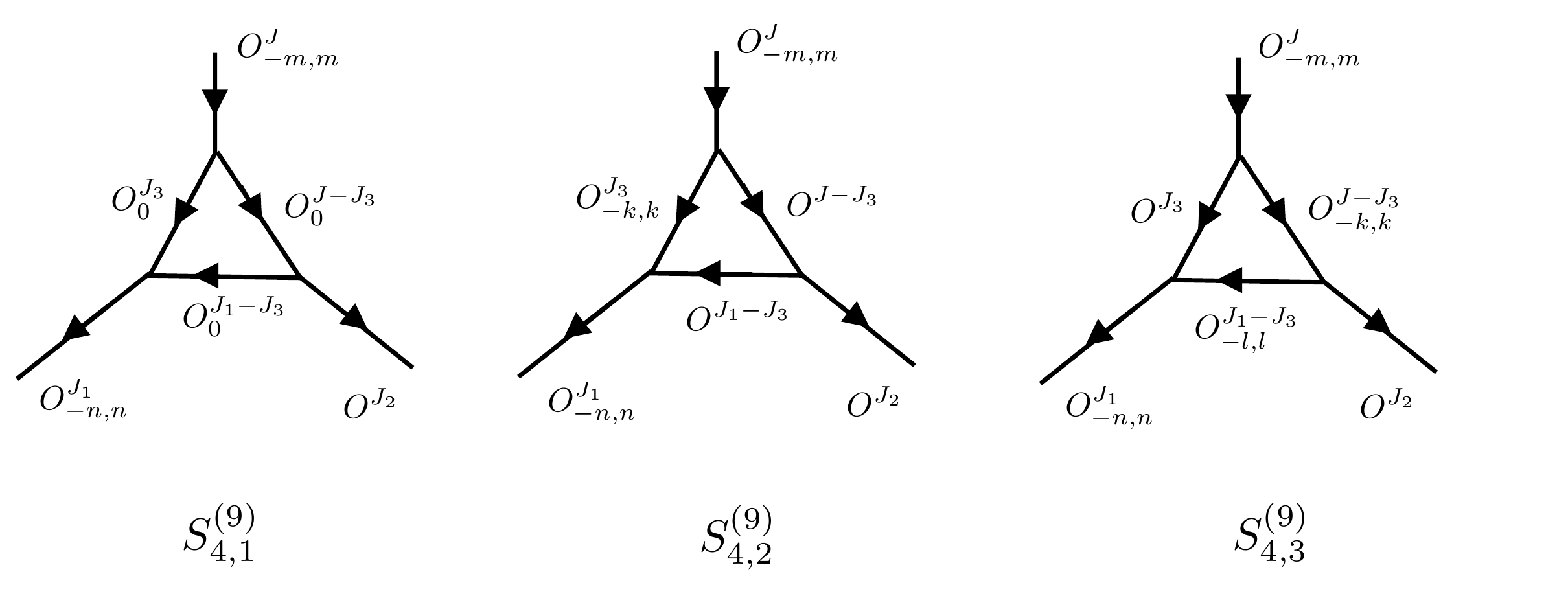} 
  \end{center}
\caption{The decorated string diagrams of $S^{(7)}_4$ in Fig. \ref{S7}.  There are 3 diagrams and we denote their contributions  $S^{(9)}_{4,1}$, $S^{(9)}_{4,2}$, $S^{(9)}_{4,3}$ respectively and the total contributions by $S^{(9)}_4$. }
\label{S94}
\end{figure}

\begin{figure}
  \begin{center}
  \includegraphics[width=6.5in]{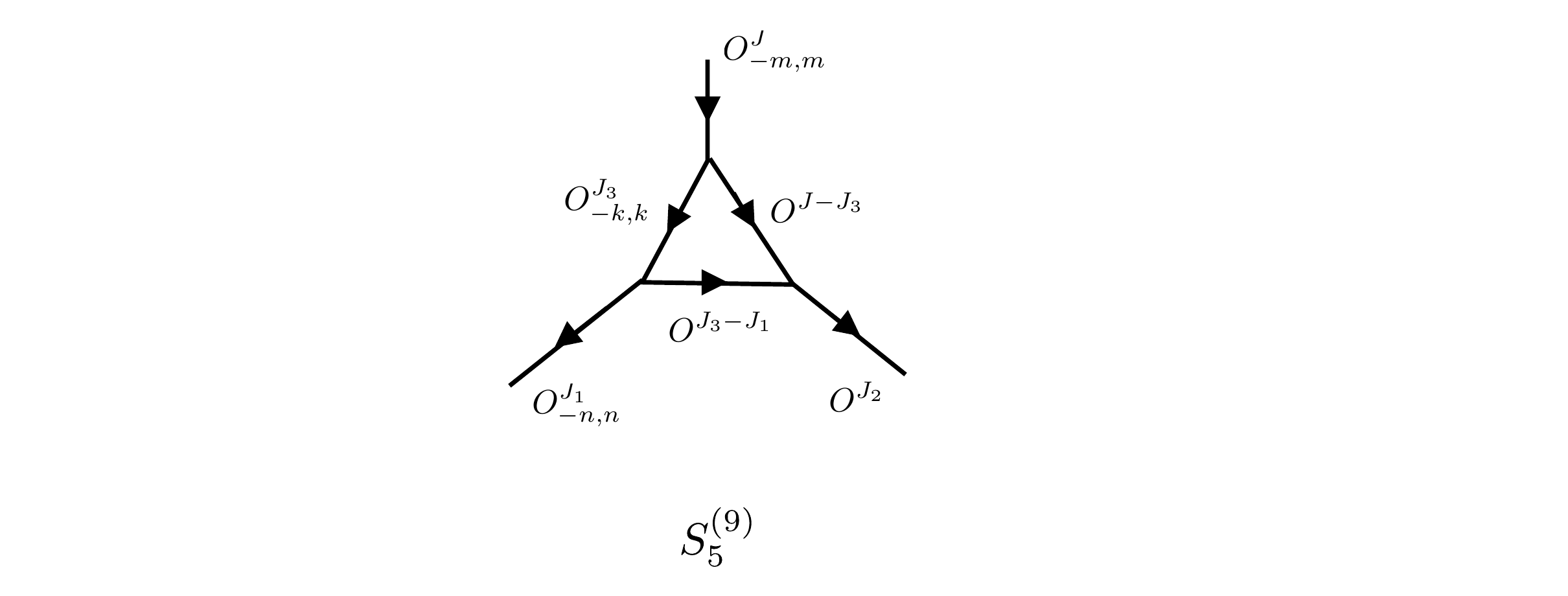} 
\end{center}
\caption{The decorated string diagrams of $S^{(7)}_5$ in Fig. \ref{S7}.  There only one diagram and we denote the  contribution by $S^{(9)}_5$. }
\label{S95}
\end{figure}

Now we consider the string diagrams, which are obtained as decoration of the vacuum diagrams in Fig. \ref{S7} with stringy excitations. For the diagrams $S^{(7)}_1$, $S^{(7)}_2$, $S^{(7)}_3$ and $S^{(7)}_5$ there is only one way to decoration, while for the case of and $S^{(7)}_4$ there are 3 decorated diagrams. We depicted these diagrams in Figs. \ref{S9}, \ref{S94}, \ref{S95}. The contributions for diagrams in Fig. \ref{S9} are 
\begin{eqnarray} \label{S9123}
S^{(9)}_1 &=& 2 \sum_{k} \bra \bar{O}^{J}_{-m,m} O^J_{-k,k}\ket _{\textrm{torus}} \bra \bar{O}^J_{-k,k} O^{J_1}_{-n,n} O^{J_2} \ket \nonumber \\
S^{(9)}_2 &=&  2 \sum_{k} \bra \bar{O}^{J}_{-m,m} O^{J_1}_{-k,k} O^{J_2} \ket\bra \bar{O}^{J_1}_{-k,k}  O^{J_1}_{-n,n} \ket _{\textrm{torus}}         \nonumber \\
S^{(9)}_3 &=&  2 \bra \bar{O}^{J}_{-m,m} O^{J_1}_{-n,n} O^{J_2} \ket \bra \bar{O}^{J_2}  O^{J_2} \ket _{\textrm{torus}} 
\end{eqnarray}
where the factor of 2 comes from the multiplicity of the one-loop string propagation diagram. Denoting the length of the intermediate operator in the loop $J_3=yJ$ and also $J_1=xJ$, we find the contributions of the diagrams in Figs. \ref{S94}, \ref{S95} as the followings 
\begin{eqnarray}  \label{eqS94}
S^{(9)}_4 &=& S^{(9)}_{4,1}+S^{(9)}_{4,2}+S^{(9)}_{4,3} \\ 
S^{(9)}_{4,1} &=& 2 \int_0^x Jdy  \bra \bar{O}^{J}_{-m,m} O^{yJ}_0 O^{(1-y)J}_0 \ket  \bra \bar{O}^{(1-y)J}_0 O^{(x-y)J}_0 O^{(1-x)J} \ket \bra \bar{O}^{(x-y)J}_0 \bar{O}^{yJ}_0 O^{xJ}_{-n,n} \ket,   \nonumber \\
S^{(9)}_{4,2} &=& \int_0^x Jdy \sum_k \bra \bar{O}^{J}_{-m,m} O^{yJ}_{-k,k} O^{(1-y)J} \ket  \bra \bar{O}^{(1-y)J} O^{(x-y)J} O^{(1-x)J} \ket \bra \bar{O}^{(x-y)J} \bar{O}^{yJ}_{-k,k} O^{xJ}_{-n,n} \ket,   \nonumber \\
S^{(9)}_{4,3} &=& \int_0^x Jdy \sum_k\sum_l \bra \bar{O}^{J}_{-m,m} O^{yJ} O^{(1-y)J}_{-k,k}   \ket  \bra \bar{O}^{(1-y)J}_{-k,k}   O^{(x-y)J}_{-l,l}  O^{(1-x)J} \ket \bra \bar{O}^{(x-y)J}_{-l,l}  \bar{O}^{yJ}_{-k,k} O^{xJ}_{-n,n} \ket,    \nonumber \\
S^{(9)}_{5} &=& \int_x^1 Jdy \sum_k \bra \bar{O}^{J}_{-m,m} O^{yJ}_{-k,k} O^{(1-y)J} \ket  \bra \bar{O}^{yJ}_{-k,k}   O^{(y-x)J} O^{xJ}_{-n,n} \ket     \nonumber \\ && \times 
 \bra \bar{O}^{(1-y)J} \bar{O}^{(y-x)J} O^{(1-x)J} \ket,   \label{eqS95}
\end{eqnarray}
where the factor of 2 in $S^{(9)}_{4,1}$ is because there are two scalar insertion fields in the BMN operator $O^J_{-m,m}$, and we are using a slightly sloppy notation for not distinguishing the different scalar insertions in the operators. We can choose any one for the operator $O^{yJ}_0$ and the other one for $O^{(1-y)J}_0$, and these two choices give the same contribution. We perform the sums and integrals for the string diagrams contributions with the helps of  the summation formulae in Appendix \ref{summationappendix}. The calculations for $S^{(9)}_{4,3} $ is the most difficult as it involves two sums over integers $k$ and $l$, besides the integral of $\int_0^x dy$, and we find it best to do the sum over $k$ first, then the sum over $l$ and the integral. We succeed in calculating the string diagrams analytically and check the factorization relation 
\begin{eqnarray}
S^{(9)}_1&=& 6 F^{(9)}_1+ 4 (F^{(9)}_2+F^{(9)}_3)+2 (F^{(9)}_4+F^{(9)}_5) \nonumber \\
S^{(9)}_2&=& 2 F^{(9)}_4 \nonumber \\
S^{(9)}_3&=& 2 F^{(9)}_5 \nonumber \\
S^{(9)}_4&=& 2F^{(9)}_2+2 F^{(9)}_4 \nonumber \\
S^{(9)}_5&=& 2F^{(9)}_3+2 F^{(9)}_5 \nonumber
\end{eqnarray}

\subsection{Genus two  correlators between two single trace operators}

\subsubsection{The vacuum diagrams and multiplicity}

\begin{figure}
  \begin{center}
  \includegraphics[width=6.5in]{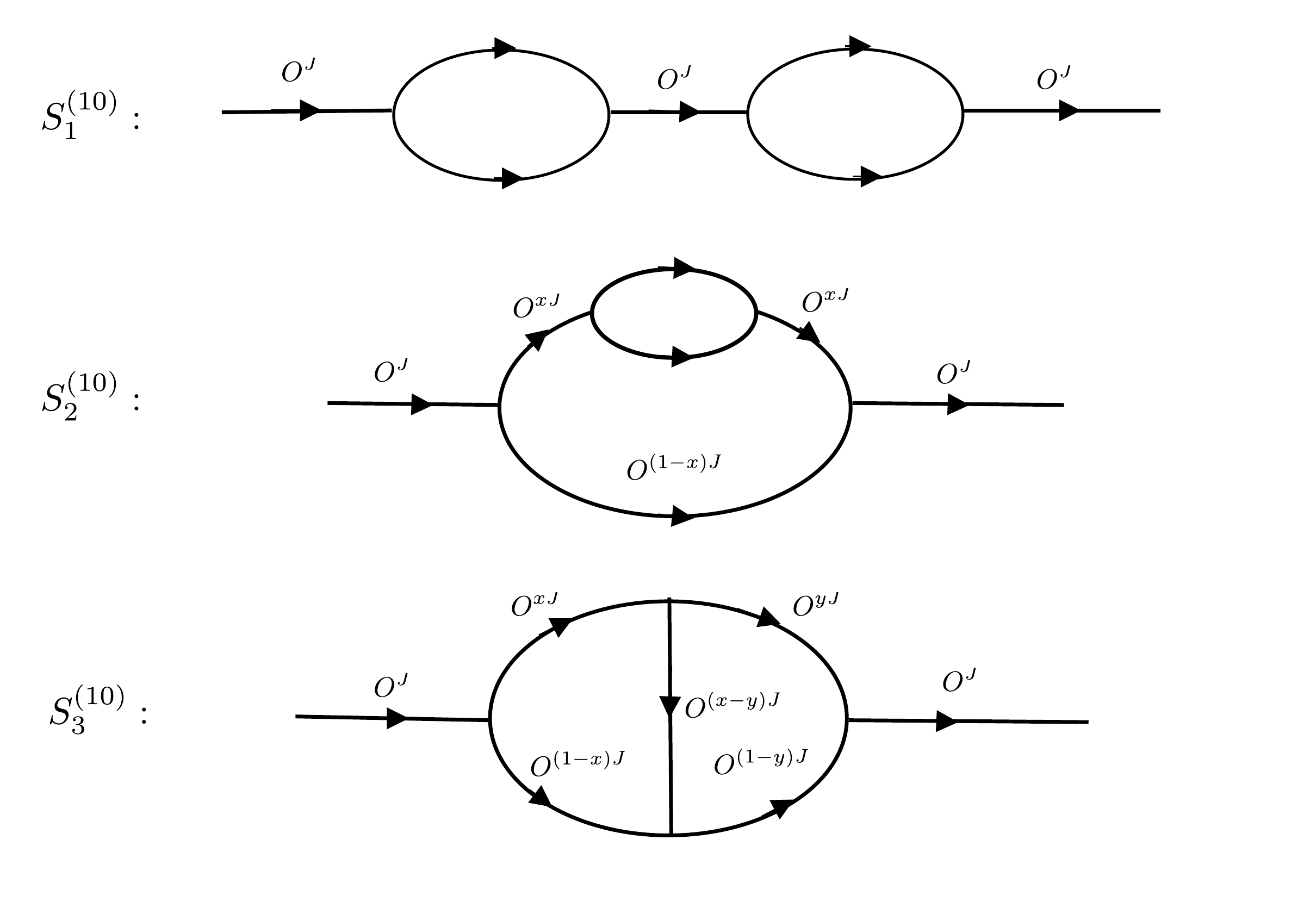} 
\end{center}
\caption{The string diagrams for vacuum operators at genus 2. There are 3 diagrams and we denote their contributions $S^{(10)}_1$, $S^{(10)}_2$ and $S^{(10)}_3$. }
\label{S10}
\end{figure}

There are 3 string diagrams for the correlator $\bra\bar{O}^J O^J\ket_{\textrm{genus 2}}$, and we depict them in Fig. \ref{S10}. These diagrams are easy to calculate
\begin{eqnarray} \label{S10contributions}
S^{(10)}_1 &=& (2 \bra\bar{O}^J O^J\ket_{\textrm{genus 1}})^2 \nonumber \\
&=& \frac{g^4}{144}, \nonumber \\
S^{(10)}_2 &=& \int_0^1 Jdx \frac{x^4g^2}{12}\bra\bar{O}^J O^{xJ}O^{(1-x)J}\ket  \bra \bar{O}^{xJ}\bar{O}^{(1-x)J} O^J \ket  \nonumber \\
&=& \frac{g^4}{504}, \nonumber \\
S^{(10)}_3 &=& J^2 \int_0^1 dx\int_0^x dy \bra\bar{O}^J O^{xJ}O^{(1-x)J}\ket  \bra \bar{O}^{yJ}\bar{O}^{(1-y)J} O^J \ket  \nonumber \\  && \times  \bra\bar{O}^{xJ} O^{yJ}O^{(x-y)J}\ket  \bra \bar{O}^{(1-x)J}\bar{O}^{(x-y)J} O^{(1-y)J} \ket   \nonumber \\
&=& \frac{g^4}{280}
\end{eqnarray}

For the calculations in field theory diagrams, we need to divide the single trace into 8 segments, and there are 21 different diagrams, i.e. short processes, which are just permutations of  $(12\cdots 8)$. Here we will not draw the diagrams again and simply use a permutation $(a_1a_2\cdots a_8)$ to represent the field theory diagram, and denote their contributions $F^{(10)}_j$ with $j=1,2,\cdots 21$. To derive the multiplicity of the string diagrams in Fig. \ref{S10}, we can start with a string $(12\cdots 8)$, and perform the splitting and joining operations according the string diagram. We keep the resulting long processes whose final states are irreducible from combining segments. Using a computer we can count the multiplicities of string diagrams, and we list them in Table. \ref{genus2multiplicity}. We do not list all the long processes here because that would take too much space. The multiplicity matrix is $m_{ij}$ where $i=1,2,3$ denote the string diagrams and $j=1,2,\cdots 21$ denote the field theory diagrams.  The factorization relation (\ref{factorization}) is 
\begin{eqnarray}
S^{(10)}_i=\sum_{j=1}^{21} m_{ij}F^{(10)}_j, ~~~\textrm{for} ~i=1,2,3 
\end{eqnarray}
For the vacuum operator, each field diagram contribute $F^{(10)}_j=\frac{g^4}{8!}$ for any $j$ due to choices of dividing the single string into 8 segments. We see the contributions of the string diagrams in (\ref{S10contributions}) agree with their respective total multiplicities with respects to the 21 field theory diagrams times $\frac{g^4}{8!}$, consistent with the factorization relation. The total contribution to the correlator $\bra \bar{O}^J O^J \ket_{\textrm{genus 2}}$ of the string diagrams is proportional that of the field theory diagrams with a factor of 24. So we can write the correlator as 
\begin{eqnarray}
\bra \bar{O}^J O^J \ket_{\textrm{genus 2}} = \sum_{j=1}^{21} F^{(10)}_j = \frac{1}{24} \sum_{i=1}^3 S^{(10)}_i = \frac{g^4}{1920}
\end{eqnarray}

\begin{table} 
\begin{center}
\begin{tabular}{| c | c | c | c | c |  }
 \hline 
$ m_{ij}$ & $S^{(10)}_1 $ & $S^{(10)}_2 $ & $S^{(10)}_3 $  Ê& Total \\ \hline
$F^{(10)}_1$:  (1,4,7,6,5,8,3,2) & 8 & 8 & 8 & 24  \\ \hline
$F^{(10)}_2$:  (1,5,8,3,7,6,4,2) & 12 & 4 & 8 & 24  \\ \hline
$F^{(10)}_3$:  (1,6,4,8,3,7,5,2) & 16 & 2 & 6 & 24  \\ \hline 
$F^{(10)}_4$:  (1,7,5,4,8,3,6,2) & 12 & 4 & 8 & 24  \\ \hline
$F^{(10)}_5$:  (1,8,3,6,5,4,7,2) & 8 & 8 & 8 & 24  \\ \hline
$F^{(10)}_6$:  (1,4,3,2,5,8,7,6) & 8 & 8 & 8 & 24  \\ \hline
$F^{(10)}_7$:  (1,4,8,7,5,3,2,6) & 12 & 4 & 8 & 24  \\ \hline
$F^{(10)}_8$:  (1,4,8,6,3,2,7,5) & 16 & 2 & 6 & 24 \\ \hline
$F^{(10)}_9$:  (1,4,7,3,2,8,6,5) & 12 & 4 & 8 & 24  \\ \hline
$F^{(10)}_{10}$:  (1,8,7,2,5,4,3,6) & 8 & 8 & 8 & 24  \\ \hline
$F^{(10)}_{11}$:  (1,8,6,4,3,7,2,5) & 12 & 4 & 8 & 24  \\ \hline
$F^{(10)}_{12}$:  (1,7,4,3,8,6,2,5) & 16 & 2 & 6 & 24  \\ \hline
$F^{(10)}_{13}$:  (1,7,6,2,5,8,4,3) & 12 & 4 & 8 & 24 \\ \hline
$F^{(10)}_{14}$:  (1,7,3,6,2,8,5,4) & 16 & 2 & 6 & 24  \\ \hline
$F^{(10)}_{15}$:  (1,5,4,2,8,7,3,6) & 12 & 4 & 8 & 24  \\ \hline
$F^{(10)}_{16}$:  (1,6,5,2,8,4,7,3) & 16 & 2 & 6 & 24  \\ \hline
$F^{(10)}_{17}$:  (1,8,4,7,2,6,5,3) & 12 & 4 & 8 & 24  \\ \hline
$F^{(10)}_{18}$:  (1,5,8,4,2,7,6,3) & 16 & 2 & 6 & 24  \\ \hline
$F^{(10)}_{19}$:  (1,5,3,8,7,4,2,6) & 16 & 2 & 6  & 24  \\ \hline
$F^{(10)}_{20}$:  (1,8,5,3,7,2,6,4) & 16 & 2 & 6 & 24  \\ \hline
$F^{(10)}_{21}$:  (1,6,3,8,5,2,7,4) & 24 & 0 & 0 & 24  \\ \hline
 Total  & 280 & 80 & 144 & \\ \hline
 \end{tabular} 
 \bigskip
  \caption{The multiplicity matrix of string diagrams in Fig. \ref{S10} with respect to the 21 short processes. These short processes are permutations of $(1,2,\cdots ,8)$, and we have used the cyclicality of the string to put the segment (1) in the first position. }
 \label{genus2multiplicity}
 \end{center}
 \end{table}

\subsubsection{The stringy BMN operators}

We consider the stringy case $\bra\bar{O}^J_{-m,m} O^J_{-n,n} \ket_{\textrm{genus 2}}$. A systematic way to do the field theory diagram calculations for higher genus single trace operators were described in \cite{Constable1}. We summarize the details in Appendix \ref{2point}. Basically the calculations of summing  over BMN phases of the scalar insertions can be expressed in terms of some standardized integrals which can be calculated recursively. We denote the contribution of a genus 2 field theory diagram by $F^{(11)}_j$ where $j=1,2,\cdots 21$, and we calculate the contribution $F^{(11)}_j$ respectively  for all the $j$'s in computer using the formula (\ref{integral2}).  

\begin{figure}
  \begin{center}
  \includegraphics[width=6.5in]{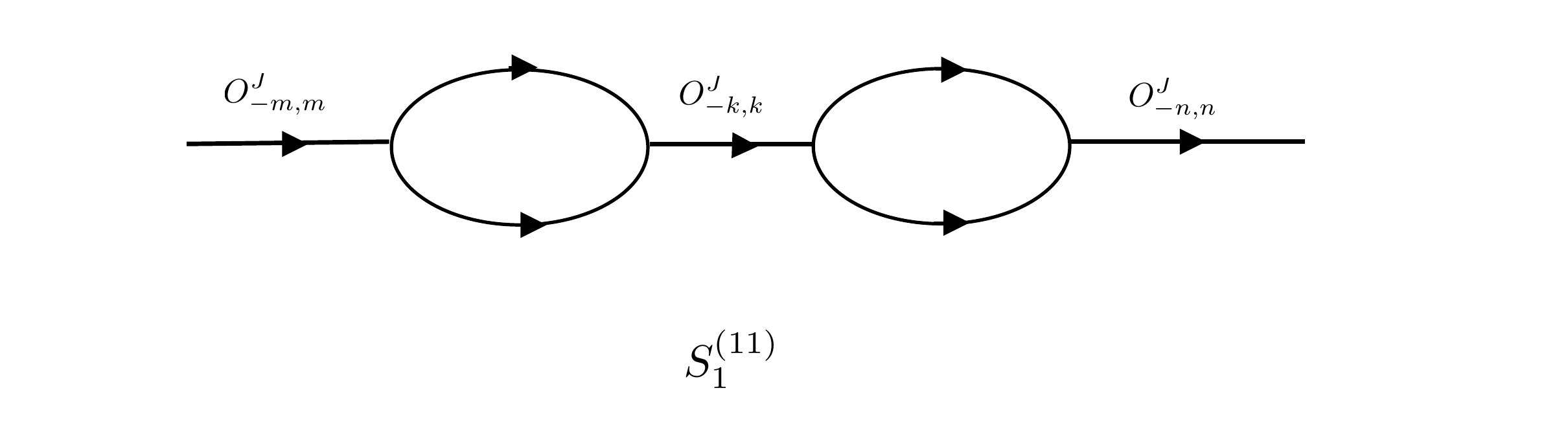} 
\end{center}
\caption{The decorated string diagrams for the first vacuum diagram $S^{(10)}_1$ in Fig. \ref{S10}. This diagram is the paste of two torus diagrams and we denote the contribution $S^{(11)}_1$. }
\label{S11-1}
\end{figure}

\begin{figure}
  \begin{center}
  \includegraphics[width=6.5in]{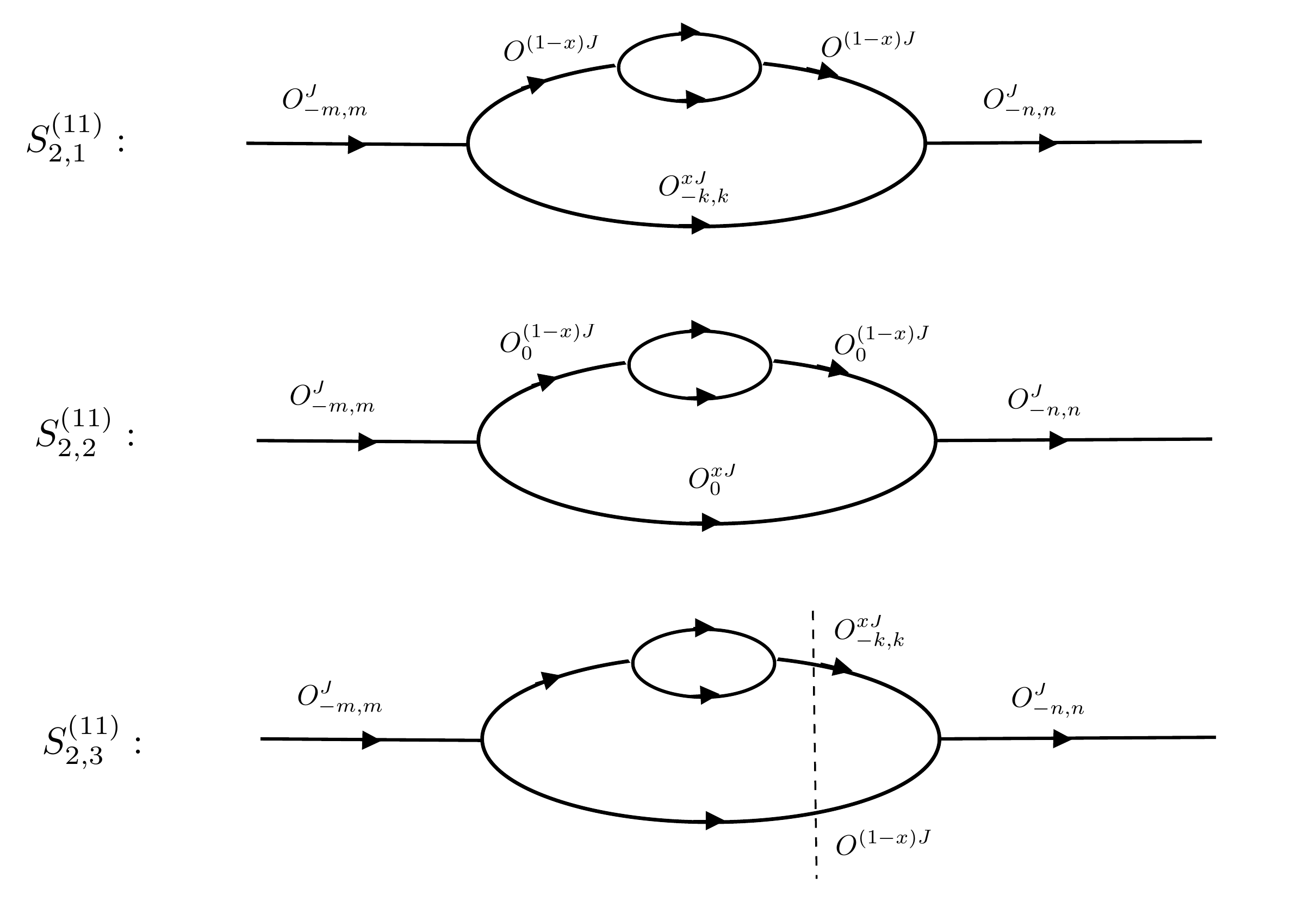} 
\end{center}
\caption{The decorated string diagrams for the first vacuum diagram $S^{(10)}_2$ in Fig. \ref{S10}. We use a dash line in the third diagram to represent it as the paste of one-loop cubic diagram with a tree level cubic vertex,  where we can use results from previous Section \ref{oneloopcubicsection} without the need for the details of the one-loop cubic part of the diagram. We denote the contributions of the 3 diagrams $S^{(11)}_{2,1}$, $S^{(11)}_{2,2}$ and $S^{(11)}_{2,3}$.}
\label{S11-2}
\end{figure}

\begin{figure}
  \begin{center}
  \includegraphics[width=6.5in]{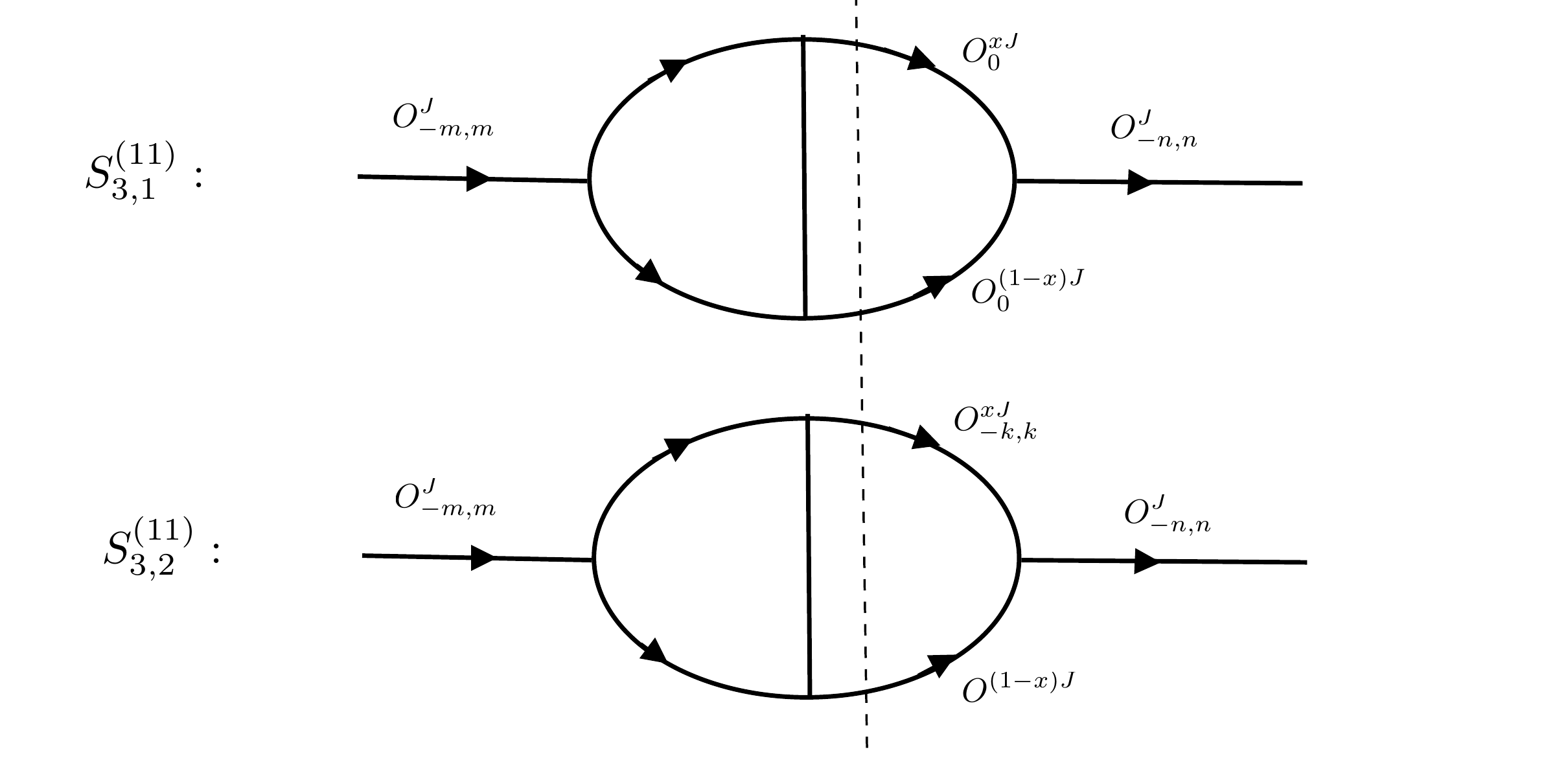} 
 \end{center}
\caption{The decorated string diagrams for the first vacuum diagram $S^{(10)}_3$ in Fig. \ref{S10}. These diagrams can be obtained by pasting a one-loop cubic diagram with a tree level cubic vertex. We do not need to draw the details in the one-loop cubic part of the diagram but simply use the results from the previous Section \ref{oneloopcubicsection}. We denote the contributions of the 2 diagrams $S^{(11)}_{3,1}$ and $S^{(11)}_{3,2}$.}
\label{S11-3}
\end{figure}

For the string diagrams, we \textit{decorate} the vacuum string diagrams in Fig. \ref{S10} with scalar insertions. We depict the decorations of the 3 string diagrams in ÊFigures \ref{S11-1}, \ref{S11-2}, \ref{S11-3}, and denote their contributions $S^{(11)}_1$, $S^{(11)}_2$ and $S^{(11)}_3$ respectively. 

We discuss the calculations of the string diagrams. The diagram in Fig. \ref{S11-1} can be calculated using the one-loop string propagation amplitude in (\ref{F6torus}) or (\ref{S6162}), and we find 
\begin{eqnarray} \label{eqS11-1}
S^{(11)}_1 &=& \sum_{k=-\infty}^{+\infty}  4\bra \bar{O}^J_{-m,m} O^J_{-k,k} \ket_{\textrm{torus}} \bra \bar{O}^J_{-k,k} O^J_{-n,n} \ket_{\textrm{torus}}  \\  \nonumber 
&=& \left\{
\begin{array}{cl}
\frac{g^4}{144},    &   m=n=0;   \\
0,              &  m=0, n\neq0,   \\
& \textrm{or}~n=0, m\neq0; \\
g^4(\frac{451}{256 \pi ^8 m^8}-\frac{145}{96 \pi ^6 m^6}+\frac{7}{40 \pi ^4 m^4}-\frac{1}{252 \pi ^2 m^2}+\frac{71}{45360}),   &  m=n\neq0; \\ & \\
\frac{g^4} {144 \pi ^2 m^2} (\frac{31185}{128 \pi ^6 m^6}-\frac{1197}{8 \pi ^4 m^4}+\frac{111}{8 \pi ^2 m^2}+1 ),  &  m=-n\neq0;   \\
& \\
\frac{g^4P_1}{360 \pi ^8 m^6 n^6 (m-n)^8 (m+n)^4} & \textrm{all~other~cases} 
\end{array}
\right.
\end{eqnarray} 
where the numerator in the last case is  
{\scriptsize
\begin{eqnarray} 
P_1 &=&  m^{16} \left(10 \pi ^6 n^6+42 \pi ^4 n^4-210 \pi ^2 n^2+315\right)-m^{15} n \left(20 \pi ^6 n^6+153 \pi ^4 n^4-795 \pi ^2 n^2+1260\right) \nonumber \\ && -3 m^{14} n^2 \left(10 \pi ^6 n^6+17 \pi ^4 n^4+125 \pi ^2 n^2-210\right)+m^{13} n^3 \left(80 \pi ^6 n^6+459 \pi ^4 n^4-2205 \pi ^2 n^2+3780\right) \nonumber \\ && +m^{12} n^4 \left(20 \pi ^6 n^6-174 \pi ^4 n^4+3225 \pi ^2 n^2-5310\right)-3 m^{11} n^5 \left(40 \pi ^6 n^6+102 \pi ^4 n^4-795 \pi ^2 n^2+855\right) \nonumber \\ && +m^{10} n^6 \left(20 \pi ^6 n^6+366 \pi ^4 n^4-2640 \pi ^2 n^2+8865\right)+m^9 n^7 \left(80 \pi ^6 n^6-306 \pi ^4 n^4-1950 \pi ^2 n^2-4815\right) \nonumber \\ && -6 m^8 n^8 \left(5 \pi ^6 n^6+29 \pi ^4 n^4+440 \pi ^2 n^2+4080\right)+m^7 n^9 \left(-20 \pi ^6 n^6+459 \pi ^4 n^4+2385 \pi ^2 n^2-4815\right) \nonumber \\ && +m^6 n^{10} \left(10 \pi ^6 n^6-51 \pi ^4 n^4+3225 \pi ^2 n^2+8865\right)-9 m^5 n^{11} \left(17 \pi ^4 n^4+245 \pi ^2 n^2+285\right) \nonumber \\ && +3 m^4 n^{12} \left(14 \pi ^4 n^4-125 \pi ^2 n^2-1770\right)+15 m^3 n^{13} \left(53 \pi ^2 n^2+252\right) \nonumber \\ && -210 m^2 n^{14} \left(\pi ^2 n^2-3\right)-1260 m n^{15}+315 n^{16}
\end{eqnarray} }

For the 3 diagrams in Fig. \ref{S11-2}, the first two are easy to handle because the one-loop propagation of the non-stringy operator just contributes a factor of $\frac{(1-x)^4}{12}$, and the calculations are  
\begin{eqnarray}
S^{(11)}_{2,1} &=& \frac{g^2}{12}\int_0^1 (1-x)^4 dx  \sum_{k=-\infty}^{\infty} 
\bra \bar{O}^J_{-m,m}O^{xJ}_{-k,k} O^{(1-x)J} \ket \bra \bar{O}^{xJ}_{-k,k} \bar{O}^{(1-x)J} {O}^J_{-n,n} \ket, \nonumber \\
S^{(11)}_{2,2} &=& \frac{g^2}{6}\int_0^1 (1-x)^4 dx 
\bra \bar{O}^J_{-m,m}O^{xJ}_{0} O^{(1-x)J}_0 \ket \bra \bar{O}^{xJ}_{0} \bar{O}^{(1-x)J}_0  {O}^J_{-n,n} \ket 
\end{eqnarray}
where there is an extra factor of 2 in front of the second diagram $S^{(11)}_{2,2}$ because there are two choices for the scalar insertion in the operator $O^{(1-x)J}_0$ that undergoes one-loop propagation. For the third diagram $S^{(11)}_{2,3}$ in Fig. \ref{S11-2}, it is much easier to use our previous results on one-loop cubic interactions. We divide the diagram into two part by a dash line and treat the one-loop cubic part on the left  as a black box, which we have calculated previously in the second equation in (\ref{S9123}). We find 
\begin{eqnarray}
S^{(11)}_{2,3} &=& \int_0^1 Jdx  \sum_{k=-\infty}^{\infty}  S^{(9)}_2(m,k,x)  \bra \bar{O}^{xJ}_{-k,k} \bar{O}^{(1-x)J}  {O}^J_{-n,n} \ket 
\end{eqnarray}
Putting the 3 contributions together we find the total contribution 
\begin{eqnarray} \label{eqS112}
S^{(11)}_2 &=& S^{(11)}_{2,1}+S^{(11)}_{2,2}+S^{(11)}_{2,3} \\  \nonumber 
&=& \left\{
\begin{array}{cl}
\frac{g^4}{504},    &   m=n=0;   \\
0,              &  m=0, n\neq0,   \\
& \textrm{or}~n=0, m\neq0; \\
g^4(\frac{1023}{256 \pi ^8 m^8}-\frac{21}{16 \pi ^6 m^6}+\frac{31}{240 \pi ^4 m^4}-\frac{1}{315 \pi ^2 m^2}+\frac{1}{2160} ),   &  m=n\neq0; \\ & \\
\frac{g^4} {504 \pi ^2 m^2} ( -\frac{76923}{512 \pi ^6 m^6}-\frac{12789}{64 \pi ^4 m^4}+\frac{987}{40 \pi ^2 m^2}+1  ),  &  m=-n\neq0;   \\ & \\
\frac{g^4P_2}{10080 \pi ^8 m^6 n^6 (m-n)^8 (m+n)^4} & \textrm{all~other~cases} 
\end{array}
\right.
\end{eqnarray} 
where the numerator in the last case is  
{\scriptsize
\begin{eqnarray} 
P_2 &=& m^{16} \left(80 \pi ^6 n^6+546 \pi ^4 n^4-4830 \pi ^2 n^2+2520\right)-m^{15} n \left(160 \pi ^6 n^6+1848 \pi ^4 n^4-15750 \pi ^2 n^2+2835\right) \nonumber \\ && -15 m^{14} n^2 \left(16 \pi ^6 n^6+28 \pi ^4 n^4-210 \pi ^2 n^2+945\right)+m^{13} n^3 \left(640 \pi ^6 n^6+5544 \pi ^4 n^4-59220 \pi ^2 n^2+17955\right) \nonumber \\ && +m^{12} n^4 \left(160 \pi ^6 n^6-3234 \pi ^4 n^4+32550 \pi ^2 n^2+19215\right)-3 m^{11} n^5 \left(320 \pi ^6 n^6+1232 \pi ^4 n^4-31430 \pi ^2 n^2+5565\right) \nonumber \\ && +m^{10} n^6 \left(160 \pi ^6 n^6+6216 \pi ^4 n^4-30870 \pi ^2 n^2-7875\right)+m^9 n^7 \left(640 \pi ^6 n^6-3696 \pi ^4 n^4-101640 \pi ^2 n^2-71505\right) \nonumber \\ &&  -6 m^8 n^8 \left(40 \pi ^6 n^6+539 \pi ^4 n^4+5145 \pi ^2 n^2+51135\right)+m^7 n^9 \left(-160 \pi ^6 n^6+5544 \pi ^4 n^4+94290 \pi ^2 n^2-71505\right) \nonumber \\ &&  +5 m^6 n^{10} \left(16 \pi ^6 n^6-84 \pi ^4 n^4+6510 \pi ^2 n^2-1575\right)-21 m^5 n^{11} \left(88 \pi ^4 n^4+2820 \pi ^2 n^2+795\right) \nonumber \\ &&  +21 m^4 n^{12} \left(26 \pi ^4 n^4+150 \pi ^2 n^2+915\right)+315 m^3 n^{13} \left(50 \pi ^2 n^2+57\right) \nonumber \\ &&  -105 m^2 n^{14} \left(46 \pi ^2 n^2+135\right)-2835 m n^{15}+2520 n^{16}
\end{eqnarray} }

For the 2 diagrams in Fig. \ref{S11-3}, we also consider them as the pastings of two diagrams which we separate by a dash line. The left parts of the diagrams have been computed before in the last two equations of (\ref{S81}) and in (\ref{eqS94}, \ref{eqS95}), so we can treat them as a black box and simply use the previous results. We note that in Section \ref{subsubcase3} we present the results for generic case $k\neq 0$, but here we also need to sum over the intermediate state with $k=0$ in the second diagram $S^{(11)}_{3,2}$, which we have calculated separately. The calculations go as the followings 
 \begin{eqnarray} 
S^{(11)}_{3,1} &=& \int_0^1 Jdx [S^{(8)}_4(m,x)+S^{(8)}_5(m,x)]  \bra \bar{O}^{xJ}_0 \bar{O}^{(1-x)J}_0 O^J_{-n,n} \ket,  \nonumber \\
S^{(11)}_{3,2} &=& \int_0^1 Jdx \sum_{k=-\infty}^{\infty} [S^{(9)}_4(m,k,x)+S^{(9)}_5(m,k,x)]  \bra \bar{O}^{xJ}_{-k,k} \bar{O}^{(1-x)J} O^J_{-n,n} \ket
\end{eqnarray}
We find the total contribution
\begin{eqnarray}  \label{eqS113}
S^{(11)}_3 &=& S^{(11)}_{3,1}+S^{(11)}_{3,2} \\  \nonumber 
&=& \left\{
\begin{array}{cl}
\frac{g^4}{280},    &   m=n=0;   \\
0,              &  m=0, n\neq0,   \\
& \textrm{or}~n=0, m\neq0; \\
g^4(-\frac{1045}{256 \pi ^8 m^8}-\frac{25}{48 \pi ^6 m^6}+\frac{7}{48 \pi ^4 m^4}-\frac{1}{210 \pi ^2 m^2}+\frac{37}{45360} ),   &  m=n\neq0; \\ & \\
\frac{g^4} {280 \pi ^2 m^2} ( -\frac{199815}{512 \pi ^6 m^6}-\frac{7665}{64 \pi ^4 m^4}+\frac{147}{8 \pi ^2 m^2}+1 ),  &  m=-n\neq0;   \\ & \\
\frac{g^4P_3}{3360 \pi ^8 m^6 n^6 (m-n)^8 (m+n)^4 } & \textrm{all~other~cases} 
\end{array}
\right.
\end{eqnarray} 
where the numerator in the last case is  
{\scriptsize
\begin{eqnarray} 
P_3 &=& 2 m^{16} \left(24 \pi ^6 n^6+133 \pi ^4 n^4-735 \pi ^2 n^2-1890\right)+m^{15} \left(-96 \pi ^6 n^7-896 \pi ^4 n^5+4970 \pi ^2 n^3+12705 n \right)\nonumber \\ && +m^{14} \left(-144 \pi ^6 n^8-308 \pi ^4 n^6+770 \pi ^2 n^4+1365 n^2\right)+3 m^{13} n^3 \left(128 \pi ^6 n^6+896 \pi ^4 n^4-6580 \pi ^2 n^2-15855\right) \nonumber \\ && +m^{12} n^4 \left(96 \pi ^6 n^6-1162 \pi ^4 n^4+11970 \pi ^2 n^2+31815\right)+m^{11} n^5 \left(-576 \pi ^6 n^6-1792 \pi ^4 n^4+36190 \pi ^2 n^2+72345\right) \nonumber \\ && +m^{10} n^6 \left(96 \pi ^6 n^6+2408 \pi ^4 n^4-11270 \pi ^2 n^2-75075\right)+m^9 n^7 \left(384 \pi ^6 n^6-1792 \pi ^4 n^4-42840 \pi ^2 n^2-68565\right) \nonumber \\ && -2 m^8 n^8 \left(72 \pi ^6 n^6+581 \pi ^4 n^4+5635 \pi ^2 n^2+12285\right)+m^7 n^9 \left(-96 \pi ^6 n^6+2688 \pi ^4 n^4+36190 \pi ^2 n^2-68565\right) \nonumber \\ && +m^6 n^{10} \left(48 \pi ^6 n^6-308 \pi ^4 n^4+11970 \pi ^2 n^2-75075\right)-7 m^5 n^{11} \left(128 \pi ^4 n^4+2820 \pi ^2 n^2-10335\right) \nonumber \\ && +7 m^4 n^{12} \left(38 \pi ^4 n^4+110 \pi ^2 n^2+4545\right)+35 m^3 n^{13} \left(142 \pi ^2 n^2-1359\right) \nonumber \\ && -105 m^2 n^{14} \left(14 \pi ^2 n^2-13\right)+12705 m n^{15}-3780 n^{16}
\end{eqnarray} }

We check the factorization relation for the 3 groups of string diagrams in a computer 
\begin{eqnarray}
S^{(11)}_i=\sum_{j=1}^{21} m_{ij}F^{(11)}_j, ~~~\textrm{for} ~i=1,2,3 
\end{eqnarray}
where the results of $S^{(11)}_i$  are written in equations (\ref{eqS11-1}, \ref{eqS112}, \ref{eqS113}), the multiplicity matrix $m_{ij}$ can be found in Table \ref{genus2multiplicity}, and we have also computed the $F^{(11)}_j$ ($j=1,2,\cdots, 21$) in computer according to the formula (\ref{integral2}) but there are too many expressions (21 of them) to write down here.  Again we can write the total contributions to the genus 2 correlator as 
\begin{eqnarray}
\bra \bar{O}_{-m,m}^J O_{-n,n}^J \ket_{\textrm{genus 2}} = \sum_{j=1}^{21} F^{(11)}_j = \frac{1}{24} \sum_{i=1}^3 S^{(11)}_i 
\end{eqnarray}

\subsection{Genus three: a test}

We consider the BMN correlator $\bra \bar{O}^J_{-m,m} O^J_{-n,n}\ket_{\textrm{genus~} 3}$. There are $\frac{11!!}{7}=1485$ different field diagrams represented by permutations of $(1,2,\cdots,12)$. We calculate these 1485 diagrams in computer using the formula (\ref{integral2}) similarly as in the previous section. Denoting the contributions as $F^{(12)}_j$, $j=1,2,\cdots 1485$, the total contribution to the correlator is 
\begin{eqnarray} 
&& \bra \bar{O}^J_{-m,m} O^J_{-n,n}\ket_{\textrm{genus} ~3} = \sum_{j=1}^{1485} F^{(12)}_j   \\ \nonumber   
&=& \left\{
\begin{array}{cl}
\frac{g^6}{322560},    &   m=n=0;   \\
0,              &  m=0, n\neq0,   \\
& \textrm{or}~n=0, m\neq0; \\
\frac{g^6}{518918400 }(\frac{8856072225}{256 \pi ^{12} m^{12}}-\frac{10877691825}{128 \pi ^{10} m^{10}}+\frac{1949592645}{64 \pi ^8 m^8}     &  m=n\neq0; \\
-\frac{26042445}{8 \pi ^6 m^6} +\frac{927355}{8 \pi ^4 m^4}-\frac{5239}{4 \pi ^2 m^2}+251 ), & \\
& \\
\frac{g^6} {215040 \pi ^2 m^2} ( -\frac{2807805}{512 \pi ^{10} m^{10}}-\frac{35315}{16 \pi ^8 m^8}+\frac{155281}{32 \pi ^6 m^6}  &  m=-n\neq0;   \\
-\frac{5461}{8 \pi ^4 m^4}+\frac{4151}{216 \pi ^2 m^2}+1),  & \\
& \\
\frac{g^6P_4(m,n) }{ m^{10} n^{10} (m-n)^{12} (m+n)^8 } & \textrm{all~other~cases} 
\end{array}
\right.
\end{eqnarray} 
where $P_4(m,n)$ is a polynomial of $m,n$ which is too long to write down here.

\begin{figure}
  \begin{center}
  \includegraphics[width=6.5in]{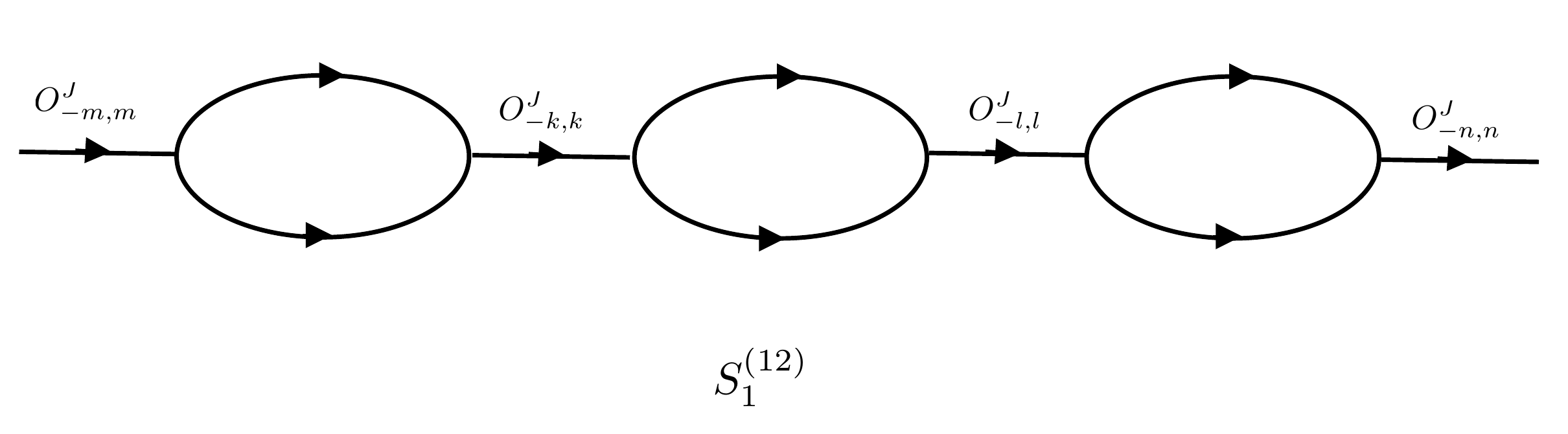} 
\end{center}
\caption{A 3-loop string diagram for $\bra \bar{O}^J_{-m,m} O^J_{-n,n}\ket_{\textrm{genus} ~3}$. We denote its contribution by $S^{(12)}_1$. We test the factorization relation for this diagram. }
\label{S12eps}
\end{figure}

We test the factorization relation for a 3-loop string diagrams shown in Fig. \ref{S12eps}. This diagram is one of simplest among 3-loop string diagrams and can be calculated as the following 
\begin{eqnarray}
S^{(12)}_1 &=& 8 \sum_{k=-\infty}^{+\infty}  \sum_{l=-\infty}^{+\infty}  \bra \bar{O}^J_{-m,m} O^J_{-k,k} \ket_{\textrm{torus}} \bra \bar{O}^J_{-k,k} O^J_{-l,l} \ket_{\textrm{torus}}   \bra \bar{O}^J_{-l,l} O^J_{-n,n} \ket_{\textrm{torus}} \nonumber \\
 &=& 2 \sum_{k=-\infty}^{+\infty}   \bra \bar{O}^J_{-m,m} O^J_{-k,k} \ket_{\textrm{torus}}  ~S^{(11)}_1(k,n),
\end{eqnarray} 
where the formula for the torus two point function can be found in (\ref{F6torus}), and we can utilize the previous result $S^{(11)}_1(k,n)$ in (\ref{eqS11-1}) of two-loop string propagation for parts of the calculations.  

We also use a computer to find the multiplicities of the string diagram in Fig. \ref{S12eps} with respect to the 1485 field theory diagrams, similarly as in the previous case of genus 2.  It turns out the multiplicity is non-vanishing with respect to all 1485 diagrams. Obviously we can not list all the multiplicities here. We provide a small sample in Table \ref{genus3multiplicity}.

\begin{table} 
\begin{center}
\begin{tabular}{| c | c |  }
 \hline 
 $j=1,2,\cdots ,1485$  & $ m_{1j}$ for $S^{(12)}_1 $   \\ \hline
(1,4,7,6,5,8,11,10,9,12,3,2) & 48 \\  \hline
 (1,4,7,11,10,8,6,5,9,12,3,2) & 72 \\  \hline
 (1,4,7,11,9,6,5,10,8,12,3,2) & 96 \\  \hline
 (1,4,7,10,6,5,11,9,8,12,3,2) & 72 \\  \hline
 (1,4,7,10,9,8,11,6,5,12,3,2) & 48 \\  \hline
 (1,4,11,10,5,8,7,6,9,12,3,2) & 48 \\  \hline
 (1,4,11,9,7,6,10,5,8,12,3,2) & 72 \\  \hline
 (1,4,10,7,6,11,9,5,8,12,3,2) & 96 \\  \hline
 (1,4,10,9,5,8,11,7,6,12,3,2) & 72 \\  \hline
 (1,4,11,6,9,8,7,10,5,12,3,2)  & 48 \\  \hline
 $\cdots$  & $\cdots$ \\ 
 $\cdots$  & $\cdots$ \\ \hline
 Total  &  277200 \\ \hline
 \end{tabular} 
 \bigskip
  \caption{The multiplicities of the string diagram in Fig. \ref{S12eps} with respect to some samples of the 1485 short processes, which are permutations of $(1,2,\cdots ,12)$, and we have used the cyclicality of the string to put the segment (1) in the first position. }
 \label{genus3multiplicity}
 \end{center}
 \end{table}

Denoting the contribution of a field theory diagram by $F^{(12)}_j$ where $j=1,2,\cdots ,1485$, the factorization relation for the string diagram in Fig. \ref{S12eps} states that 
\begin{eqnarray} \label{S12F12}
S^{(12)}_1 = \sum_{j=1}^{1485} m_{1j} F^{(12)}_j
\end{eqnarray}
We calculate both the left hand side and the right hand side analytically, and check the factorization relation with the following result 
\begin{eqnarray} 
&&  S^{(12)}_1 = \sum_{j=1}^{1485} m_{1j} F^{(12)}_j  \\ \nonumber   
&=& \left\{
\begin{array}{cl}
\frac{g^6}{1728},    &   m=n=0;   \\
0,              &  m=0, n\neq0,   \\
& \textrm{or}~n=0, m\neq0; \\
\frac{g^6}{926640 }(\frac{73345119705}{1024 \pi ^{12} m^{12}}-\frac{1955422755}{64 \pi ^{10} m^{10}}+\frac{397910799}{64 \pi ^8 m^8}-\frac{33182721}{56 \pi ^6 m^6}    &  m=n\neq0; \\
+\frac{176891}{8 \pi ^4 m^4}-\frac{5109}{28 \pi ^2 m^2}+83 ), & \\
& \\
\frac{g^6} {1152 \pi ^2 m^2} ( \frac{90781119}{1024 \pi ^{10} m^{10}}-\frac{4409493}{128 \pi ^8 m^8}+\frac{486819}{80 \pi ^6 m^6}   &  m=-n\neq0;   \\
-\frac{147149}{280 \pi ^4 m^4}+\frac{34583}{2520 \pi ^2 m^2}+1 ),  & \\
& \\
\frac{g^6P_5(m,n) }{ m^{10} n^{10} (m-n)^{12} (m+n)^8 } & \textrm{all~other~cases} 
\end{array}
\right.
\end{eqnarray} 
where $P_5(m,n)$ is a polynomial of $m,n$ too long to write down here.

The sums and integrals in both the string diagrams and the field theory diagrams become more and more difficult to do analytically as we go up in genus and also include multi-trace operators. But it is certainly possible to check the factorization relation further numerically since all sums and integrals are convergent in this paper.

\section{Correlators of BMN operators with more stringy modes }  \label{othertype}

In the previous sections we considered correlators of BMN operators with at most two excitations, where the first stringy mode can appear due to the closed string level matching condition. One can certainly add more stringy modes to the BMN operators, which corresponds to more field insertions in the trace operators with phases. One can also consider the case that some of the scalar insertions are identical, which we do not expect to make a qualitative change to the factorization rules. To illustrate that the factorization relation also works for these cases, in this section we study some correlators involving  BMN operators with 3 different scalar insertions. 

\subsection{The operator and vertices}

We use 3 different scalar fields $\phi^1$, $\phi^2$ and $\phi^3$ to insert into the single trace operator $O^J=\Tr (Z^J)$ with phases. The resulting properly normalized BMN operator is 
\begin{eqnarray}  
O^{J}_{(m_1,m_2,m_3)} = \frac{1}{\sqrt{N^{J+2}}J} \sum_{l_1, l_2=0}^{J-1}  e^{\frac{2\pi im_2l_1}{J}} e^{\frac{2\pi im_3l_2}{J}}  \Tr(\phi^1 Z^{l_1} \phi^2 Z^{l_2-l_1} \phi^3 Z^{J-l_2}),  \label{operator3}
\end{eqnarray}
where the integers $m_i$'s satisfy the level matching condition $m_1+m_2+m_3=0$, and we have used the cyclicality of the trace to put the scalar $\phi^1$ in the first positions. Similar to the case of 2 excitations, the summing over the position of $\phi^1$ make the operator vanish if the level matching condition $m_1+m_2+m_3=0$ is not satisfied. From now on we use a subscript to denote the string modes when confusion may arise. For example, we denote the BMN operator with 2 excited modes as $O^J_{(-n,n)_{(1,3)}}$ with the scalar insertions of modes $-n$ and $n$ from $\phi^1$ and $\phi^3$. For the BMN operator with 3 string modes this is not necessary since there is no confusion.

It is straightforward to compute the vertices with the operator (\ref{operator3}) by summing over the scalar insertions into diagram in Fig. \ref{3point} with phases. We find the vertices 
\begin{eqnarray} \label{vertices3-1}
 \bra \bar{O}^J_{(m_1,m_2,m_3)} O^{xJ}_{(-n,n)_{(1,2)}} O^{(1-x)J}_{(0)_3} \ket
&=& - \frac{g}{\sqrt{J}} x^{\frac{3}{2}} \frac{\sin(\pi m_1 x) \sin (\pi m_2 x)\sin(\pi m_3 x)}{\pi^3m_3 (m_1x+n)(m_2x-n)} \nonumber \\   \nonumber \\ 
\bra \bar{O}^J_{(m_1,m_2,m_3)} O^{xJ}_{(n_1,n_2,n_3)} O^{(1-x)J} \ket
&=&  \frac{g}{\sqrt{J}} x^2 (1-x)^{\frac{1}{2}} \frac{\sin(\pi m_1 x) \sin (\pi m_2 x) \sin(\pi m_3 x)}{\pi^3 (m_1x-n_1)(m_2x-n_2)(m_3x-n_3) } \nonumber \\
\end{eqnarray}
The above correlators are valid as long as the denominator is not zero. For the special cases when the denominator vanishes, we have the following correlators  
\begin{eqnarray} \label{vertices3-2}
 \bra \bar{O}^J_{(-m,m, 0)} O^{xJ}_{(-n,n)_{(1,2)}} O^{(1-x)J}_{(0)_3} \ket
&=& \frac{g}{\sqrt{J}} x^{\frac{3}{2}}(1-x) \frac{\sin ^2(\pi m x) }{\pi^2 (mx-n)^2} \nonumber \\  
 \bra \bar{O}^J_{(0,-m, m)} O^{xJ}_{(0,0)_{(1,2)}} O^{(1-x)J}_{(0)_3} \ket
&=& - \frac{g}{\sqrt{J}} x^{\frac{1}{2}} \frac{\sin ^2(\pi m x) }{\pi^2 m^2} \nonumber \\  
 \bra \bar{O}^J_{(0,0, 0)} O^{xJ}_{(0,0)_{(1,2)}} O^{(1-x)J}_{(0)_3} \ket
&=&  \frac{g}{\sqrt{J}} x^{\frac{3}{2}}(1-x)  \nonumber \\  
 \bra \bar{O}^J_{(-m,m, 0)} O^{xJ}_{(-n,n,0)} O^{(1-x)J} \ket
&=& \frac{g}{\sqrt{J}} x^2 (1-x)^{\frac{1}{2}} \frac{\sin^2 (\pi m x) }{\pi^2 (mx-n)^2} \nonumber \\  
 \bra \bar{O}^J_{(0,0, 0)} O^{xJ}_{(0,0,0)} O^{(1-x)J} \ket
&=& \frac{g}{\sqrt{J}} x^2 (1-x)^{\frac{1}{2}} 
\end{eqnarray}

\subsection{The case of $\bra \bar{O}^J_{(m_1,m_2,m_3)} O^{x_1J}_{(0)_1} O^{x_2J}_{(0)_2} O^{x_3J}_{(0)_3} \ket$}

\begin{figure}
  \begin{center}
  \includegraphics[width=6.5in]{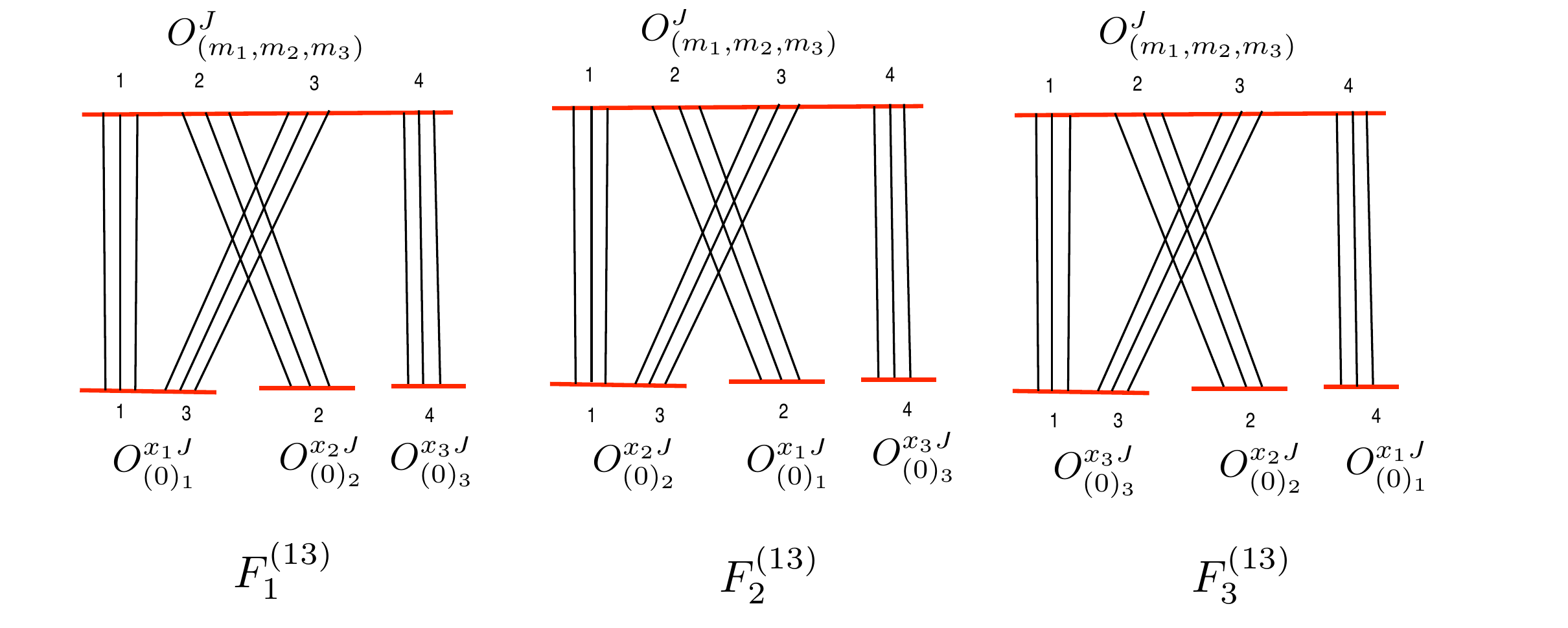} 
\end{center}
\caption{ The field theory diagrams for the correlators $\bra \bar{O}^J_{(m_1,m_2,m_3)} O^{x_1J}_{(0)_1} O^{x_2J}_{(0)_2} O^{x_3J}_{(0)_3} \ket$. We denote the contributions $F^{(13)}_1$, $F^{(13)}_2$ and $F^{(13)}_3$. }
\label{F13eps}
\end{figure}

We study a simple case of the correlator $\bra \bar{O}^J_{(m_1,m_2,m_3)} O^{x_1J}_{(0)_1} O^{x_2J}_{(0)_2} O^{x_3J}_{(0)_3} \ket$ to illustrate the factorization relation with more than 2 stringy mode excitations. It is implicit that the parameters satisfy  $m_1+m_2+m_3=0$ and $x_1+x_2+x_3=1$. The field theory diagrams are basically the same as the case of 2 string modes in Figs. \ref{F1}, \ref{F2} and we draw them in Fig. \ref{F13eps}.

We calculate the contributions by summing over the 3 scalar insertions with phases. 
\begin{eqnarray}
F^{(13)}_1 &=& \frac{g^2}{J} \int_0^{x_1} dy_1 (\int_0^{y_1}+\int _{y_1+x_2}^{x_1+x_2})dy_2e^{-2\pi i m_1 y_2}
\int_{y_1}^{y_1+x_2} dy_3 e^{-2\pi i  m_2 y_3}\int_{x_1+x_2}^1 dy_4 e^{-2\pi i m_3y_4} \nonumber \\
&=& \frac{\sin(m_2x_2\pi)\sin(m_3x_3\pi)}{2\pi^4m_1m_2^2m_3^2} \{ m_1\cos[\pi(m_2x_2-m_3x_3)] \nonumber \\ && +(-1)^{m_2}m_3 \cos[\pi(m_1x_3+m_2x_1)]+(-1)^{m_3}m_2\cos[\pi (m_1x_2+m_3x_1)] \} \nonumber \\
F^{(13)}_2 &=& F^{(13)}_1 (x_1 \leftrightarrow x_2, m_1\leftrightarrow m_2) \nonumber \\
F^{(13)}_3 &=& F^{(13)}_1 (x_1 \leftrightarrow x_3, m_1\leftrightarrow m_3)
 \end{eqnarray}

\begin{figure}
  \begin{center}
  \includegraphics[width=6.5in]{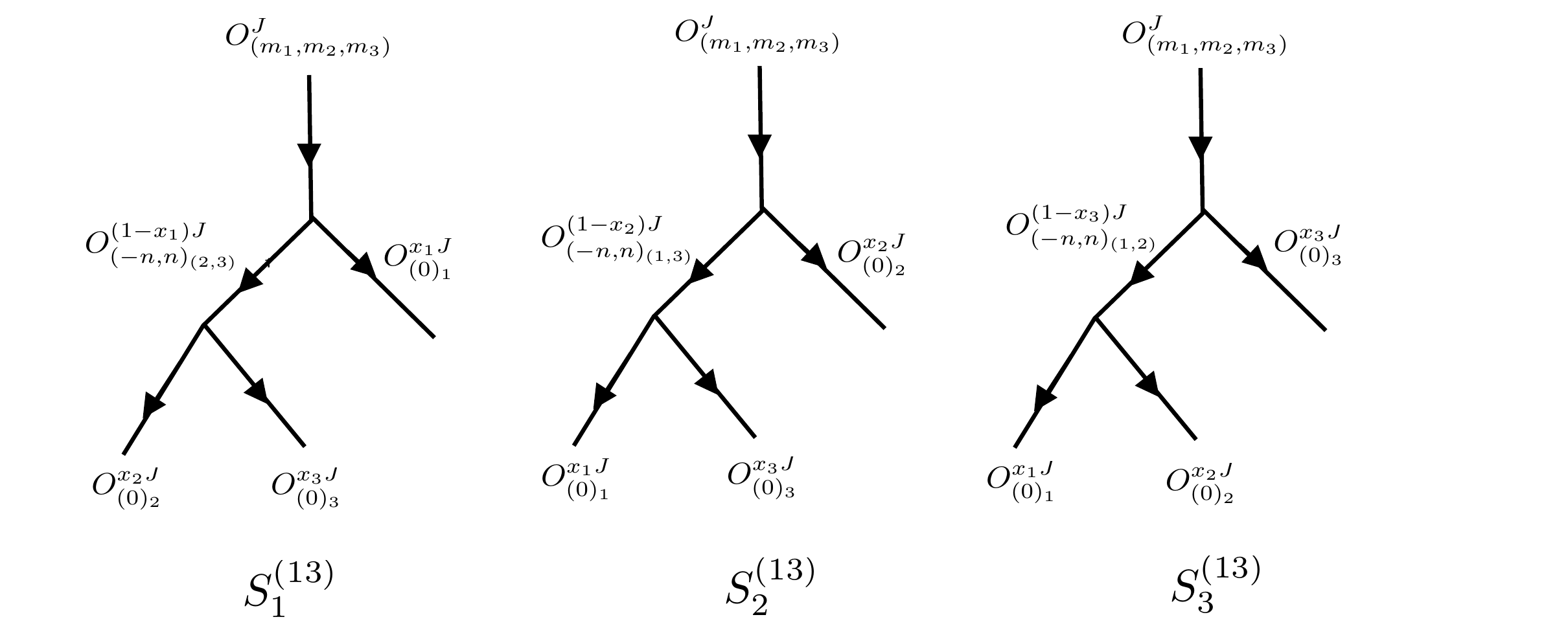} 
  \end{center}
\caption{ The string diagrams for the correlators $\bra \bar{O}^J_{(m_1,m_2,m_3)} O^{x_1J}_{(0)_1} O^{x_2J}_{(0)_2} O^{x_3J}_{(0)_3} \ket$. We denote the contributions $S^{(13)}_1$, $S^{(13)}_2$ and $S^{(13)}_3$. }
\label{S13eps}
\end{figure}

For the string diagrams, we draw them in Fig. \ref{S13eps}. we calculate the diagrams using the vertices and summing over intermediate states 
\begin{eqnarray}
S^{(13)}_1 &=& \sum_{n=-\infty}^{\infty}  \bra \bar{O}^J_{(m_1,m_2,m_3)} O^{(1-x_1)J}_{(-n,n)_{(2,3)}} O^{x_1J}_{(0)_1} \ket \bra \bar{O}^{(1-x_1)J}_{(-n,n)_{(2,3)}}  O^{x_2J}_{(0)_2} O^{x_3J}_{(0)_3}\ket  \nonumber \\
S^{(13)}_2 &=& S^{(13)}_1 (x_1 \leftrightarrow x_2, m_1\leftrightarrow m_2) \nonumber \\
S^{(13)}_3 &=& S^{(13)}_1 (x_1 \leftrightarrow x_3, m_1\leftrightarrow m_3)
\end{eqnarray}
We perform the sum and check the factorization relation 
\begin{eqnarray}
S^{(13)}_1 &=& F^{(13)}_2+F^{(13)}_3 \nonumber \\
S^{(13)}_2 &=& F^{(13)}_1+F^{(13)}_3 \nonumber \\
S^{(13)}_3 &=& F^{(13)}_1+F^{(13)}_2 
\end{eqnarray}

\subsection{One-loop string propagation}

We study one more example of $\bra \bar{O}^J_{(m_1,m_2,m_3)} O^J_{(n_1,n_2,n_3)} \ket_{\textrm{torus}}$ where it is implicit that $m_1+m_2+m_3=n_1+n_2+n_3=0$ due to the closed string level matching condition. First we consider the generic case that none of  $m_i$, $n_i$, $m_i-n_j$, $m_i+n_j$ ($i,j=1,2,3$) is zero.  There is only one field theory diagram as depicted in Fig. \ref{F6}. We sum over 3 scalar insertions into the diagram with phases
\begin{eqnarray} \label{integralF14}
&& \bra \bar{O}^J_{(m_1,m_2,m_3)} O^J_{(n_1,n_2,n_3)} \ket_{\textrm{torus}} \\ \nonumber
&=& \int_0^1 dx_1dx_2dx_3dx_4 \delta(x_1+x_2+x_3+x_4-1) \int_0^{x_1} dy_3 e^{2\pi i (n_3-m_3)y_3} \times \\ \nonumber 
&&  \prod_{i=1}^2  (\int_0^{x_1} +e^{2\pi i n_i (x_3+x_4)}\int_{x_1}^{x_1+x_2}+e^{2\pi i n_i(x_4-x_2)}\int_{x_1+x_2}^{1-x_4}+e^{-2\pi i n_i(x_2+x_3)}\int_{1-x_4}^1) dy_i e^{2\pi i (n_i-m_i)y_i }
\end{eqnarray}  
This is a 7-dimensional integral. The integration variables $x_1, x_2, x_3, x_4$ are the lengths of 4 segments in the single trace operator, and the integration variables $y_1, y_2, y_3$ are the positions of the scalar insertion where we have used the cyclic symmetry to put $y_3$ in the first segment. The integration variables $y_1, y_2, y_3$ further divide the 4 segments into 7 segments and the integral can be reduced into sums of the standard integrals (\ref{integral1}) but it is more complicated than the case of 2 scalar insertions. We find the expression in terms of the standard integral (\ref{integral1}) as the followings 
\begin{eqnarray}
F^{(14)}\equiv \bra \bar{O}^J_{(m_1,m_2,m_3)} O^J_{(n_1,n_2,n_3)} \ket_{\textrm{torus}} = F^{(14)}_1+F^{(14)}_2+F^{(14)}_3+F^{(14)}_4,
\end{eqnarray}
where 
\begin{eqnarray} \label{F14}
 F^{(14)}_1 &\equiv & \sum_{i\neq j} I_{(1,1,5)}(2\pi i (m_i-n_i), -2\pi i (m_j-n_j),0) \nonumber \\
 F^{(14)}_2 &\equiv & \sum_{i\neq j} I_{(1,2,2,2)}(2\pi i (m_i-n_i), -2\pi i (m_j-n_j),-2\pi i m_j, 0)  \nonumber \\ && + 
 I_{(1,2,2,2)}(2\pi i (m_i-n_i), -2\pi i (m_j-n_j),2\pi i n_j, 0)  \nonumber \\
 F^{(14)}_3 &\equiv & \sum_{i\neq j} I_{(1,1,1,2,2)}(2\pi i (m_i-n_i), -2\pi i m_j, 2\pi i n_j ,-2\pi i (m_j-n_j),0) \nonumber \\
 F^{(14)}_4 &\equiv & \sum_{i\neq j} I_{(1,1,1,1,1,1,1)}(2\pi i m_i, 2\pi i n_i, -2\pi i m_j, -2\pi i n_j  \nonumber \\
&&  , 2\pi i (m_i-n_j) ,-2\pi i (m_j-n_i),0)  
\end{eqnarray} 
For the generic case of $m_i, n_i$, it turns out that $F^{(14)}_2=F^{(14)}_3=F^{(14)}_4=0$. There seems to be some hidden symmetries which are not obvious the integral expression (\ref{integral1}).  The contribution vanishes for each term in $F^{(14)}_4$, but only the total contributions vanish in the cases $F^{(14)}_2$ and $F^{(14)}_3$. So the only non-vanishing contribution is $F^{(14)}_1$ and we find the correlator
\begin{eqnarray} \label{F14result}
 \bra \bar{O}^J_{(m_1,m_2,m_3)} O^J_{(n_1,n_2,n_3)} \ket_{\textrm{torus}}   
= \frac{\sum_{i=1}^3(m_i-n_i)^2    }{32\pi^4 \prod_{i=1}^3 (m_i-n_i)^2}
\end{eqnarray}

\begin{figure}
  \begin{center}
  \includegraphics[width=6.5in]{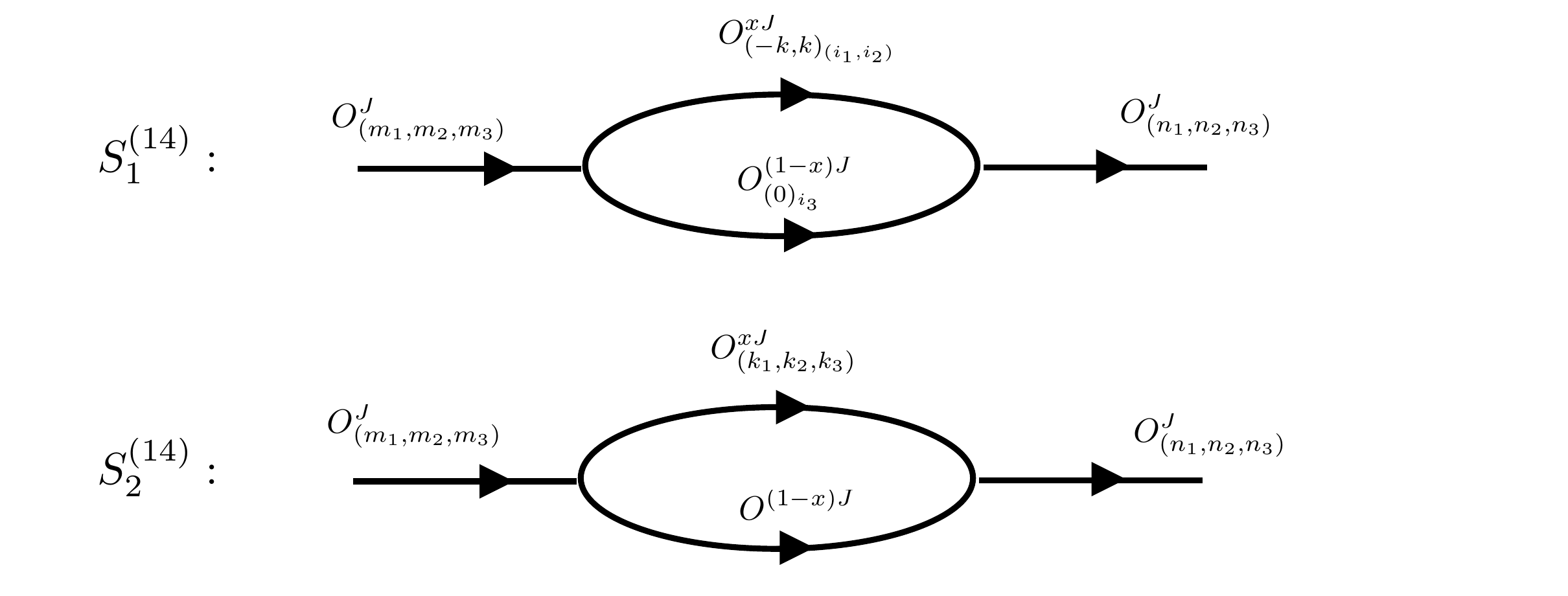} 
  \end{center}
\caption{ The string diagrams for the correlators $\bra \bar{O}^J_{(m_1,m_2,m_3)} O^J_{(n_1,n_2,n_3)} \ket_{\textrm{torus}}$, where there are 3 cases for the first diagrams $(i_1,i_2,i_3) = (1,2,3), (2,3,1), (3,1,2)$. We denote the contributions of these 2 diagrams $S^{(14)}_1$ and $S^{(14)}_2$. }
\label{S14pdf}
\end{figure}

We consider the string diagrams. The 2 diagrams are drawn in Fig. \ref{S14pdf} and we denote the contributions $S^{(14)}_1$ and $S^{(14)}_2$. The computations are carried out by summing over the intermediate states 
\begin{eqnarray} 
S^{(14)}_1 &=& \sum_{i_3=1}^3 \int_0^1 Jdx \sum_{k=-\infty}^{\infty} \bra \bar{O}^J_{(m_1,m_2,m_3)} O^{xJ}_{(-k,k)_{(i_1,i_2)}} O^{(1-x)J}_{(0)_{i_3}} \ket 
\bra \bar{O}^{xJ}_{(-k,k)_{(i_1,i_2)}} \bar{O}^{(1-x)J}_{(0)_{i_3}} O^J_{(n_1,n_2,n_3)} \ket  \nonumber \\
S^{(14)}_2 &=& \int_0^1 Jdx \sum_{k_1+k_2+k_3=0}  \bra \bar{O}^J_{(m_1,m_2,m_3)} O^{xJ}_{(k_1,k_2,k_3)} O^{(1-x)J} \ket 
\bra \bar{O}^{xJ}_{(k_1,k_2,k_3)} \bar{O}^{(1-x)J} O^J_{(n_1,n_2,n_3)} \ket \nonumber \\
\end{eqnarray} 
We perform the the sums and integrals for the contributions. It turns out for the generic case of $m_i,n_i$, the first diagram vanishes $S^{(14)}_1=0$. The vanishing is due to an antisymmetry $x\rightarrow 1-x$ of the integrand, which is not present at the vertex level, but only appears after summing over the string modes $k$ of the intermediate states. So the total contribution  $S^{(14)}\equiv S^{(14)}_1 + S^{(14)}_2$ only come from the second diagram $ S^{(14)}_2$. We do the calculations and check the factorization relation with the field theory contribution (\ref{F14result}),
\begin{eqnarray}
S^{(14)}= 2 \bra \bar{O}^J_{(m_1,m_2,m_3)} O^J_{(n_1,n_2,n_3)} \ket_{\textrm{torus}}  = \frac{\sum_{i=1}^3(m_i-n_i)^2    }{16\pi^4 \prod_{i=1}^3 (m_i-n_i)^2}
\end{eqnarray}

\

The formulae (\ref{F14result}) for the correlator is valid for the generic case when the arguments in the equations (\ref{F14}) are not degenerate. When some arguments are identical, we need to combine them according to (\ref{combine}) before we can use (\ref{integral3}, \ref{integral4}) to compute them. It can be easily checked that the degeneracy only happens when some of  the $m_i$, $n_i$, $m_i-n_j$, $m_i+n_j$ ($i,j=1,2,3$) vanish. We discuss these various cases in the followings. Needless to say, one can check that the factorization relation $S^{(14)}=2F^{(14)}$ is fulfilled for all these cases.

\begin{enumerate}
\item $m_i=0$, $n_i\neq 0$ or $m_i\neq 0$, $n_i=0$ for some $i\in \{1,2,3\}$. We find the correlator  vanish $F^{(14)}=0$ regardless whether there are further degeneracies in the other parameters. The vanishing can be directly seen from the integral (\ref{integralF14}). For example, if $n_1=0$ and $m_1\neq 0$, then one integral contributes a factor $\int_0^1 dy_1e^{-2\pi i m_1y_1}=0$. For the string diagrams, we find both $S^{(14)}_1$ and $S^{(14)}_2$ no longer vanish but their contributions cancel each others. 

\item $m_i=n_i=0$ for  some $i\in \{1,2,3\}$. The correlator reduces to the case of correlator with 2 scalar insertions (\ref{F6torus}) studied before, since there is no phase factor in summing over the scalar insertion $\phi^i$ and it contributes just a constant factor which is properly cancelled. 

\item $n_3=m_3$ (without loss of generality) and everything else generic. The level matching conditions are $m_2=-m_1-m_3$, $n_2=-n_1-n_3$, and we can express the correlator using 3 parameters $m_1,n_1,m_3$. In this case we find all $F^{(14)}_i$ ($i=1,2,3,4$) in (\ref{F14}) are non-vanishing, and the answer looks more complicated than the generic case 
\begin{eqnarray} \label{F14n3m3}
F^{(14)}  &=& \frac{1}{48\pi^2 (m_1-n_1)^2} +\frac{1}{16\pi^4(m_1-n_1)^4}  \\ && 
-\frac{m_3^4+(m_1+n_1 )m_3^3  +m_1n_1 m_3^2 -m_1n_1(m_1+n_1) m_3 -m_1^2 n_1^2 }{16\pi^4 (m_1-n_1)^2 m_1n_1m_3^2(m_3+m_1)(m_3+n_1)} \nonumber \\ &&
+ \frac{1}{16\pi^4 (m_1-n_1)^2 m_1^2n_1^2m_3^2(m_3+m_1)^2(m_3+n_1)^2}\{  m_3^6 (m_1^2+n_1^2)
\nonumber \\ && +2 m_3^5 (m_1^2 n_1+m_1 n_1^2+m_1^3+n_1^3)+m_3^4 (4 m_1^3 n_1+2
   m_1^2 n_1^2+4 m_1 n_1^3+m_1^4+n_1^4)\nonumber \\ &&
   +2 m_3^3 m_1n_1 (m_1^3+n_1^3)-8  m_3^2m_1^3 n_1^3-4
    m_3 m_1^3 n_1^3 (m_1+n_1)-2 m_1^4 n_1^4
   \} \nonumber 
\end{eqnarray}  
 
\item $n_3=m_2$ (without loss of generality) and everything else generic.  It turns out in this case the generic formula (\ref{F14result}) is still valid even though some arguments in (\ref{F14}) are degenerate. One can simply plug in the parameters with $n_3=m_2$. 

\item $n_3= -m_3$ (without loss of generality) and everything else  generic.  It turns out in this case the generic formula (\ref{F14result}) is also still valid even though some arguments in (\ref{F14}) are degenerate.

\item $n_3= -m_2$ (without loss of generality) and everything else  generic. The correlator is different from the generic formula. We find  
\begin{eqnarray} \label{F14n3-m2}
F^{(14)} = \frac{2m_1^2-3m_1n_1+2n_1^2}{16\pi^4m_1^2n_1^2(m_1-n_1)^2}
\end{eqnarray}
We note that there are 3 free parameters after taking into account the level matching conditions, but the correlator only depends on 2 parameters. 

\item $n_3=m_3$, $n_2=m_2$  (without loss of generality) and everything else  generic. The level match conditions also require $n_1=m_1$. We find the correlator 
\begin{eqnarray}
F^{(14)} = \frac{1}{120} +\frac{5}{16\pi^4 }(\frac{1}{m_1^4}+\frac{1}{m_2^4}+\frac{1}{m_3^4})-\frac{(m_1^2+m_2^2+m_3^2)^2}{192\pi^2m_1^2m_2^2m_3^2} 
\end{eqnarray}

\item \label{case8} $n_3=m_3$, $n_2=m_1$  (without loss of generality) and everything else generic. The level match conditions also require $n_1=m_2$. We find the correlator 
\begin{eqnarray}
F^{(14)} &=& \frac{1}{48 \pi ^4 m_1^2 n_1^2
   (m_1-n_1)^4 (m_1+n_1)^2} 
   [9 (m_1^6+n_1^6)  -12m_1n_1(m_1^4+n_1^4) \nonumber \\ && 
   -9m_1^2n_1^2(m_1^2+n_1^2)  +36m_1^3n_1^3 -\pi^2 m_1^2n_1^2(m_1-n_1)^2 ]
   \end{eqnarray}
We note that  we would get an incorrect result by simply setting $m_3=-m_1-m_2=-m_1-n_1$ in (\ref{F14n3m3}), though there is no apparent singularity in doing so. In other words, the condition  $n_2=m_1$ generates more degeneracies and further modifies the result.  

\item $n_3=m_2$, $n_2=m_1$  (without loss of generality) and everything else  generic. The level match conditions also require $n_1=m_3$. In this case the generic formula (\ref{F14result}) is valid.

\item $n_3=-m_3$, $n_2=-m_2$  (without loss of generality) and everything else  generic. The level match conditions also require $n_1=-m_1$. In this case the generic formula (\ref{F14result}) is valid. 

\item $n_3=-m_2$, $n_2=-m_3$  (without loss of generality) and everything else  generic. The level match conditions also force $n_1=-m_1$. In this case the generic formula (\ref{F14result}) is not valid, and we find 
\begin{eqnarray}
F^{(14)} = \frac{15}{64\pi^4m_1^4}
\end{eqnarray}
We note that this can not be obtained from the more generic formula (\ref{F14n3-m2}) for the case of $n_3=-m_2$ by further setting $n_1=-m_1$.

\item $n_3=-m_2$, $n_2=-m_1$  (without loss of generality) and everything else  generic. The level match conditions also force $n_1=-m_3$. In this case the generic formula (\ref{F14result}) is not valid, and we find 
\begin{eqnarray}
F^{(14)} = \frac{m_1^2-m_1n_1+n_1^2}{4 \pi^4m_1^2n_1^2 (m_1-n_1)^2}
\end{eqnarray}
This is also different from the more generic formula (\ref{F14n3-m2}).

\item $n_3=m_3$, $n_2=-m_2$ (without loss of generality) and everything else  generic. In this case the correlator can be obtained from the case of $n_3=m_3$ discussed above by further setting   $n_2=-m_2$ in (\ref{F14n3m3}). In other words, the condition  $n_2=-m_2$ does not further change the correlator through degeneracies. 

\item $n_3=m_3$, $n_2=-m_1$ (without loss of generality) and everything else  generic. In this case the correlator can not be obtained from the previous case of $n_3=m_3$ by further setting $n_2=-m_1$.  In other words, the condition  $n_2=-m_1$ changes the correlator through more degeneracies in the integral.   We find the correlator
\begin{eqnarray}
F^{(14)} = \frac{3m_1^6 + 3m_1^2m_3^4 + 6m_3^6 + \pi^2 m_1^2m_3^2(m_1^2-m_3^2)^2  }{48
   \pi ^4 m_1^2 m_3^4 (m_1^2-m_3^2)^2}
\end{eqnarray}    

\item $n_3= -m_3$, $n_2= m_1$ (without loss of generality) and everything else  generic. In this case the generic formula (\ref{F14result}) is valid. 

\item $n_3= - m_2$, $n_2= m_3$ or $n_2=m_1$ (without loss of generality) and everything else  generic. In this case the correlator can be obtained from the more generic previous case of $n_3= - m_2$  (\ref{F14n3-m2}). In fact, the condition  $n_2= m_3$ or $n_2=m_1$ does not change the correlator at all, since the formula (\ref{F14n3-m2}) only depends on two parameters $m_1$ and $n_1$.

\end{enumerate}

One general pattern in these discussions is that the degeneracies $m_i=-n_i$ and $m_i=n_j$ ($i\neq j$) are more benign than the other cases of $m_i=n_i$ and $m_i= -n_j$ ($i\neq j$) , and one can often obtain the correlator by directly plugging these benign conditions in the formula of a more generic case. There is only  one exception to this pattern encountered in case \ref{case8} where the condition $n_2=m_1$ does modify the formula for the correlator from a more generic situation. 

It is also possible to derive the factorization relation $S^{(14)}=2F^{(14)}$ using the integral form of the vertices as we did in Section \ref{integralsection} for BMN operators with 2 stringy modes. This would be much more complicated than the previous case but the derivation would apply to all the degenerate cases without the need to discuss each case separately.

\section{Conclusion}

In this paper we check the factorization relation (\ref{factorization}) in many examples where the initial state is a single string state. However, we expect the factorization to also work for certain string diagrams in the cases where both initial and final states are multi-string states, so long as no string in the intermediate steps of the string diagram is longer than all the external initial and final strings in terms of the number of $Z$ fields in the corresponding trace operator. This is supported by the study of tree level  $2\rightarrow 2$ process in Sec \ref{22section}, where we find the factorization works for $T,U$ channels, but fails for the $S$ channel because the string propagating in the $S$ channel is the longest string.

It is well known that in flat Minkowski space, the string amplitude at $h$-loop level goes like $(2h)! g_s^{2h}$ for large $h$, while the Yang-Mills field theory amplitude goes like $h! g_{YM}^{2h}$, where $g_s$ and $g_{YM}$ are the coupling constants of the string theory and field theory. The perturbation theory is divergent but is Borel summable. There is an ambiguity in performing the Borel summation of asymptotic series, which is of the order $e^{-A/g_s}$ for string theory and $e^{-A/g_{YM}^2}$, where $A$ represents a positive number. These ambiguities come from non-perturbative effects not captured by the perturbation theory, and they come from D-branes in the case of string theory and instantons in the case of gauge theory. In our case, the effective coupling constant is $g=\frac{J^2}{N}$. At genus $h$ level, there are $\frac{(4h-1)!!}{2h+1}$ field theory diagrams \cite{HZ}. For the vacuum operator each diagram contributes $1/(4h)!$, so the perturbation series is actually convergent and can be summed up
\begin{eqnarray}
\bra \bar{O}^J O^J \ket = \sum_{h=0}^{\infty} \frac{(4h-1)!!}{(2h+1) (4h)!} g^{2h}= \frac{2\sinh(g/2)}{g}
\end{eqnarray} 
We do not expect a qualitative change to convergence property for the correlators of general stringy BMN operators. To explain the convergence, we note that in the free field limit $\lambda^\prime = 0$, we effectively ``zoom in" an infinitesimal patch of the spacetime where the corresponding string theory lives, so that the spacetime becomes infinitely curved and the strings are infinitely long. We conjecture that in this limit we have decoupled the D-branes and their non-perturbative effects, so the string perturbation theory is complete and convergent.

\appendix 
\section{Some useful summation formulae} \label{summationappendix}

Some useful summation formulae for many of the calculations of string diagrams is 
\begin{eqnarray} \label{summation1}
\sum_{p=-\infty}^{\infty}\frac{1}{(p-\alpha_1)(p-\alpha_2)} &=& -\pi \frac{\cot(\alpha_1\pi)-\cot(\alpha_2\pi)}{\alpha_1-\alpha_2} \nonumber \\
\sum_{p=-\infty}^{\infty}\frac{\sin^2(p\pi\beta)}{(p-\alpha_1)(p-\alpha_2)} &=& 
\frac{\pi}{(\alpha_1-\alpha_2)}[\frac{\sin(\alpha_1\pi(1-\beta))\sin(\alpha_1\pi\beta)}{\sin(\alpha_1\pi)}
\nonumber \\ && -\frac{\sin(\alpha_2\pi(1-\beta))\sin(\alpha_2\pi\beta)}{\sin(\alpha_2\pi)}]   \\
\sum_{p=-\infty}^{\infty}\frac{\sin(2p\pi\beta)}{(p-\alpha_1)(p-\alpha_2)} &=& 
\frac{\pi}{(\alpha_1-\alpha_2)}[\frac{\sin(\alpha_1\pi(1-2\beta))}{\sin(\alpha_1\pi)}
 -\frac{\sin(\alpha_2\pi(1-2\beta))}{\sin(\alpha_2\pi)}]   \nonumber
\end{eqnarray}
Here we assume $\alpha_1, \alpha_2$ are not integers, and $0<\beta<1$ in the second and third equations. Since the series is absolute convergent, we can take derivative with respect to $\alpha_1$ or $\alpha_2$ and generate more formula with higher power of $p-\alpha_i$ in the denominator. Sometime we need to take the limit where one of the $\alpha_i$'s is an integer, in this case the summation formulae are still valid but we need to exclude $p=\alpha_i$ in the summation on both sides carefully. We can also subtract the formulae with each others to generate summation formulae with more factors in the denominator. For example, we can see 
\begin{eqnarray}
\frac{1}{(p-\alpha_1)(p-\alpha_2)(p-\alpha_3)}=\frac{1}{(\alpha_2-\alpha_3)}[ \frac{1}{(p-\alpha_1)(p-\alpha_2)}  -\frac{1}{(p-\alpha_1)(p-\alpha_3)}],
\end{eqnarray} 
which can generate a summation formula with 3 factors in the denominator from formulae in (\ref{summation1}).

The formulae in (\ref{summation1}) can be also thought of as coming from the following simpler formulae
\begin{eqnarray}
\sum_{p=-\infty}^{\infty}\frac{1}{p-\alpha} &=& -\pi  \cot(\alpha \pi) \nonumber \\
\sum_{p=-\infty}^{\infty}\frac{e^{2\pi i p\beta}}{p-\alpha} &=& -\pi  \frac{e^{-\pi i \alpha(1-2\beta)}}{\sin(\alpha \pi)}, ~~~(0<\beta<1)
\end{eqnarray}
However, the sums in these formulae are not by themselves convergent, though they are Borel summable. They should be only thought of as ``seed formulae" for formal manipulations to generate convergent summation formulae such as (\ref{summation1}). All the sums in the string diagrams are absolute convergent without the need for regularization.

Another useful formula is about the Dirac delta function,  
\begin{eqnarray} \label{deltafunction}
\sum_{p=-\infty}^{\infty} e^{2\pi ipx}= \sum_{k=-\infty}^{+\infty}\delta(x-k) 
\end{eqnarray}
This is special case of the Poisson resummation formula.  This formula would be useful for performing the sum over intermediate string states in string diagrams with the integral form of the 3-string vertex.

\section{Field theory calculations for $\bra \bar{O}^J_{-m,m} O^J_{-n,n} \ket_h$  } \label{2point}

Here we recapitulate the methods in \cite{Constable1} for computing free field correlator $\bra \bar{O}^J_{-m,m} O^J_{-n,n} \ket$ at genus $h$. At genus $h$ there are $\frac{(4h-1)!!}{2h+1}$ cyclically different diagrams \cite{HZ}. Each diagram can be represented by a permutation $\sigma: (1,2,\cdots ,4h)\rightarrow (\sigma(1), \sigma(2),\cdots, \sigma(4h))$. In our terminology this is an irreducible short process, extendable into long processes and string diagrams.  

To compute the contributions of a field theory diagram, one defines the following standardized integral
\begin{eqnarray} \label{integral1}
I(u_1,u_2,\cdots,u_r) \equiv \int_0^1  dx_1\cdots dx_r \delta(x_1+\cdots+x_r-1) e^{ u_1x_1+\cdots u_rx_r}
\end{eqnarray} 
It is clear that the integral is unchanged if we add an integer multiple of $2\pi i$ to all the arguments. If some of the $u_i$'s are identical, one uses the following notation 
\begin{eqnarray} \label{combine}
I_{(a_1,\cdots,a_r)} (u_1,u_2,\cdots ,u_r)\equiv I(u_1,\cdots, u_1, u_2,\cdots ,u_2, \cdots ,u_r,\cdots ,u_r), 
\end{eqnarray}
where $a_i$'s are integers representing the numbers of the $u_i$'s in the right hand side, and for $a_i=0$ we can just eliminate the corresponding argument. The integral can be calculated by the following recursion relation  
\begin{eqnarray} \label{recursive}
&& (u_i-u_j) I_{(a_1,\cdots,a_r)} (u_1,u_2,\cdots ,u_r)  \nonumber \\
&=& I_{(a_1,\cdots,a_j-1,\cdots, a_r)} (u_1,u_2,\cdots ,u_r) -I_{(a_1,\cdots,a_i-1,\cdots, a_r)} (u_1,u_2,\cdots ,u_r)  , 
\end{eqnarray}
If $u_i\neq u_j$ then this equation can be used to reduce the number of arguments, but the relation is also valid and both sides are zero when $u_i=u_j$. From the recursion relation one can obtain the formulae for the integral 
\begin{eqnarray} \label{integral3}
I(u_1,u_2,\cdots u_r) &=& \sum_{i=1}^r e^{u_i} \prod_{j\neq i} (u_i-u_j)^{-1}, \\
 \label{integral4}
I_{(a_1+1,\cdots,a_r+1)}(u_1,\cdots,u_r) &=& \prod_{i=1}^r \frac{(\partial / \partial u_i)^{a_i}}{a_i!} I(u_1,\cdots,u_r),
\end{eqnarray}
where the $u_i$'s are different.  

Now the contribution of a field theory diagram of permutation $\sigma\in S_{4h}$ can be expressed in terms of the integrals (\ref{integral1}). First one adds a fixed point $4h+1$ to the permutation $\sigma$ to obtain another permutation $\tilde{\sigma}\in S_{4h+1}$, and for $1\leq i\leq 4h+1$ one defines the  following numbers 
\begin{eqnarray}
ll_i(\sigma) &=&  \textrm{number of } \{j ~|~ j<i, \tilde{\sigma}(j)<\tilde{\sigma}(i) \},   \nonumber \\
lr_i(\sigma)  &=&  \textrm{number of } \{j ~|~ j<i, \tilde{\sigma}(j)>\tilde{\sigma}(i) \},  \nonumber \\
rl_i(\sigma)  &=&  \textrm{number of } \{j ~|~ j>i, \tilde{\sigma}(j)<\tilde{\sigma}(i) \},  \nonumber \\
rr_i(\sigma)  &=&  \textrm{number of } \{j ~|~ j>i, \tilde{\sigma}(j)>\tilde{\sigma}(i) \},  \nonumber 
\end{eqnarray} 
then the contribution of a field theory diagram represented by permutation $\sigma$ to the correlator $\bra \bar{O}^J_{-m,m} O^J_{-n,n} \ket_h$ can be expressed as 
\begin{eqnarray} \label{integral2}
F_\sigma(m,n) = \sum_{i=1}^{4h+1} I_{(ll_i(\sigma)+1, lr_i(\sigma), rl_i(\sigma), rr_i(\sigma)+1)}(2\pi i (m-n), 2\pi i m,-2\pi i n,0)
\end{eqnarray}
To understand this formula, we note that since $ll_i(\sigma)+lr_i(\sigma)+rl_i(\sigma)+rr_i(\sigma)=4h$, this is a $4h+2$ dimensional integral with $r=4h+2$ in terms of (\ref{integral1}).  The integration variables come from the division of the single trace into $4h$ segments, and also there are two scalar insertions in the BMN operators which further add 2 integration variables when we sum over the positions of the scalar insertions. The exponential factor in (\ref{integral1}) corresponds to the BMN phases in the operators.  For single traces operators, we can use the cyclicality to fix one scalar insertion in one of the $4h$ segments, then the other scalar insertion can run in any of the resulting $4h+1$ segments, generating a sum over $4h+1$ terms in (\ref{integral2}). If one of operators are multi-trace, we can no longer use the cyclicality and the integrals would be more complicated.

\end{document}